\documentclass{aa}  
\usepackage{graphicx,xcolor,natbib,hyperref}
\usepackage{epstopdf}
\usepackage{txfonts}
\usepackage{ulem} 
\usepackage{soul}
\usepackage{hyperref}
\usepackage{longtable}
\usepackage{lscape}
\usepackage{caption}
\usepackage{subcaption}
\usepackage{listings}

\definecolor{dkgreen}{rgb}{0,0.6,0}
\definecolor{gray}{rgb}{0.5,0.5,0.5}
\definecolor{mauve}{rgb}{0.58,0,0.82}

\newcommand\gaia{\textit{Gaia}\xspace}

\newcommand\msun{M_\odot\xspace}
\newcommand{\orcid}[1]{\protect\href{https://orcid.org/#1}{\protect\includegraphics[width=8pt]{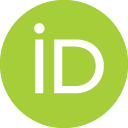}}}

\begin{document} 

   \title{$^{12}$C/$^{13}$C of \textit{Kepler} giant stars: \\
  the missing piece of the mixing puzzle\thanks{Result of Table~\ref{observations_table} are available at the CDS via anonymous ftp to cdsarc.u-strasbg.fr}}

   \author{
  Lagarde, N.\inst{1}\orcid{0000-0003-0108-3859}, Minkevi{\v c}i{\= u}t{\. e}, R. \inst{2}, Drazdauskas, A. \inst{2}, Tautvai{\v s}ien{\. e}, G. \inst{2}, Charbonnel, C. \inst{3,4}, Reyl\'e,  C.\inst{5}, \\ Miglio, A.\inst{6,7}, Kushwahaa, T. \inst{8} and Bale, B.\inst{2}
          }
\institute{Laboratoire d'Astrophysique de Bordeaux, Universit\'e Bordeaux, CNRS, B18N, All\'ee Geoffroy Saint-Hilaire, 33615 Pessac, France 
	\email{nadege.lagarde@u-bordeaux.fr}
\and
Astronomical Observatory, Institute of Theoretical Physics and Astronomy, 
Vilnius University, \\ Sauletekio av. 3, 10257 Vilnius, Lithuania
\and
Department of Astronomy, University of Geneva, Chemin de P\'egase 51, 1290 Versoix, Switzerland 
\and 
IRAP, UMR 5277 CNRS and Universit\'e de Toulouse, 14, Av. E.Belin, 31400 Toulouse, France
\and
Institut UTINAM, CNRS UMR 6213, Univ. Franche-Comt\'e, OSU THETA Franche-Comt\'e-Bourgogne, Observatoire de Besan\c con, BP 1615, 25010 Besan\c con Cedex, France
\and 
Dipartimento di Fisica e Astronomia, Università degli Studi di Bologna, Via Gobetti 93/2, I-40129 Bologna, Italy
\and
INAF – Osservatorio di Astrofisica e Scienza dello Spazio di Bologna, Via Gobetti 93/3, I-40129 Bologna, Italy
\and
Cardiff Hub for Astrophysics Research and Technology (CHART), School of Physics \& Astronomy, Cardiff University, The Parade, CF24 3AA Cardiff, UK
            }
    
\date{Accepted for publication on december, 2023}
\authorrunning{Lagarde, N et al.} \titlerunning{$^{12}$C/$^{13}$C of \textit{Kepler} giant stars}
 
  \abstract 
   {Despite a rich observational background, few spectroscopic studies have dealt with the measurement of the carbon isotopic ratio in giant stars. However, it is a key element in understanding the mixing mechanisms that occur in the interiors of giant stars.}     
   { We present the CNO and $^{12}$C/$^{13}$C abundances derived for 71 giant field stars. Then, using this new catalogue and complementary data from the \textit{Kepler} and \gaia satellites, we study the efficiency of mixing occurring in the giant branch as a function of the stellar properties (e.g., mass, age, metallicity).}
   {We have determined the abundances of CNO and more specifically the carbon isotopic ratio using the high-resolution FIbre-fed Echelle Spectrograph on the Nordic Optical Telescope, for 71 giant field stars. In addition, asteroseismology from \textit{Kepler} satellite is available for all stars, providing their stellar mass, age as well as the evolutionary states. Finally, astrometry from \gaia data is also available for the majority of the sample. We compare these new determinations with stellar evolution models taking into account the effects of transport processes. To exploit the complete potential of our extensive catalogue and considering both the Milky Way’s evolution and the impact of stellar evolution, we built mock catalogues using the Besan\c con Galaxy model in which stellar evolution models taking into account the effects of thermohaline instability are included.}
   {We confirm that the carbon isotopic ratio at the surface of core He-burning stars is lower than that of first ascent RGB stars. The carbon isotopic ratio measured at the surface of the core He-burning stars increases with [Fe/H] and stellar mass while it decreases with stellar age. These trends are all very well explained by the thermohaline mixing that occurs in red giants. We have shown that our models can explain the behaviour of $^{12}$C/$^{13}$C versus N/O, although the observations seem to show a lower N/O than the models. We also note that more constraints on the thick disc core He-burning stars are needed to understand this difference.} 
   {Overall, the current model including thermohaline mixing is able to reproduce very well the $^{12}$C/$^{13}$C with the stellar metallicity as well as with the stellar mass and age.} 
  

   \maketitle
%

\section{Introduction} 

Mechanisms of formation and evolution of our Galaxy are strongly encoded into kinematics, chemistry, and ages of its stars. The chemical elements synthesised in stellar interiors and ejected through stellar winds and late-stage ejection mechanisms like explosions add new ingredients to the interstellar medium (ISM) for the next stellar generations, driving the chemical evolution of the Milky Way. 
The so-called stellar yields vary as a function of the nucleosynthesis paths stars of different initial masses and metallicities go through, of the mass loss efficiency  along their life, as well as of how much the ejected layers are contaminated by the nucleosynthesis products, which strongly depends on the way chemicals are transported (mixed) in stellar interiors.
As a matter of fact, stars rarely exhibit their initial chemical compositions, because of different types of in situ (magneto-) hydrodynamical mixing processes that can modify their internal and surface chemical properties along their nucleosynthetic life as well as the main stellar parameters and lifetimes \citep[e.g., ][]{Vauclair78,Charbonnel95,Palacios03,Talon06,Lagarde12a,Maeder14,Pignatari16,LiCh18,Deal20,Bouret21,Dumont21,Guerco22}. 
Understanding these mixing processes is a key for stellar physics and Galactic archeology, with unique and complementary constraints coming from the large Galactic astrometric, spectroscopic, and asteroseismic surveys that are presently revolutionizing our knowledge of the links between Galactic and stellar evolution  \citep[e.g.,][]{Miglio21,Lagarde21}. 

High-precision astrometric data, including trigonometric parallaxes as well as abundances of certain chemical elements for nearly 5.6 million stars provided by Gaia \citep{GaiaEDR3, RecioBlanco23}, offer exclusive information on the position of stars in the Hertzsprung-Russel diagram (HRD) and on their membership to different Galactic populations. Spectroscopic surveys provide chemical abundances and radial velocities for stars in different populations in our Galaxy (e.g., Gaia-ESO survey \citealt{GilmoreGES22,RandichGES22}, APOGEE \citealt{APOGEE}, GALAH \citealt{GALAHDR3}). 
Finally, asteroseismology provides key information on the internal structure of the stars, hence on their age, evolutionary stage and mass, through the detection of rich oscillation spectra \citep[including non-radial oscillation modes, e.g.,][]{Baglin06,Borucki10,ChMi13}.  
For the first time, we have the ability to determine: (1) very precise stellar gravity (typically one order of magnitude more precise than what is currently achievable by high resolution spectroscopy of bright stars); (2) precise estimates of stellar masses and radii of giant stars, which can be used to infer precise distances for field stars; (3) more accurate (model-dependent) age estimates of giant stars even several kilo parsec away from the sun. The major spectroscopic surveys have taken care to observe the stars in the asteroseismic surveys, for example with the APOKASC consortium pointing to the \textit{Kepler} stars observed by APOGEE \citep{Pinsonneault18}, CoRoGEE \citep[CoRoT+APOGEE,][]{Anders17a,Anders17b}, RAVE+K2 \citep{Valentini19}, CoRoGES \citep[CoRoT+Gaia-ESO,][]{Valentini16}, or APO-K2 \citep{APOK2}. For both (and related) stellar and galactic objectives, we can thus start to use golden stellar samples for which astrometric, spectroscopic, and seismic data can be combined for stars of different masses, metallicities, and evolutionary stages \citep[e.g.,][]{Lagarde15,Lagarde21,Montalban21}. \\

In this study, we built for the first time a golden sample to probe the mixing processes that have been revealed to occur in low- and intermediate-mass red-giant stars (RGB) by previous spectroscopic observations of light elements such as lithium, carbon, and nitrogen in samples without asteroseismic coverage \citep[e.g.,][]{ChBa00,Smiljanic09,Smiljanic10,ChaLag20,Mikolaitis10,Tautvaisiene13,Lagarde15,Takeda19,Magrini21}. We focus on the best chemical indicator to constrain the mixing efficiency in giant stars, namely the carbon isotopic ratio \citep[e.g.,][]{ Dearborn76,BrWa89,SmSu89,Bell90,GiBr91,Gratton00, Charbonnel98,Tautvaisiene10,Tautvaisiene13,Morel14,Takeda19, McCormick23,AguileraGomez23}. To the best of our knowledge, only four giant field stars with both measured carbon isotopic ratios and asteroseismic diagnostics coming from CoRoT satellite are available in the literature \citep{Morel14}.
Previous spectroscopic studies have shown that the atmospheric value of this quantity first changes during the so-called first dredge-up (1DUP) on the subgiant branch, when the deepening convective envelope engulfs the $^{13}$C peak that was build inside the stars through CNO-cycle during the main sequence \citep[e.g.][]{Iben67,Charbonnel94}. The amplitude of the decrease of the $^{12}$C/$^{13}$C ratio then depends on the stellar mass, metallicity, as well as on rotation-induced mixing while the star was on the main sequence \citep[e.g., ][]{ChaLag10}. 
The surface $^{12}$C/$^{13}$C ratio drops again later, together with the Li abundance and the C/N ratio, when the stars move across the so-called RGB bump 
\citep[e.g.,][]{FusiPecci90, 
Charbonnel94}.
As of today, the only explanation for this second pattern that is not predicted in classical stellar evolution models is the so-called thermohaline instability or double-diffusive mixing process \citep{ChaZah07a}.
Simplified prescriptions for this mechanism \citep{Kippen80} in 1D stellar evolution models explain the observed behaviour of the carbon isotopic ratio together with that of Li, C, and N in field and open cluster red-giant stars with various masses and metallicities \citep[][]{ChaLag10,Lagarde12a}. Some other 1D models challenge its ability to explain simultaneously the behaviour of these three indicators, in particular in low-metallicity globular cluster red giants \citep[e.g.][but see also \citealt{Fraser22}]{Angelou11,Henkel17,Tayar22}. Additionally, multi-D hydrodynamical models points towards lower efficiency of this mechanism in stellar interiors than expected to explain the observations \citep[e.g.,][]{Denissenkov10,DenissenkovMerryfield10,Traxleretal11,Brown13}, while in specific conditions other effects induced by rotation and magnetic field might significantly enhance the fingering transport rate \citep{SenGar18,Harrington19,Fraser23}.  
Since the thermohaline instability is also expected to significantly  lower the amount of $^{3}$He released by low- and intermediate-mass stars as required by Galactic evolution models \citep[][see also \citealt{Eggleton06}]{ChaZah07a,ChaZah07b,Lagarde11,Lagarde12b}, it is fundamental to better understand if, how, and by how much this mechanism actually modifies the surface composition of red giants and their chemical yields. \\

Our study deals with both observational and theoretical aspects. First, we present the carbon isotopic ratios and CNO abundances we determine from our spectroscopic observations of 71 giant stars previously observed by the \textit{Kepler} satellite during its long 4-year run. The asteroseismic properties (such as $\Delta\nu$, $\nu_{max}$ and $\Delta\Pi_{\ell=1}$) of these objects are also available, allowing a precise determination of their mass, radius, age and evolutionary stage. This sample is therefore the largest sample of stars for which surface properties from spectroscopy including the carbon isotopic ratio and internal properties from asteroseismology are available. Thanks to the \gaia satellite, the kinematics of these stars is also accessible, making it the best 
characterised RGB sample so far. 
Then, we compare our observations and additional data from the literature to the predictions of stellar evolution models computed with the code  STAREVOL \citep[e.g.,][]{Lagarde12a,Amard19}, taking into account the thermohaline instability. To do the comparison, we use the  Besan\c con Galaxy model (hereafter BGM), a state-of-the-art Galactic stellar population synthesis model \citep{Lagarde17}, to compute mock catalogues that we statistically compare to our data. \\

The paper is structured as follows. In Sect.~\ref{obs} we present the spectroscopic and atmospheric characterisation for our sample stars. In Sect.~\ref{combined} we present complementary properties of our sample coming from the \textit{Kepler} and \gaia satellites and we gather literature data to probe the behaviour of the carbon isotopic ratio over the largest [Fe/H] and mass range possible. In Sect.~\ref{BGMvsObs}, we discuss the efficiency of extra-mixing on the red giant branch comparing our observations qith theoretical trends provide by BGM simulations described in Sect/~\ref{section:BGM}. Conclusions are presented in Sect.~\ref{conclu}.


\section{Observations and spectral analysis} 
\label{obs}

To construct our golden sample, we chose to observe field stars at two different evolutionary phases: during the central helium burning phase and on the RGB at a lower luminosity than the RGB bump (see Sect.~\ref{sismo}).  

\subsection{Observations}

We performed two observing runs using the high-resolution FIbre-fed Echelle Spectrograph \citep[FIES]{Telting2014} on the Nordic Optical Telescope (NOT). We used the medium resolution configuration ($R=46000$), which provided a large continuous wavelength range from about 4100 to 8500~\AA. It contains the forbidden [O\,{\sc i}] line, C$_2$ bands, strong $^{12}\rm C^{14}\rm N$ features, and the $^{13}\rm C^{14}\rm N$ line at 8004~\AA\, suitable to determine C/N and $^{12}\rm C/^{13}\rm C$ ratios (see Subsection~\ref{analysis}). 
The first run, from June 27 to July 2 2018, under the program P57-105, was dedicated to the observation of 41 stars with masses <1.5$\msun$. On the second run, from July 25 to 28, 2019, under the program P59-103, we completed our observations targeting more massive (1.5--2.2\,$\msun$) and low-metallicity stars ([Fe/H]<$-0.2$). The observed targets are listed in Table~\ref{observations_table}. The exposure times were chosen to get a signal to noise ratio equal to 80 or higher. We performed the data reduction with the dedicated software (FIEStool) available at the telescope. FIEStool provides a fully reduced spectrum using the nightly calibration frames.


\subsection{Spectral analysis}
\label{analysis}

In this work, we used an equivalent widths method for the determination of the main stellar atmospheric parameters and a spectral synthesis approach for the determination of abundances of chemical elements. Both analysis techniques make use of the one-dimensional, plane-parallel, the local thermodynamical equilibrium (LTE) model stellar atmospheres MARCS \citep{Gustafsson2008}, the Gaia-ESO linelist (\citealt{Heiter2015}), and Solar abundances by \citet{Grevesse07}.


\subsubsection{Atmospheric parameters}

\begin{figure}
  \resizebox{\hsize}{!}{\includegraphics{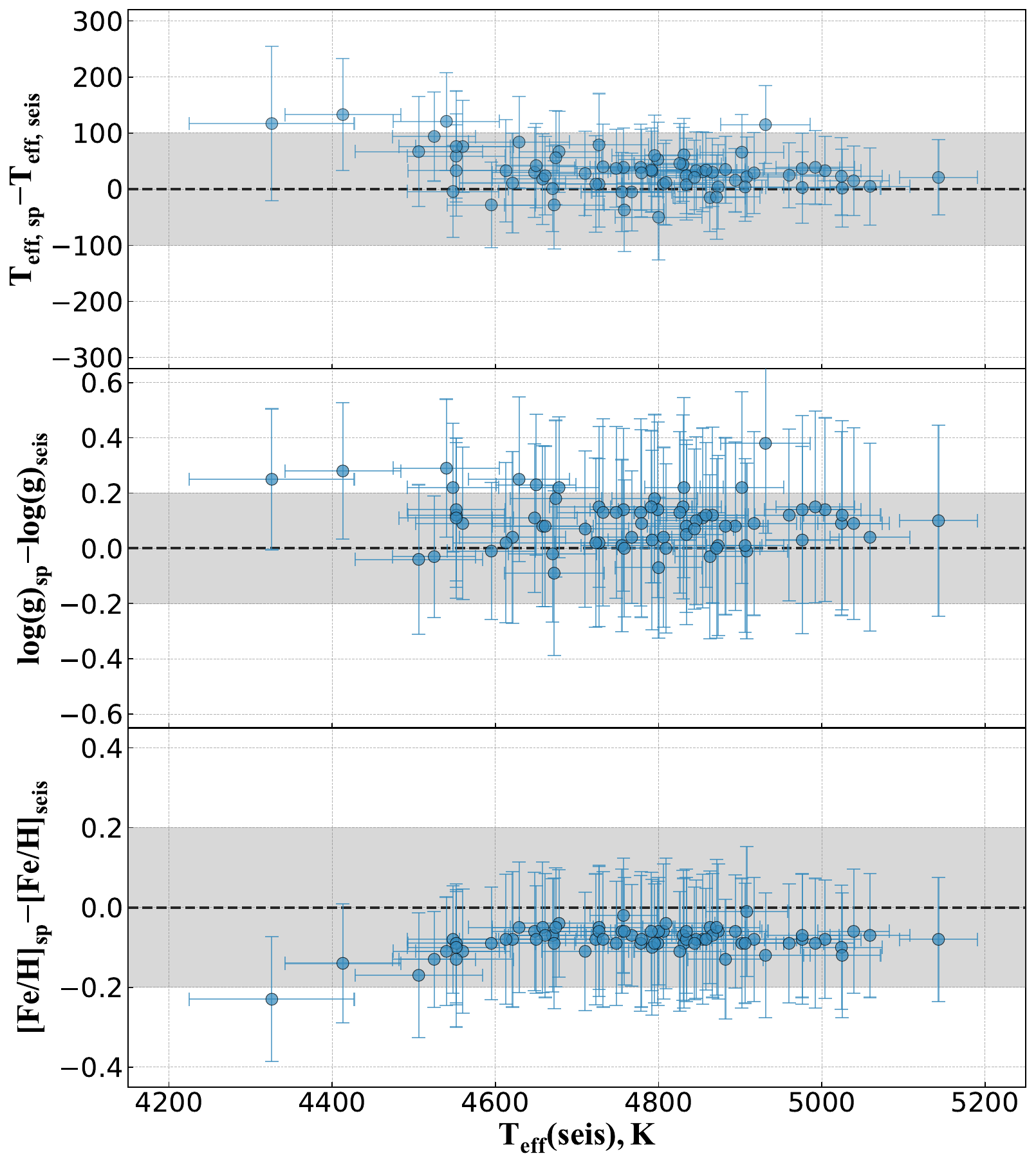}}
  \caption{Comparison of effective temperatures, surface gravities and metallicities determined during the first (spectroscopic) and the second (asteroseismic) steps of the parameter determination process.}
  \label{param_diff}
\end{figure}

For the determination of the main atmospheric parameters, we adopted a two-step approach. At first, we used the classical equivalent widths method which is based on the analysis of Fe\,{\sc i} and Fe\,{\sc ii} atomic lines. The equivalent widths were measured with a DAOSPEC code (\citealt{Stetson2008}), and then the abundances were calculated using a MOOG code (\citealt{Sneden1973}). The initial iron line-list included 112 Fe\,{\sc i} and 15 Fe\,{\sc ii} lines. The line-list was carefully compiled to avoid blended and strong lines (we discarded lines which were larger than 150\,m\AA\ in observed stars). On average 103 Fe\,{\sc i} and 13 Fe\,{\sc ii} lines remained and were used for the analysis. The available Fe\,{\sc ii} lines are usually scarce and weak in comparison to the Fe\,{\sc i}, and their measurements greatly influence the determination of log\,$g$. Surface gravities are considered to be more accurate when determined from asteroseismic properties \citep{Morel2012, Creevey2013}.
Thus, in the second step, we determined stellar surface gravities using the \textit{Kepler} asteroseismic data from \citet{Mosser12a} and the spectroscopic effective temperatures calculated in our work during the first step. Then we fixed the asteroseismic log\,{$g$} value and re-determined the other parameters to get the required balance of abundances from the Fe\,{\sc i} lines in relation to their excitation potential and equivalent widths. 
This way, we eliminated the weakest point of spectroscopic analysis which is the accurate log\,{$g$} determination. The difference between purely spectroscopic parameters and those which had the fixed asteroseismic log\,$g$ values are plotted in Fig.~\ref{param_diff}. No clear trends can be observed neither in the surface gravity, nor in the effective temperature comparisons. The average differences are $-33 \pm 36\, (\mathrm{K})$ for effective temperatures and $-0.10 \pm 0.09\, (\mathrm{dex})$ for the surface gravity.

\subsubsection{Abundance determination}
\label{abundance}

For the spectral analysis we used a differential model atmosphere technique. The spectral synthesis was performed for determination of carbon, nitrogen,  oxygen, and magnesium abundances, as well as in determining carbon isotopic ratios. The calculations were performed using a TURBOSPECTRUM code (\citealt{Alvarez1998}). For the carbon abundance determination we used two regions: ${\rm C}_2$ Swan (0,1) band head at 5135.5~{\AA} and ${\rm C}_2$ Swan (0, 1) band head at 5635.2~{\AA}. The $^{12}{\rm C}/^{13}{\rm C}$\ ratio 
was obtained from the $^{13}{\rm C}/^{12}{\rm N}$ feature at 8004.7~{\AA}. An interval of 6470-6490~{\AA} and of  7980–8005~{\AA}, with up to seven strong $^{12}{\rm C}/^{14}{\rm N}$
features was used to determine nitrogen abundances. The forbidden [O\,{\sc i}] line at 6300.31~\AA\ was used for the oxygen abundance derivation. This oxygen line is blended by \textsuperscript{58}Ni and \textsuperscript{60}Ni. The oscillator strength values for these two nickel lines were taken from \citet{Johansson03}. In several cases the oxygen line at 6300.31~\AA\ was contaminated by the telluric absorption line and could not be used for the investigation of oxygen abundance. As C and O abundances are bound together, in such situations where the oxygen abundance was unavailable, the magnesium abundance was used as a  proxy to determine carbon and nitrogen abundances. Spectral synthesis of three spectral lines at 5711.1, 6318.7, 6319.2~\AA\ was used for the Mg abundance determination.

\begin{figure}
\centering
\resizebox{\hsize}{!}{
\includegraphics[]{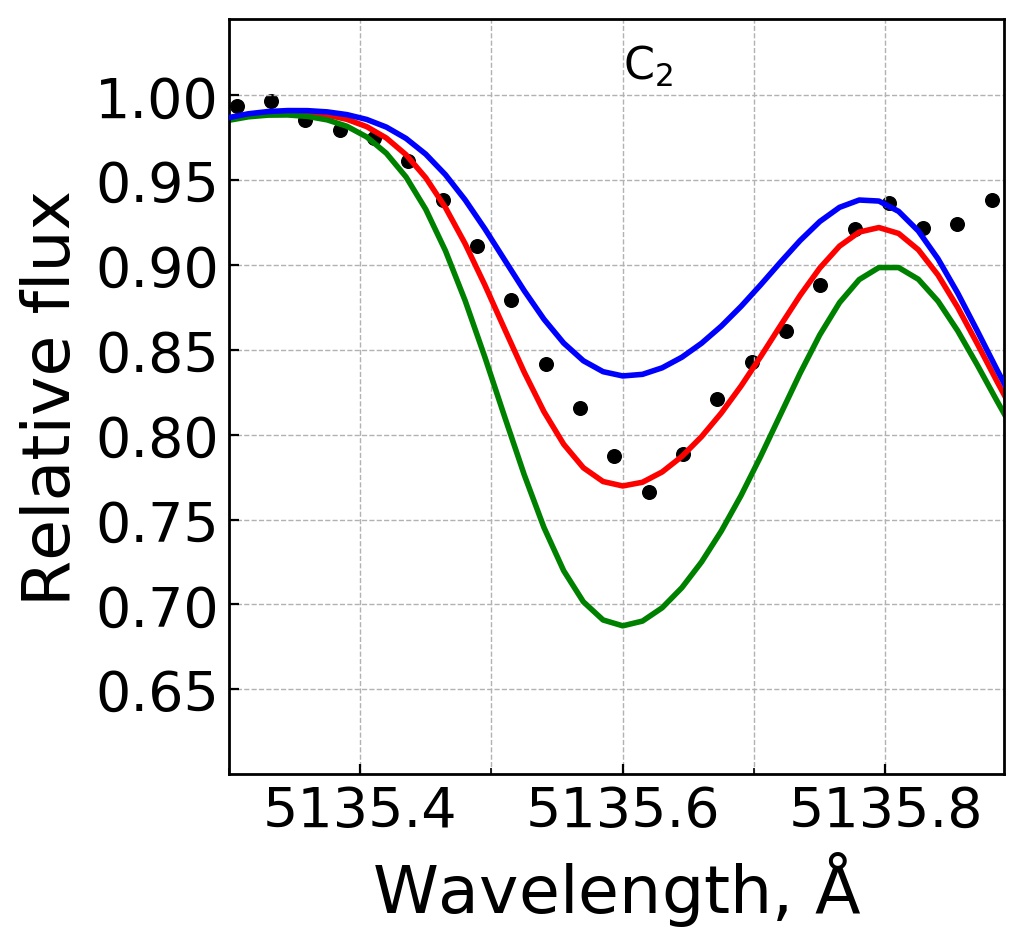}
\includegraphics[]{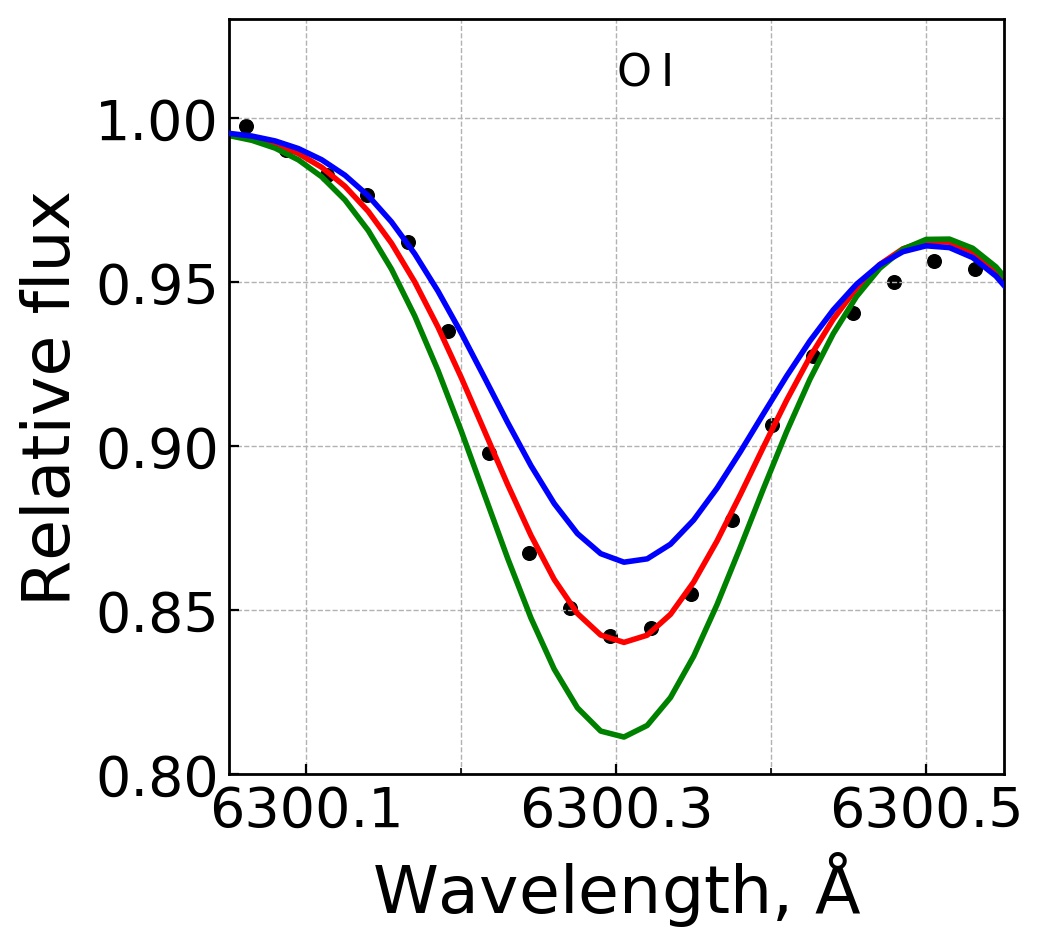}
}
\caption{Examples of fits to the carbon and oxygen lines in a spectrum of KIC4457200. The middle line represents the best fit for [C/H]=0.01 and [O/H]=0.12, whereas the other two lines indicate $\pm 0.10$~dex for the corresponding
abundances
\label{fig:fit_examples1}}
\end{figure}

\begin{figure}
\centering
\resizebox{\hsize}{!}{
\includegraphics[]{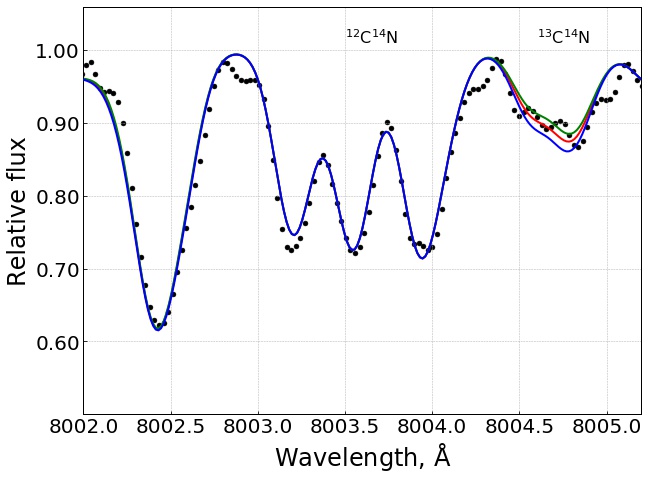}
}
\caption{Examples of fits to the $^{12}{\rm C}^{14}{\rm N}$ and $^{13}{\rm C}^{14}{\rm N}$ molecular lines in a spectrum of KIC4457200. The middle line represents the best fit for $^{12}{\rm C}/^{13}{\rm C}$=12, whereas the other two lines indicate $\pm 2$ in the $^{12}{\rm C}/^{13}{\rm C}$ ratio.
\label{fig:fit_examples2}}
\end{figure}

\begin{figure}
\centering
\resizebox{\hsize}{!}{
\includegraphics[]{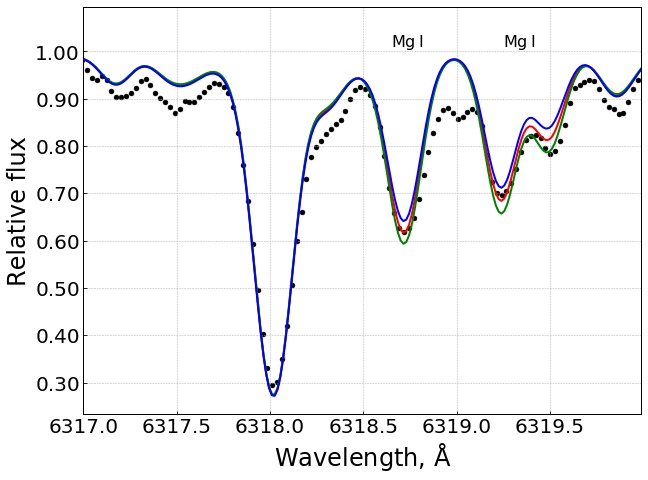}}
\caption{Example of fits to the magnesium lines for KIC4457200. The middle line represents the best fit for [Mg/H]=0.22, whereas the other two lines indicate $\pm 0.10$~dex for the magnesium abundance.
\label{fig:fit_examples3}}
\end{figure}

Synthetic spectra have been calibrated to the Solar spectrum by \citet{Kurucz05} with log~$A_{\odot}$(C)~=~8.39, log~$A_{\odot}$(N) = 7.78, log~$A_{\odot}$(O) = 8.66, log~$A_{\odot}$(Mg) = 7.53, and log~$A_{\odot}$(Fe) = 7.45 taken from \citet{Grevesse07}. Solar atmospheric parameters were taken from \citet{Pavlenko2012}: $T\mathrm{_{eff}}$ = 5777\,K, log $g$ = 4.44, [Fe/H] = 0.00, $v\mathrm{_{t}}$ = 0.75 km s$^{-1}$, $v$ sin i = 1.6 km s$^{-1}$. The calibration required slight changes to the log\,\textit{gf} values of the original line list. Several examples of the synthetic spectra fits are presented in Figs.~\ref{fig:fit_examples1}--\ref{fig:fit_examples3}.

 \begin{figure}
  \centering
     \includegraphics[width=0.98\hsize,clip=true,trim= 0cm 0cm 0cm 0cm]{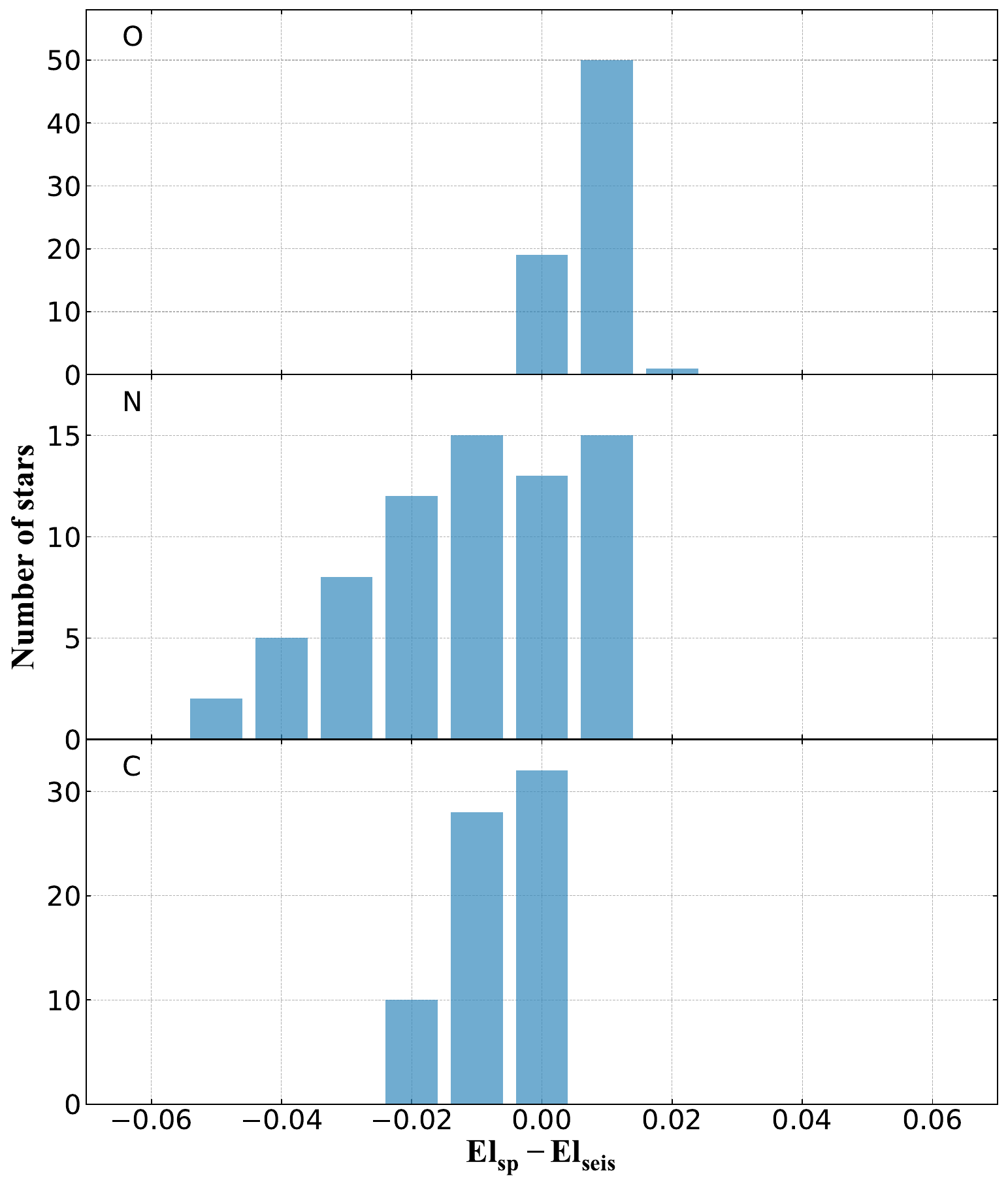}
     \caption{Differences in abundances of C, N, O 
     derived using the spectroscopic and asteroseismic atmospheric parameters.}
   \label{abund_diff_loggspect_seism}
 \end{figure}

Figure~\ref{abund_diff_loggspect_seism} shows the differences for A(C), A(N) and A(O) abundances determined from spectroscopic and seismic atmospheric parameters. The differences do not exceed the uncertainties of the method of abundance determinations (lower than 0.5\%) with no visible impact on the $^{12}{\rm C}/^{13}{\rm C}$ value. 

\subsubsection{Evaluation of uncertainties}

The calculated medians of asteroseismic atmospheric parameter determination errors in our stellar sample are $\sigma_{T_{\rm eff}}$ = 51~K, $\sigma_{{\rm log}~g}$ = 0.22 dex, $\sigma_{{\rm [Fe/H]}}$=0.11 dex, and $\sigma_{v_{\rm t}}$ = 0.22~${\rm km~s}^{-1}$. In Table~\ref{tab:effects}, we present changes in abundances caused by the median values of atmospheric parameter determination errors of each individual atmospheric parameter, calculated keeping other parameters fixed. For this investigation, we took  KIC7366121 which has  $T_{\rm eff}$ = 4806~K, ${\rm log}~g$ = 2.56~dex, ${\rm [Fe/H]}=-0.10$~dex, and ${v_{\rm t}}$ = 1.4~${\rm km~s}^{-1}$. 

Carbon, nitrogen, and oxygen are bound by the molecular equilibrium in the stellar atmospheres, therefore these chemical elements require a more detailed analysis. We started from derivations of carbon and oxygen abundances and iterated this operation until the determinations of carbon and oxygen abundances converged. Finally the abundance of nitrogen were derived, using both carbon and oxygen values.
An error in one of the elements bounded by the molecular equilibrium -- carbon, nitrogen or oxygen -- typically influences the abundance determination of another as well as measurement of C/N and $^{12}{\rm C}/^{13}{\rm C}$\ ratios. These effects are also presented in Table~\ref{tab:effects}. 

As expected, only the oxygen abundance exhibits a larger sensitivity to surface gravity variations as well as the C/N ratio -- to varying C and N abundances. Other values are not very sensitive to uncertainties. 

\begin{table*}
\centering
\caption{Effects on the derived chemical abundances and abundance ratios resulting from uncertainties in atmospheric parameters and abundances for the programme star KIC7366121.}
 \label{tab:effects}
 \begin{tabular}{cccccccccccccc}
  \hline
  \hline
 Species & $\Delta{T_{\rm eff}}$  & $\Delta{\rm log}\,g$  & $\Delta$[Fe/H] & $\Delta v_{\rm t}$ & $\Delta$C & $\Delta$N  & $\Delta$O  \\
 & $\pm51 $ K & $\pm0.22$ & $\pm0.11$ dex  & $\pm0.22 $ ${\rm km~s}^{-1}$ & $\pm0.10$\,dex  & $\pm0.10$\,dex  & $\pm0.10$\,dex \\
  \hline
C (C$_{2}$) & $\mp0.01$ &  $\pm0.03$  & $\pm0.03$ & $ 0.00 $ & -- & $ 0.00 $ & $ \pm0.04 $\\
N (CN) &  $\pm0.02$ &  $\pm0.05$  & $\pm0.05$ & $ 0.00 $ & $ \mp0.12 $ & -- & $ \pm0.09 $\\
O {[O\,{\sc i}]  } & $0.00$ &  $\pm0.10$  & $\pm0.01$ & $0.00$ & $\pm0.01$ & $0.00$ & -- \\
C/N &  $\mp0.11$ &  $\mp0.08$  & $\mp0.08$ & $0.00$ & $\pm0.35$  & $\mp0.35$ & $\mp0.16$\\
$^{12}{\rm C}/^{13}{\rm C}$\ & $\mp1$ &  $\mp1$  & $\mp2$ &   $0$ & +3/-5 &  +5/-3 & $\pm1$\\
 \hline
 \end{tabular}
\end{table*}

\subsection{Comparisons with other studies}
\label{section:comparison}

In this section, we present a comparison between our derivations and abundances published in literature.

\paragraph{APOGEE DR17 catalogue:} To compare with the APOGEE data, we used the clean and calibrated abundances which verify: 
\noindent 
\begin{verbatim} 
FE_H_FLAG=0 AND 
C_FE_FLAG=0 AND 
N_FE_FLAG=0 AND 
O_FE_FLAG=0 AND 
MG_FE_FLAG=0.
\end{verbatim}

\noindent 53 stars of our sample are in common with the APOGEE data release 17 catalogue \citep{APOGEEDR17}. \citet{Grevesse07} are used as solar abundances to derived surface abundances such as [C/Fe], [N/Fe], [O/Fe], and [M/Fe]. 

\paragraph{{\it Gaia} catalogue:} We have 70 stars in common with the {\it Gaia} DR3 catalogue \citet{Vallenari23}. Using the quality flags on global parameters and on abundances as given in \citet{RecioBlanco22} and \citet{RecioBlanco23} we selected only stars with the best  determinations of [N/Fe], [Mg/Fe], and [Fe/H]\footnote{We deduce [Fe/H] by adding [M/H] and [Fe/M]}. We use the following criteria (\texttt{ADQL} queries used in this study are presented in Appendix~\ref{ADQL}): 

\noindent 
\begin{verbatim}
3500<Teff<7000 & 0<log\,$g$<5 AND
vbroadT=0 & vbroadG=0 & vbroadM=0 AND
vradT=0 & vradG=0 & vradM=0 AND 
fluxNoise=0 & extrapol<3 & KMtypestars<2 AND
NegFlux=0 & nanFlux=0 & emission=0 & nullFluxErr=0 

For [N/Fe]: 
NUpLim<2 & NUncer=0 AND
nfe_gspspec_nlines>=2 AND
nfe_gspspec_linescatter<0.1 AND
[N/Fe]_unc<0.15 

For [Mg/Fe]: 
MgUpLim<2 & MgUncer=0 AND
[Mg/Fe]_unc<0.2 

For [Fe/H]:
FeUpLim<2 & FeUncer=0 AND
[M/H]_unc<0.2 
\end{verbatim}

\noindent  
62 stars verify the selection criteria for the iron abundance, 19 stars for the magnesium abundance and 15 for the nitrogen abundance.  Abundances are derived using \citet{Grevesse07} solar abundances. As proposed by \citet{RecioBlanco23} we apply a correction to the {\it Gaia} published abundances of [N/Fe], [Mg/Fe], and [Fe/H] as well as for log\,$g$ (see Appendix \ref{ADQL}).  

\paragraph{PASTEL sample:} 20 stars of our sample are in common with the last release of PASTEL catalogue \citep{Soubiran22}.

\paragraph{LAMOST DR7 catalogue:} 8 stars of our sample are in common with the LAMOST data release 7 catalogue \citep{LAMOSTDR7}. \\

\begin{figure}
  \centering
     \includegraphics[width=0.98\hsize,clip=true,trim= 4.2cm 1cm 7.5cm 4cm]{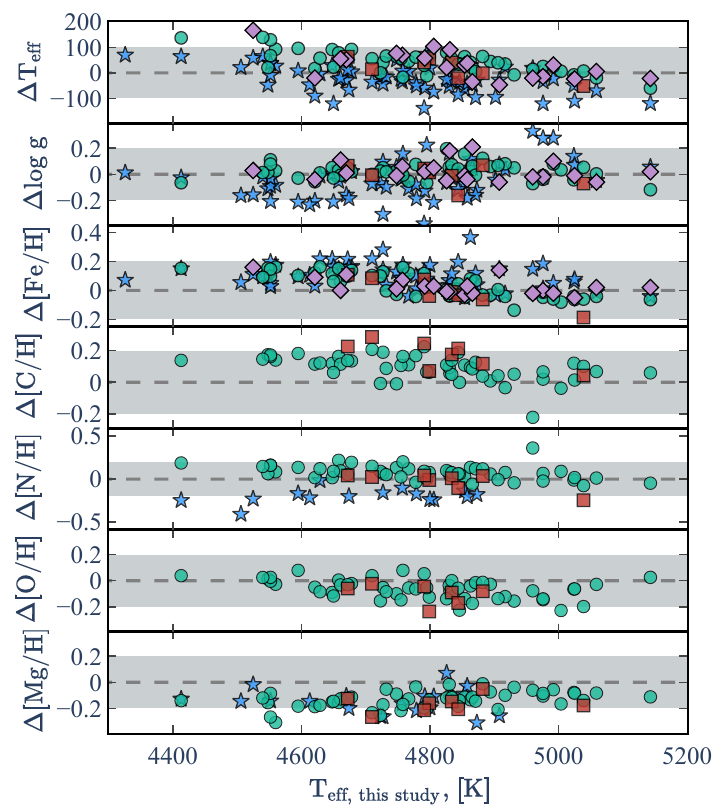}
     \caption{Differences between [X/H], log\,$g$ and T$_{\rm eff}$ coming from literature and derived in this study. The comparisons are done using the \gaia DR3 (blue stars), APOGEE DR17 (green circles), LAMOST DR7 (orange squares), and PASTEL (purple diamonds) catalogues. The $\pm$0.2dex and $\pm$100K are shown with the shadow regions.
     }
   \label{compa}
 \end{figure}

Figure \ref{compa} shows differences between the abundance determinations published in the literature using spectra coming from different large surveys and those derived in this study. This comparison is done for [Fe/H], [C/H], [N/H], [O/H], and [Mg/H] abundance ratios as well as for log\,$g$ and $T_{\rm eff}$. We note a good agreement between our log\,$g$ and $T_{\rm eff}$ determination with previous studies excepted for the \gaia DR3 catalogue. This implies a quite well agreement within $\pm$0.2~dex for the abundances. However, we can note that our [O/H] and [Mg/H] abundances are systematically higher than those of the other surveys. Considering  [Fe/H] ratio, the median differences are around 0.01--0.02~dex according to the APOGEE DR17, LAMOST DR7, and PASTEL surveys while the difference is larger considering the \gaia DR3 catalogue (0.095~dex). A slight overabundance of carbon is visible in the LAMOST survey. The biggest difference is visible between [N/H] determined by \gaia DR3 and our study with a median difference around $-0.20$~dex. Nevertheless, our study seems to be in agreement with the APOGEE and LAMOST spectroscopic surveys for this element. Thus, the \gaia [N/Fe] values should therefore be treated with caution, and an additional calibration would probably be necessary (c.f. de Laverny, P. et al. private communication). In the {\it Gaia} survey, the nitrogen calibration is based on a small number of reference stars. 

\section{Overview of the stellar sample}
\label{combined}

\subsection{Asteroseismic properties}
\label{sismo}

The first asteroseismic quantities available for our sample is the average large separation, $\Delta\nu$, which is expected to be proportional to the square root of the mean stellar density \citep[e.g.][]{Ulrich86}. 
The frequency $\nu_{max}$ at which the oscillation modes reach their strongest amplitudes is approximately proportional to the acoustic cut-off frequency \citep[e.g.,][]{Brown91,Belkacem11} is also available. 
We use the large frequency separation $\Delta\nu$, and the frequency corresponding to the maximum observed oscillation power $\nu_{max}$ to determine the stellar radius and mass of our solar-like oscillating giants. 
We used the code PARAM \citep{Rodrigues17} to infer the radius, mass, and age of the red giant component by using a combination of seismic and non-seismic constraints. On the one hand, the average large frequency separation is computed using the radial mode frequencies of the models in the grid, not added as an a posteriori correction to the scaling relation between $\Delta\nu$ and the square root of the stellar mean density. On the other hand, $\nu_{max}$ in the model grid is computed using a simple scaling relation \citep{KjeBed95}, by considering $\nu_{max,\odot}$=3090 $\mu$Hz. The grid of stellar evolution models used in PARAM is the same as the reference grid adopted in \citet{Khan19} and \citet{Miglio21} (grid G2). We obtained seismic masses for 69 sample stars.

\begin{figure}
  \includegraphics[width=\hsize,clip=true,trim= 0cm 0.2cm 1.8cm 2cm]{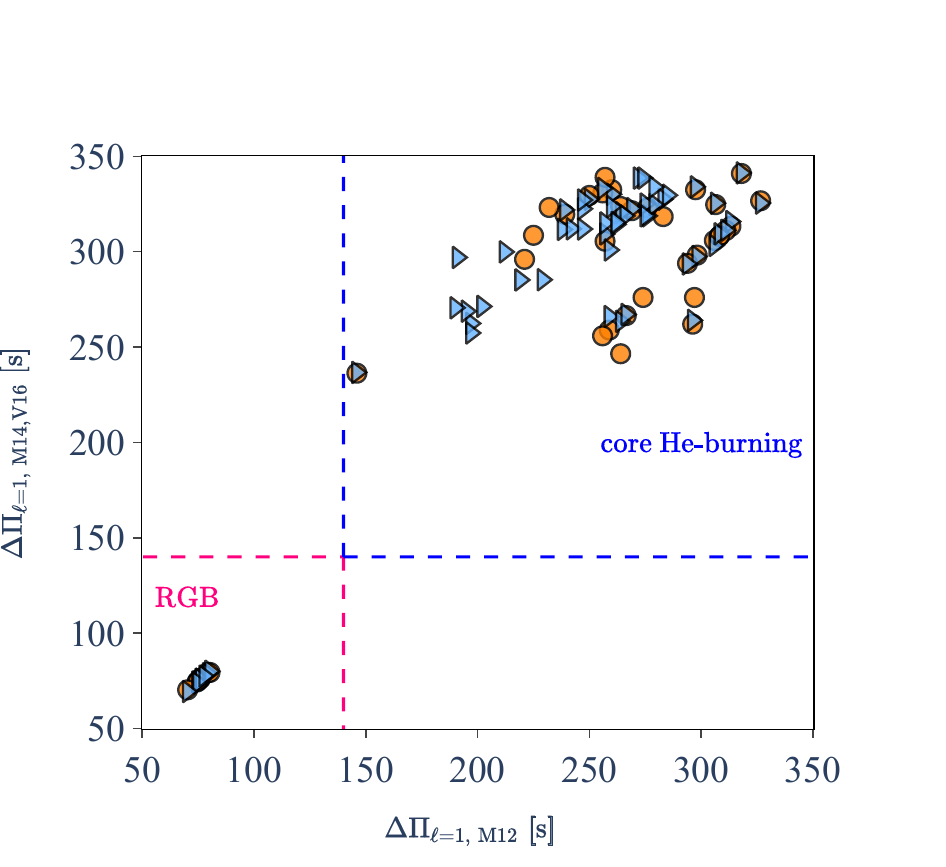}
  \caption{The asymptotic period spacing of g-modes, $\Delta \Pi_{\ell=1}$, computed by \citet{Mosser14} (orange circles) and by \citet{Vrard16} (blue triangle) as a function $\Delta \Pi_{\ell=1}$ derived by \citet{Mosser12a}. We considere all stars with $\Delta \Pi_{\ell=1}\leq$140\,s as first-ascent RGB stars and core He-burning stars with $\Delta \Pi_{\ell=1}$ >140\,s. The error bars on the $\Delta \Pi_{\ell=1}$ determination are smaller than the size of the symbols.}
  \label{compa_deltapi}
\end{figure}

Finally, the asymptotic period spacing of gravity modes for $\Delta\Pi_{\ell=1}$ provides information on stellar structure. This quantity allows us to distinguish two giants having the same luminosity, one being at the RGB bump and the other one being at the clump undergoing central He burning \citep{Bedding11,Mosser11}. As the stellar structure, and more particularly the presence of the convective core affects the domain where the g-modes are trapped, $\Delta\Pi_{\ell=1}$ is larger in clump stars than in RGB stars. Figure \ref{compa_deltapi} presents the period spacing of g-modes for our sample derived by \citet{Mosser12a} compared to the one derived by \citet{Mosser14} and \citet{Vrard16}. The figure shows good agreement between the different  $\Delta \Pi_{\ell=1}$ determinations. We adopt the values derived by \citet{Mosser12a} to classify the stars in our sample. Using this value, we find that our sample contains 9 and 62 RGB and core He-burning stars, respectively. This classification remains correct whatever the three different values considered here.

\subsection{Masses derived using SPInS} 
\label{spins}

\begin{figure}
 \includegraphics[width=\hsize,clip=true,trim= 4.2cm 3.5cm 5.5cm 5cm]{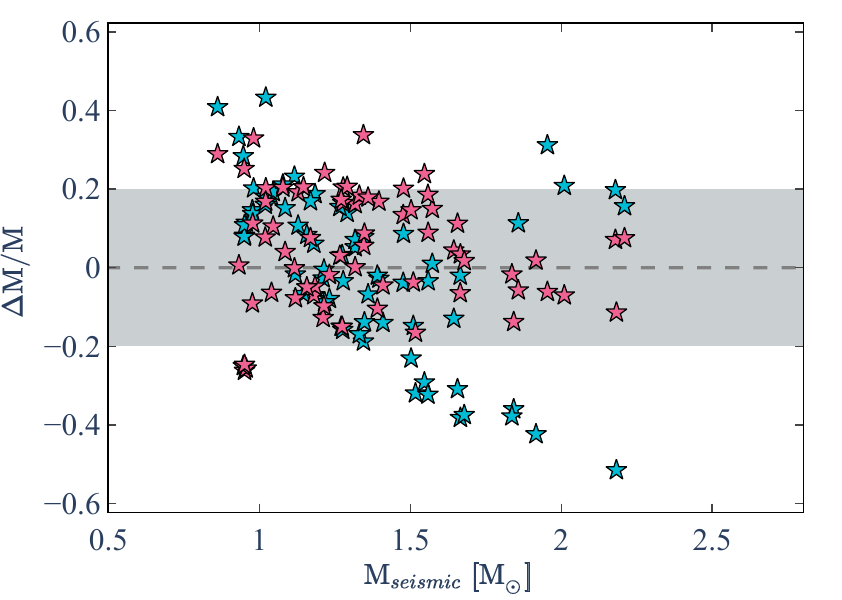}
  \caption{Relative difference of the stellar masses derived from SPiNS and the seismic masses derived with PARAM ($\Delta \rm{M/M}=(M_{\rm{SPINS}}-M_{\rm{seismic}})/M_{\rm{seismic}}$) using or not asteroseismic diagnostics (red and blue stars, respectively). Dashed-lines represents a relative difference at $\pm$20\%.}
  \label{compa_Mass}
\end{figure}

In order to better appreciate the possible uncertainties in the determination of the mass and age of our sample stars, we independently determine these quantities with 
the Stellar Parameters INferred Systematically (SPInS) tool \citep{SPINS20}.  We use the stellar evolution models computed with the STAREVOL code by \citet{Lagarde17, Lagarde19}. The same grid will be used in the Besan\c con Galaxy model in \S~\ref{section:BGM} for consistency. 

SPInS characterises stars through an MCMC approach, with different possible sets of input observable quantities. We choose to use two different combinations: 
(1) \textit{\textbf{Spectroscopic observables.} }We consider the spectroscopic observables derived from our spectra: $T_{\rm{eff}}$, the surface gravity log\,$g$, and the iron abundances [Fe/H]; (2) \textit{\textbf{Spectroscopic + asteroseismic observables.}} We add to the above the asteroseismic quantities  $\Delta\nu$, $\nu_{max}$, and $\Delta \Pi_{\ell=1}$. In this case, the  $\Delta\nu$ in the stellar evolution models is computed from scaling relations.
In both cases we assume a Kroupa Initial Mass Function \citep[][IMF]{Kroupa01,Kroupa13} as a prior as well as a truncated uniform Stellar Formation Rate between 0 and 13.8~Gyr that is roughly the age of the Universe. \\ 
Figure \ref{compa_Mass} shows the relative difference between the stellar mass we determined with SPiNS and the seismic mass we computed with PARAM. While differences up to $\sim50~\%$ can be obtained when we use only the spectroscopic input in  SPInS, the masses derived with PARAM and SPInS agree within $\sim 20~\%$, when we include the asteroseismic input. This is mainly thanks to $\Delta \Pi_{\ell=1}$ that lifts the degeneracy between red giant stars and core He-burning stars in the HR-diagram. 
We want to stress that this is a very good agreement, considering the uncertainties of the input physics that can affect the modelling of red giant stars. In \S~\ref{cir BGM vs Kepler} we will discuss in particular the impact of the assumptions on the mass loss rate.

For the rest of the study, we will use the seismic masses derived by PARAM. In addition, for the purpose of consistency and further discussion, we determine the stellar properties ($M$, age) of all the literature samples discussed in Sect.~\ref{compil}, using the same SPInS configuration and using only spectroscopic observables.

 \subsection{Kinematics from \gaia data}
 \label{kine}
 
\begin{figure}
  \centering
     \includegraphics[width=\hsize,clip=true,trim= 0cm 0cm 0cm 5cm]{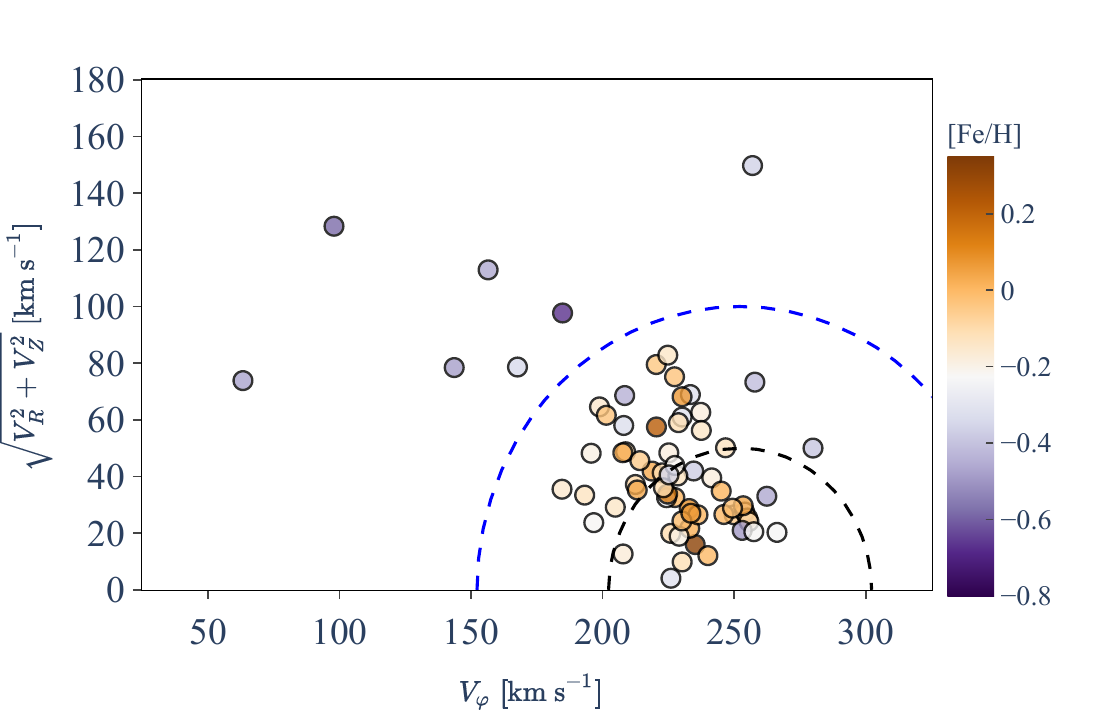} 
   \caption{Toomre diagram for the stars in our sample. The dashed curves indicate constant space motion. Stars are colour-coded according to their metallicity.}
   \label{Toomre}
 \end{figure}

In order to determine whether the stars in our sample belong to a specific Galactic stellar population, we use the kinematics derived by \citet{RecioBlanco22} (see more details in their Sect.\ 2.4). They computed the stellar positions (Galactocentric Cartesian X, Y, and Z positions, and cylindrical radius $R$) as well as the Galactocentric cylindrical velocities ($V_r$, $V_\varphi$ and $V_Z$) using the astrometric parameter  from Gaia EDR3 \citep{GaiaEDR3} as well as distances computed by \citet{BailerJones21}. 
Fig.~\ref{Toomre} displays the velocity components in a Toomre diagram. This diagram is used extensively to identify the main components of our Galaxy, such as the thin and thick disks and the halo, using only kinematics. We assume that the stars outside the blue circle belong to the thick disc, while the others belong to the thin disc \citep[e.g.,][]{Yoshii82, GiRe83, RecioBlanco14}. Using these criteria and for guidance, our sample is therefore composed of 7 stars from the thick disc and 63 stars from the thin disc.
 
   \begin{figure}
  \centering
     \includegraphics[width=\hsize,clip=true,trim= 2cm 0cm 2cm 2cm]{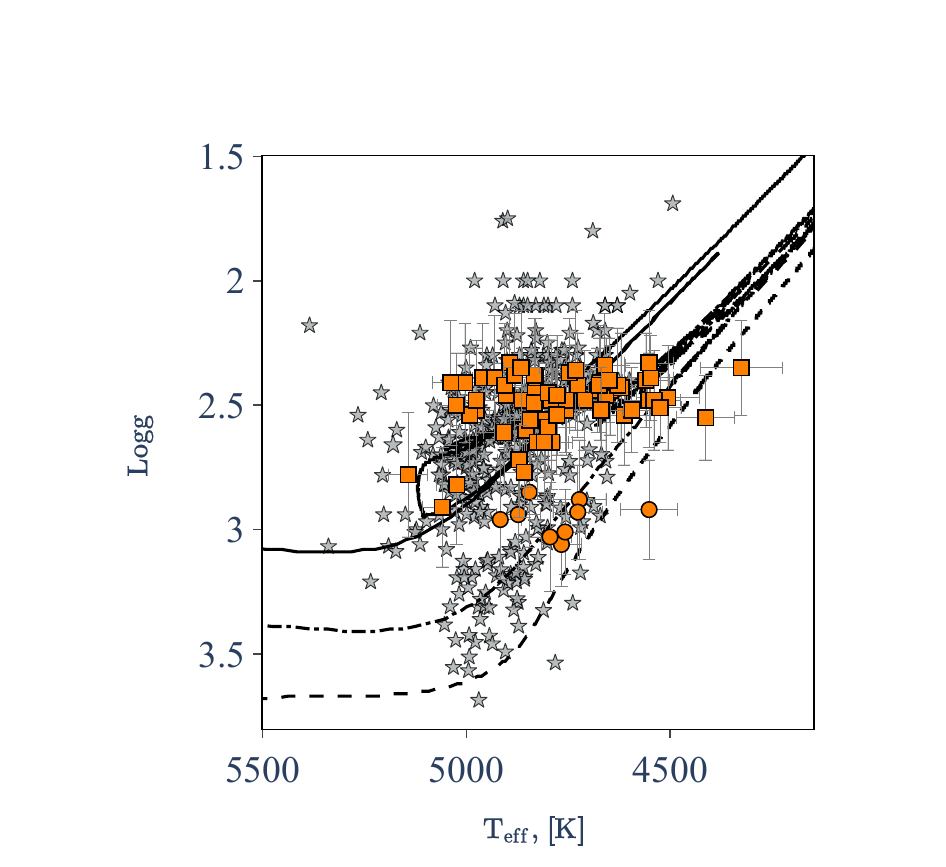} 
   \caption{Kiel diagram for stars in our sample (orange symbols). RGB and core He-burning stars are represented by circles and squares, respectively. Stars compiled from the literature for which the carbon isotopic ratio (and log\,$g$) is determined are  represented by grey star symbols (see Sect.\ref{compil}). Stellar evolutionary tracks computed with STAREVOL for 1.4M$_\odot$ (dashed line) and 2.0M$_\odot$ (dash-dotted line) models at solar metallicity and 2.5\,$M_\odot$ at [Fe/H]=$-0.23$ (the solid line) are also plotted.   }
   \label{Kieldiagram}
 \end{figure}

\begin{figure*}
  \centering
     \includegraphics[width=0.32\hsize,clip=true,trim= 0cm 0.3cm 4.5cm 2.5cm]{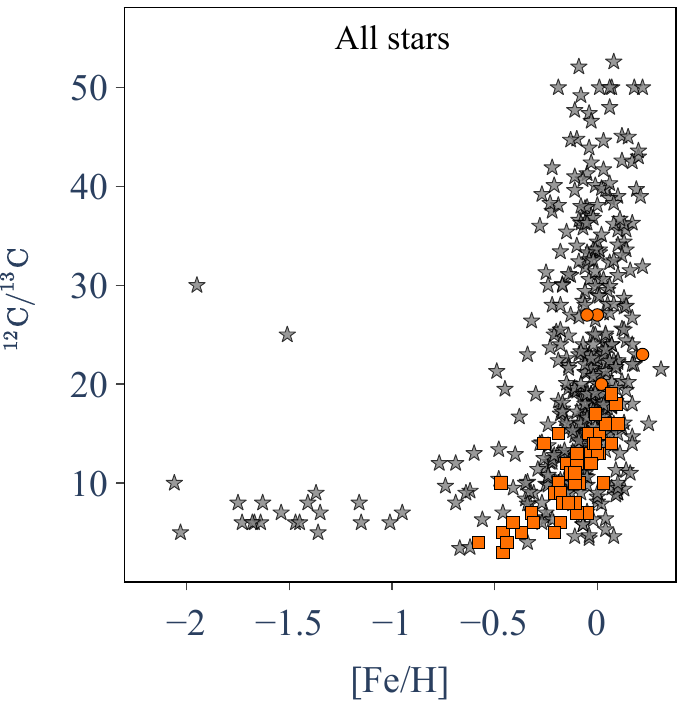}
     \includegraphics[width=0.32\hsize,clip=true,trim= 0cm 0.3cm 4.5cm 2.5cm]{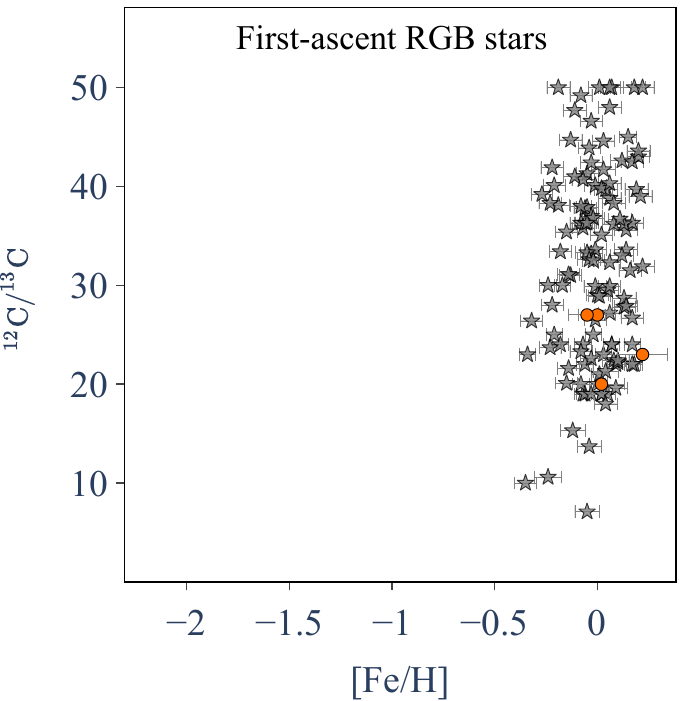}
     \includegraphics[width=0.32\hsize,clip=true,trim= 0cm 0.3cm 4.5cm 2.5cm]{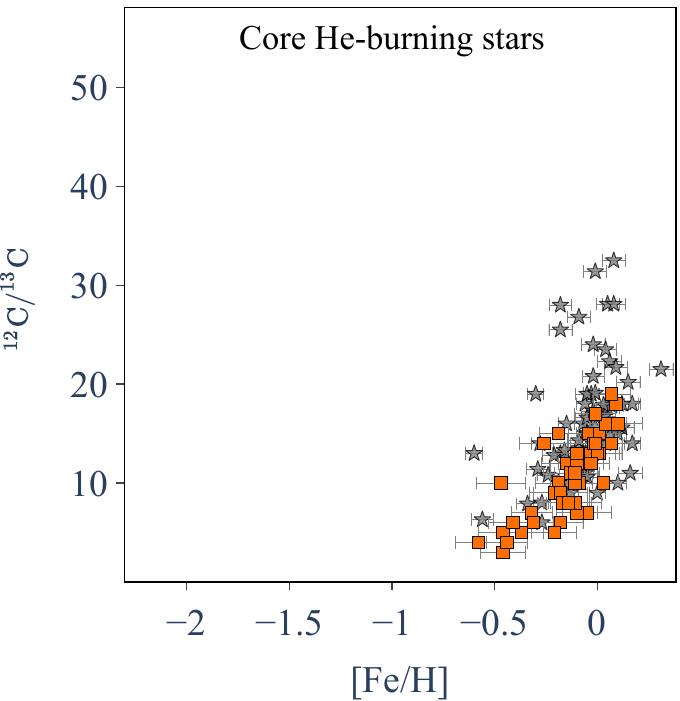}\\
     \includegraphics[width=0.32\hsize,clip=true,trim= 0cm 0.3cm 4.5cm 2.5cm]{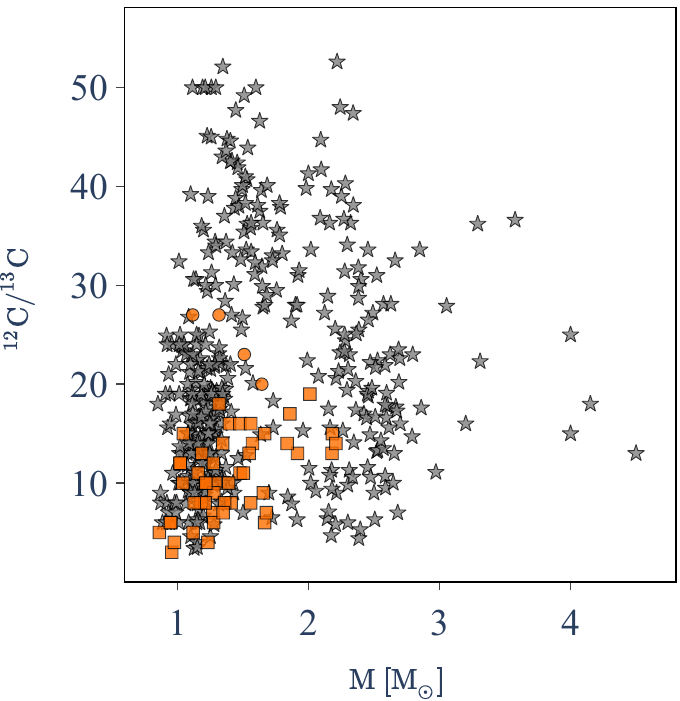}
     \includegraphics[width=0.32\hsize,clip=true,trim= 0cm 0.3cm 4.5cm 2.5cm]{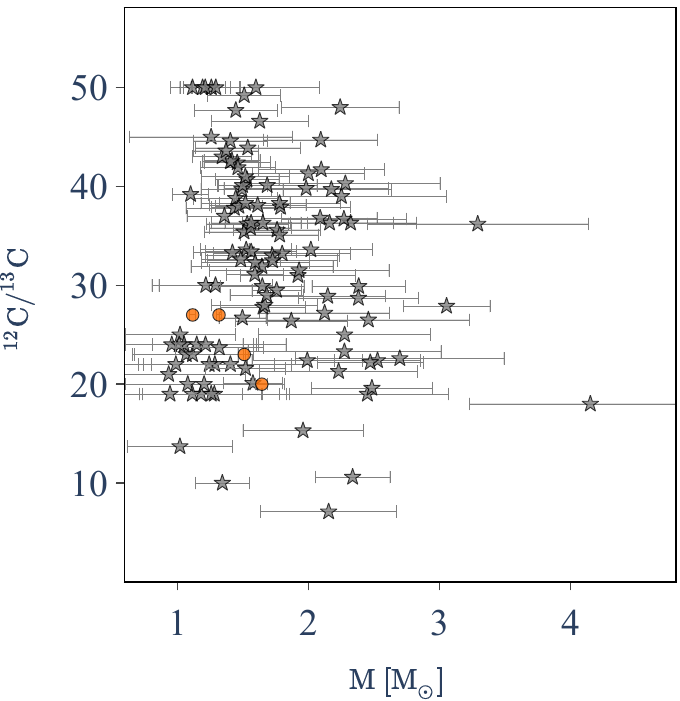}
     \includegraphics[width=0.32\hsize,clip=true,trim= 0cm 0.3cm 4.5cm 2.5cm]{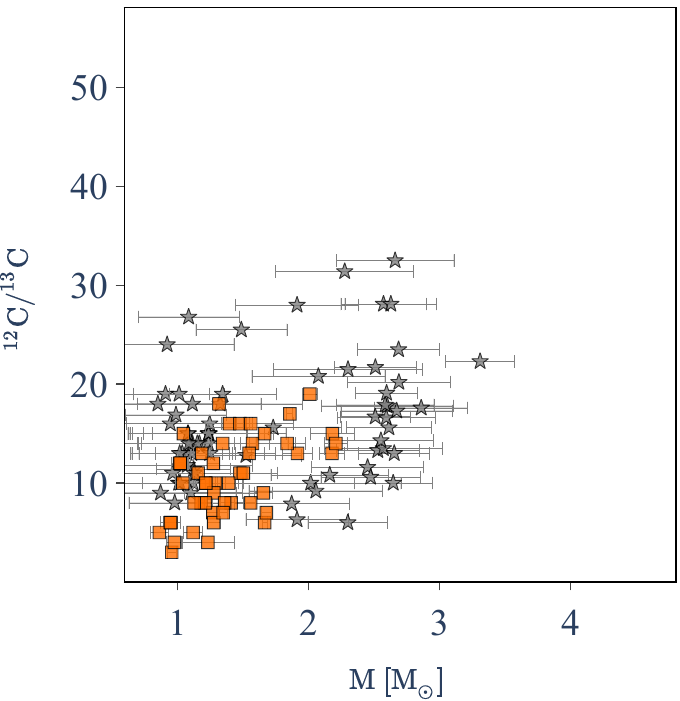}\\
     \includegraphics[width=0.32\hsize,clip=true,trim= 0cm 0.3cm 4.5cm 2.5cm]{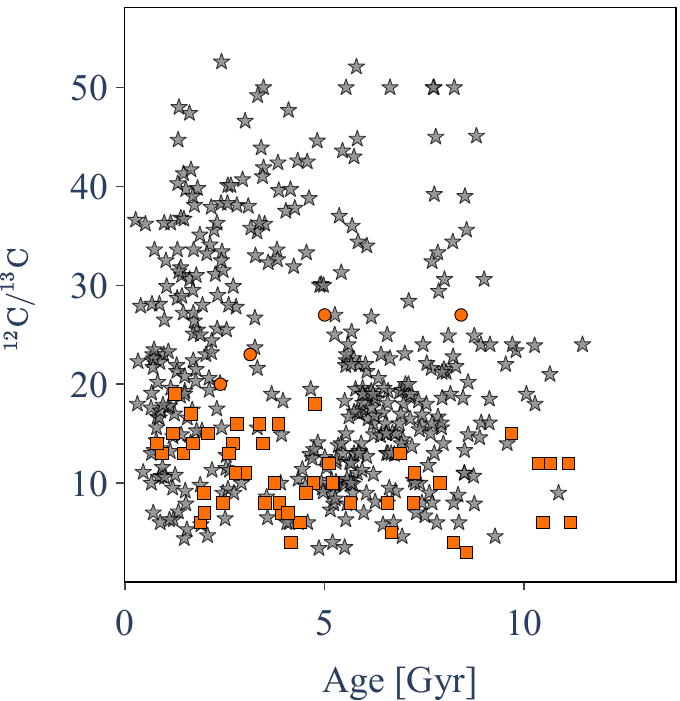}
     \includegraphics[width=0.32\hsize,clip=true,trim= 0cm 0.3cm 4.5cm 2.5cm]{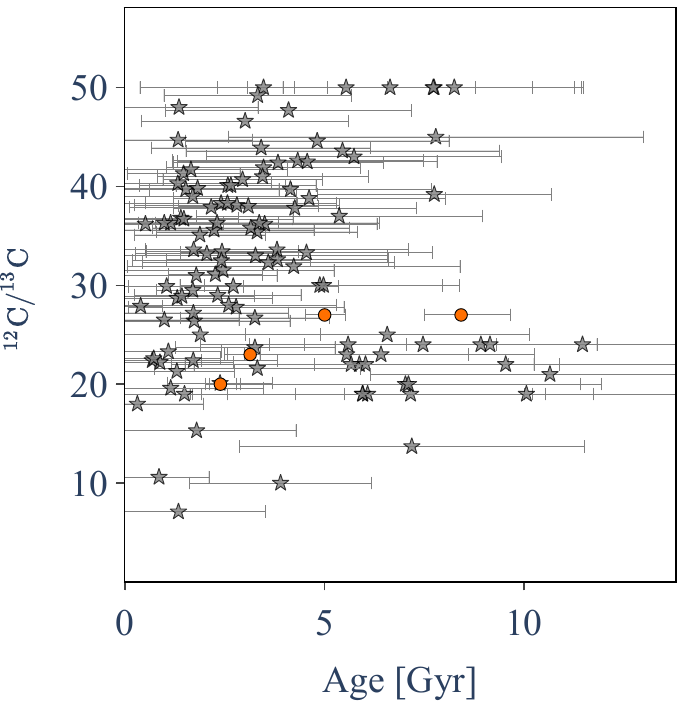}
     \includegraphics[width=0.32\hsize,clip=true,trim= 0cm 0.3cm 4.5cm 2.5cm]{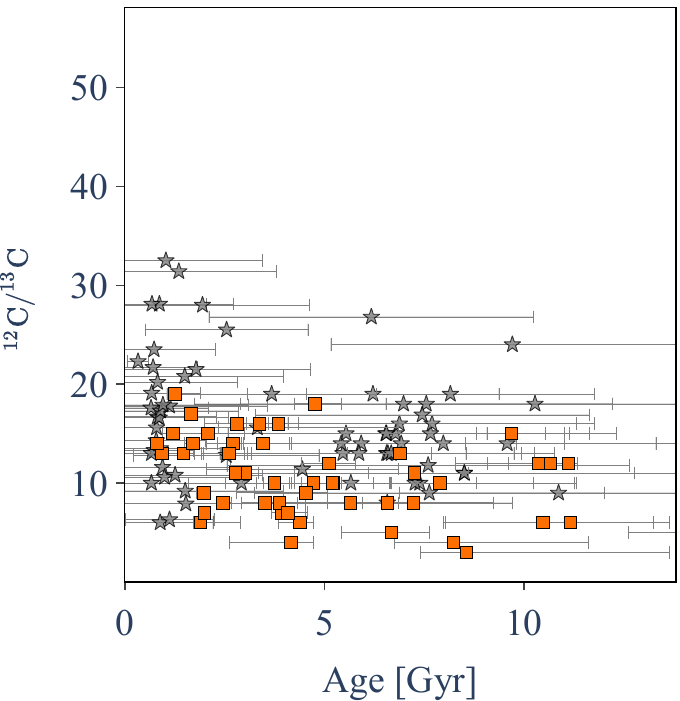}
  \caption{$^{12}$C/$^{13}$C versus [Fe/H] (top panel), mass (middle panel), and age (bottom panel) for stars in our sample (orange symbols). Abundances published by \citet{Gratton00} 
  , \citet{Tautvaisiene10}   , \citet{Tautvaisiene13}
  , \citet{Morel14}   , \citet{Takeda19}   , respectively), \citet{ChBa00}   , and  \citet{AguileraGomez23}   are also plotted. When the information is available in the original papers, first ascend giant and core He-burning stars are plotted on the middle and right column, respectively. For our golden sample stars, we show the asteroseismic  masses and ages we obtained with PARAM. For the literature sample, we use the masses and ages we obtained with SPInS with only spectroscopic but no asteroseismic constraints.  
  }
  \label{c1213_lit}
\end{figure*}

\subsection{Compilation of $^{12}$C/$^{13}$C form others studies}
\label{compil}

Because of the limitations of the {\it{Kepler}} field and of our selection criteria, our golden sample contains only relatively metal-rich stars (i.e., with [Fe/H] higher than $\sim -0.6$~dex), with luminosities around that of the RGB bump and of the clump (see Fig.\ref{Kieldiagram}). We thus miss carbon isotopic ratios for brighter RGB stars as well as for metal-poor field giants to test possible correlations between the metallicity, the position on the RGB, and the mixing efficiency.  
We thus collect this information from the literature studies that are listed in Table~\ref{tab:compil}  \citep{Gratton00,Tautvaisiene10,Tautvaisiene13,Morel14,Takeda19,AguileraGomez23}. For this \textit{extended sample}, however, asteroseismic constraints are missing, making the determination of the stellar masses and evolution phases more uncertain than in our golden sample, as described below. Figure \ref{Kieldiagram} presents the Kiel diagram for both our golden and extended samples. We derived the stellar mass of all of the stars of the literature sample using SPInS tool and the list of input parameters of set (1) described in \S\ref{spins}; we use the values of $T_{\rm eff}$, log\,$g$, and [Fe/H] published in the original papers. We are aware of the systematics that can be induced by the sample inhomogeneity. However, this allows us to explore a mass range between 0.6 and 4.2~$M_\odot$ and a [Fe/H] range between $-2.1$ and 0.4~dex, thus scanning different Galactic stellar populations, and to probe the  dependency of the mixing with the stellar mass and metallicity. \\

Figure~\ref{c1213_lit} shows the carbon isotopic ratio as a function of [Fe/H], stellar mass, and age, for both our \textit{golden and extended samples}. In the first raw from the left, we show the data for all the stars. In the middle and the third raws, we eliminate the literature stars for which no evolution stage indication was provided in the original papers. We split the remaining stars with respect to their evolution phase, and we show the error bars on the masses, which are much higher for the \textit{extended sample} than for \textit{the golden sample} stars for which asteroseismic data provide more reliable estimates.
Our \textit{golden sample} contains mostly clump stars, and only four RGB stars with luminosity close to that of the clump for which $^{12}$C/$^{13}$C is derived. The literature sample contains more supposedly RGB stars, for which definitive information about their actual evolution phase would require asteroseismic information. In any case, the observed trends are very similar between \textit{the golden and the extended samples}.
As can be seen in Fig.~\ref{c1213_lit}, \textit{the golden sample} undoubtedly confirms that the carbon isotopic ratio is lower at the surface of core He-burning stars compared to RGB stars with slightly lower luminosity than the RGB bump. 
The dispersion of $^{12}$C/$^{13}$C observed at the surface of RGB stars in \textit{the extended sample} is greater than for clump stars (see middle panel of Fig.\ref{c1213_lit}). This depicts the combined effects of 1DUP (all the RGB stars have a luminosity high enough to have already undergone the 1DUP), and of additional mixing in decreasing the carbon isotopic ratio while the stars are climbing along the red giant branch until it reaches a 'plateau' value where the stars lie during the helium burning phase. Future asteroseismic studies are however required to definitively pinpoint the actual phase where the extra-mixing occurs along or at the tip of the RGB.
Finally, $^{12}$C/$^{13}$C decreases clearly when the stellar metallicity and mass decrease, while it seems to slowly decrease with the stellar age (see Fig.~\ref{c1213_lit}). These behaviours are consistent with other previously published studies, but they will have to be confirmed in the future when asteroseismology studies become feasible for low-metallicity red giants. 

\begin{table*}
\centering
\caption{Properties of the star samples used in this study.  }
 \label{tab:compil}
 \begin{tabular}{cccc}
  \hline
  \hline
  Reference & Number of stars  & [Fe/H] range & Mass\tablefootmark{a} range\\
     &with $^{12}$C/$^{13}$C& dex & $M_\odot$ \\
  \hline
\citet{Gratton00} & 54 & [$-1.75$;-0.64] & [0.87;1.70]\\
\citet{ChBa00} & 13 & [$-0.46$;0.09] & [1.1;4.5]\tablefootmark{b}\\
\citet{Tautvaisiene10} & 30 & [$-0.60$;0.18] & [1.02;2.45]\\
\citet{Tautvaisiene13} & 25 & [$-0.30$;0.17] & [0.85;1.32]\\
\citet{Morel14} & 4 & [$-0.67$;0.11] & [1.0;1.13]\\
\citet{Takeda19} & 214\tablefootmark{c} & [$-0.74$;0.19] & [0.92;3.5] \\
\citet{AguileraGomez23} & 134 & [$-0.56$;0.31] & [0.92;4.15]\\
\hline
 &  & & \\
This study & 71 & [$-0.67$;0.32] & [0.86;2.2] \\
 \hline
 \end{tabular}
 \tablefoot{\tablefoottext{a}{Stellar masses are computed with SPiNS when asteroseismic are not available} ; \tablefoottext{b}{ Mass are derived by authors using isochrone fitting} ; \tablefoottext{c}{91 with class A \& B}}
\end{table*}


\section{Forward modelling using the Besançon Galaxy Model} 
\label{section:BGM}

The stars we consider in both \textit{the golden and the extended sample} belong to different stellar populations of the Milky Way; each with a different formation and history.
Additionally, they cover a large range in terms of mass, [Fe/H], and they lie either on the RGB or the clump. In order to exploit the full potential of these observations by taking into account both the evolution of the Milky Way and the evolution of stars, which can affect the stellar properties at different degrees, we run simulations with a Galactic stellar population synthesis model. We use the Besan\c con Galaxy model \citep[BGM, e.g.][]{Lagarde17} that provides forward modelling and realistic data simulations (mock catalogues) where selection biases on observables from specific surveys can be accurately reproduced.\\

The BGM is built on a Galactic formation and evolution scenario that reflects our present understanding of the Milky Way. We consider four stellar populations: a thin disc, a thick disc, a bar, and a halo, and each stellar population has a specific density distribution. The stellar content of each population is modelled through an initial mass function (IMF) and a star formation history (SFH) \citep{Czekaj14}, and follows stellar evolutionary tracks \citep[implemented in][]{Lagarde17, Lagarde19} taking into account different transport processes (see Sect.~\ref{thermoh}). The resulting stellar properties are used to compute the observational properties using atmosphere models and assuming a 3D extinction map. A Galactic dynamical model is used to compute radial velocities and proper motions \citep{Robin17}.\\
Our sample stars belong mostly to the thin and thick discs. The main ingredients to describe these two populations are the following: 

\begin{itemize}
\item The initial mass functions for both stellar populations are taken from the analysis of the Tycho-2 data \citep{Mor18}. 
\item The star formation history (SFH) of the thin disc is from \citet{Mor18}. The SFH of the thick disc is modelled assuming a two-episode formation \citep{Robin14} with a Gaussian age distributions from 8 to 12~Gyr and from 10 to 13~Gyr for the young and the old thick discs, respectively. 
\item Considering the thin disc, the iron abundance [Fe/H] and its dispersion are estimated assuming the age-metallicity relation deduced from \citet[][for more details see \citealt{Czekaj14}]{Haywood06}: 

\begin{equation}
\begin{array}{l}
{\rm [Fe/H]}=-0.016\times {\rm age} +0.01\\
\sigma_{\rm [Fe/H]}=0.010\times {\rm age} +0.1, 
\end{array}
\end{equation}

with the stellar age given in gigayears, and a radial metallicity gradient of $-0.07$~dex\,kpc$^{-1}$ limited to Galactocentric radii of between 5 and 12~kpc.
Considering the thick disc, a mean metallicity is assumed for the young thick disc and the old thick disc ($-0.5$ and $-0.8$~dex, respectively) with a dispersion of 0.3~dex.

\item The adopted [$\alpha$/Fe] versus [Fe/H] relations follow the trend observed in the DR17 of APOGEE for both stellar populations. For [Fe/H]$\leq0.1$, 
\begin{equation}
      [\alpha/{\rm Fe}] =
     \left\{
     \begin{array}{rl}
     -0.121\times\text{[Fe/H]}+0.0259\\
      \text{for\ the\ thin\ disc\ stars}, \\
    -0.2074\times\text{exp}(1.8354\times\text{[Fe/H]})+0.3397\\
    \text{ for the thick disc stars}.
     \end{array}
     \right.
     \end{equation}

For [Fe/H]$>$0.1, [$\alpha$/Fe] is assumed to be solar. An intrinsic Gaussian dispersion of 0.02~dex is added to these relations.

\item The adopted velocity dispersions as a function of age have been constrained from the RAVE survey \citep[DR4, ][]{Kordopatis13a} and proper motions from the TGAS part of the \textit{Gaia} DR1 \citep{Gaia16} by \citet[][see their Table 4]{Robin17}. 

\item The rotation curve is given by \citet{Sofue15}. The asymmetric drift is also taken into account following \citet{Robin17}. The dynamical statistical equilibrium is ensured by assuming the St\"{a}ckel approximation of the potential from \citet{Bienayme15, Bienayme18}. 

\item The 3D extinction map adopted is a combination of the 3D maps published by \citet{Marshall2006} and by \citet{Lallement2018}. Simulations were conducted using this extinction, but to account for small-scale variations in the distribution of interstellar matter, a dispersion around the average value was introduced on a star-by-star basis. This dispersion was equivalent to 10\% of the mean extinction.

\end{itemize}

Stellar properties of the mock sample are directly deduced from stellar evolution models. \citet{Lagarde17} improved the BGM including stellar evolutionary models computed with the STAREVOL code \citep[e.g.,][]{Lagarde12a,Amard19} for stars between 0.2 and 6.0 M$_\odot$ at six metallicities ([Fe/H]=-2.14,-1.8,-0.54,-0.23, 0 and 0.51) with different $\alpha$-enhancement (0, 0.15 and 0.3) to simulate properties of stars in different stellar populations \citep{Lagarde19}. Thanks to this improvement, the BGM provides the global properties (e.g. surface gravity, effective temperature), chemical abundances (for 54 stable and unstable species) and asteroseismic properties (e.g., $\delta \nu$, $\nu_{max}$, $\Delta\Pi_\ell=1$). Importantly for our study, we compare the observed abundances of our \textit{golden and extended samples} to the predictions of the BGM for two sets of stellar models (Sect.~\ref{BGMvsObs}). The first one corresponds to ``classical" models that include no transport process for chemicals except convection (hence, only the effect of the 1DUP is accounted for). The second one takes into account the additional effects of thermohaline instability during the red giant branch (see \S~\ref{thermoh}). We use a diffusion coefficient for thermohaline mixing that is based on \citet{Ulrich72} prescription corrected for non-perfect gas and with an aspect ratio of 6 for the instability fingers as in \citet[][see also \citealt{Lagarde12a,Lagarde17}]{ChaZah07a}.
 
For this study, we computed simulation for the \textit{Kepler} fields and according to the following criteria: 

\begin{itemize}
\item Apparent magnitude range of 4.5$<$H$<$10.5; 
\item Stars with an effective temperature lower than 5300~K to select only giant stars; 
\item Stars within the observed domains of the large separation ($\Delta\nu$) and the frequency of maximum power ($\nu_{\text{max}}$) ($3<\Delta\nu<12~\mu$~Hz; $25<\nu_{max}<150~\mu$~Hz);
\item We select core He-burning stars in the simulation according to the period spacing of g-modes $\Delta\Pi_{\ell=1} > 140$~s. 
\end{itemize}

Finally, we apply an error to the surface abundances predicted by the BGM. We assume the observational errors for C, N, O, [Fe/H], $^{12}$C/$^{13}$C, and $^{12}$C/$^{14}$N, and the theoretical error bars are generated with an uniform noise. 

\section{Comparison between BGM predictions and observations}
\label{BGMvsObs}
 
We first recall the theoretical impact of thermohaline mixing on the surface abundances along stellar evolution tracks. We then use simulations of the \textit{Kepler} field computed with the BGM which include stellar models taking into account or not thermohaline mixing and study its effects on the surface abundances of stars in the {\it{Kepler}} field and its dependency on stellar metallicity, mass, and age.

\subsection{Surface abundance variations predicted by the stellar evolution models}
\label{thermoh}

\begin{figure}
  \centering
\includegraphics[width=0.49\hsize,clip=true,trim= 0cm 0.5cm 13.5cm 5cm]{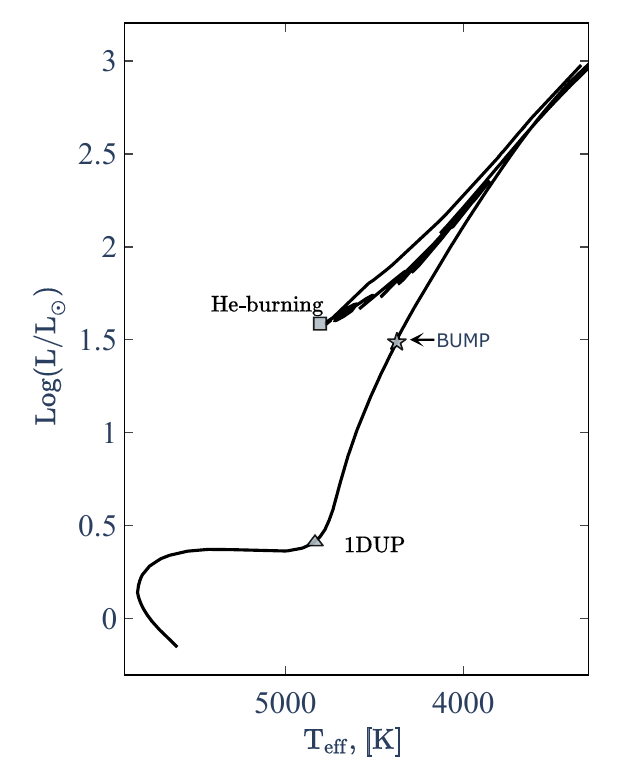} 
\includegraphics[width=0.49\hsize,clip=true,trim= 0cm 0.5cm 13.5cm 5cm]{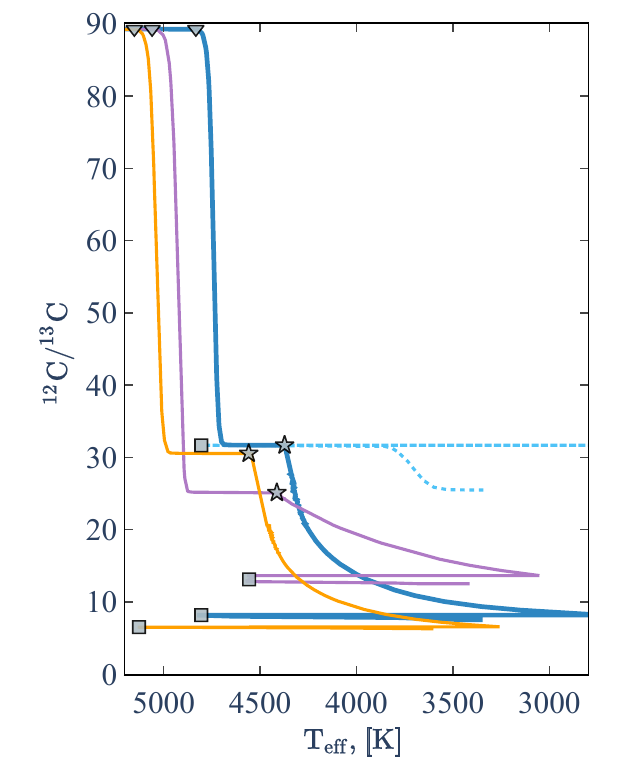} \\
\includegraphics[width=0.49\hsize,clip=true,trim= 0cm 0.5cm 13.5cm 5cm]{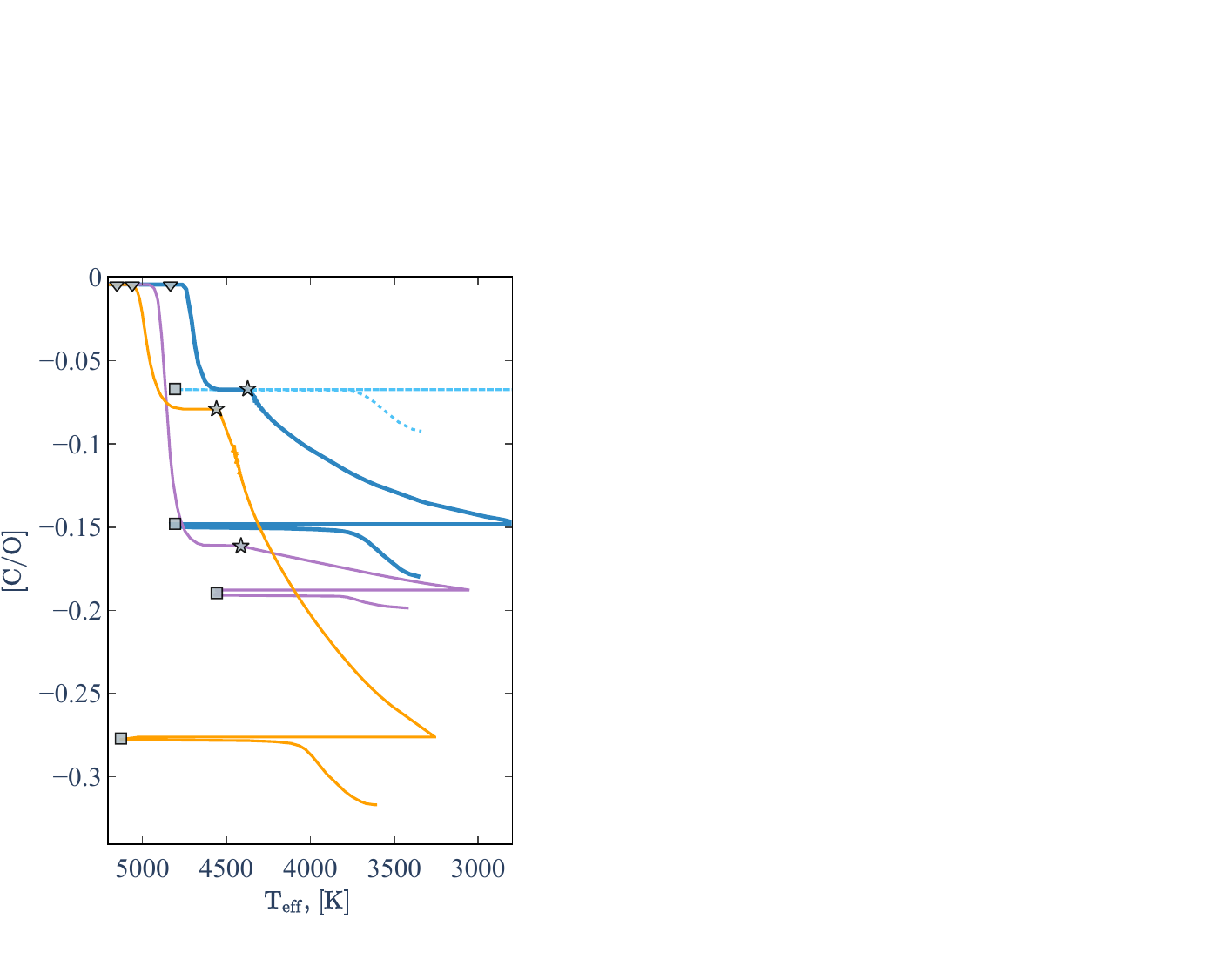} 
\includegraphics[width=0.49\hsize,clip=true,trim= 0cm 0.5cm 13.5cm 5cm]{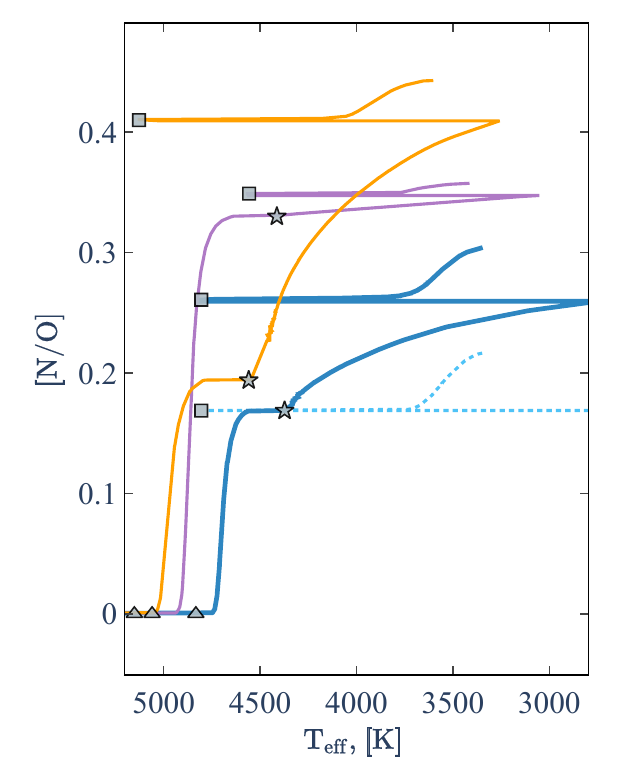} 
  \caption{Hertzsprung-Russel diagram of a 1.0M$_\odot$ model at solar metallicity (\textit{Top-left panel)}.The evolution of $^{12}$C/$^{13}$C (\textit{Top-right panel}), [C/O] (\textit{bottom-left panel}), and [N/O] (\textit{bottom-left panel}) for 1.0M$_\odot$ and 1.5M$_\odot$ model at solar metallicity (blue and purple lines, respectively) and for 1.0M$_\odot$ at [Fe/H]=-0.54 (orange line) from the subgiant branch to the RGB tip and later from the core He-burning to AGB phase. Tracks show the predictions behaviour from stellar evolution models taking into account or not the effects of thermohaline mixing (solid and dashed lines, respectively). The position of the beginning of the first dredge-up, the RGB bump, and the core He-burning phase are indicated on each panel (triangle, star, and square, respectively). }
  \label{evolution}
\end{figure}

Properties of stars simulated by the BGM are deduced from STAREVOL evolutionary tracks, which may or may not account of the impact of thermohaline mixing on surface properties. Figure \ref{evolution} presents some of these tracks and the impacts of thermohaline instability on the surface values of $^{12}$C/$^{13}$C, [C/O], and [N/O] as a function of effective temperature. After the main sequence, once a star has burnt its hydrogen in the centre, it evolves into the red giant branch. The core of the star contracts and the convective envelope expands and dives towards deeper regions where H-burning occurred while the star was on the main sequence. This episode, called the first dredge-up (1DUP), changes the surface abundances of chemical elements such as carbon and nitrogen. The 1DUP induces a decrease of $^{12}$C/$^{13}$C and of [C/O] while [N/O] increases (see Fig.~\ref{evolution}). The post-1DUP values of $^{12}$C/$^{13}$C and [C/O] are lower when mass increases and metallicity decreases while [N/O] has the opposite behaviour.

When the effects of thermohaline mixing are not taken into account, the surface abundances do not change along the red giant branch and during the central helium burning phase until the 2DUP (see dashed lines on Fig.~\ref{evolution}). In particular, the post-1DUP carbon isotopic ratio range between $\sim$ 25 and 30, which is higher than its observational value in clump stars and in RGB stars that have passed the RGB bump (see Fig.~\ref{c1213_lit}). Only after the end of core He-burning, the surface abundance ratios change again slightly when the convective envelope deepens again as the star reaches the early-AGB.

On the other hand, we see in Fig.~\ref{evolution} that the double diffusive thermohaline instability has a significant impact on the surface abundances of the light elements, such as the carbon isotopes and nitrogen (also on lithium, not shown here, but see e.g. \citealt{ChaZah07a,ChaLag10,2020A&A...633A..34C,Magrini21} who use models with the same prescription and assumptions for the thermohaline diffusion coefficient).
From the RGB-bump luminosity to the RGB tip, the surface abundance of $^{12}$C decreases while $^{13}$C and N increase, and O stays constant, resulting in a decrease of the $^{12}$C/$^{13}$C and C/O ratios and an increase of the N/O ratio (see the solid lines on Fig.~\ref{evolution}). The surface abundances do not change during the core He-burning phase until the second dredge-up occurs, similarly to the classical case. With the prescription we use in the models, the efficiency of this instability varies with the stellar mass and metallicity \citep[for more details see][]{Lagarde12a}. The less massive the star and the lower its metallicity, the more efficient the mixing. We discuss below how these differences affect the predictions of the BGM model.

\begin{figure*}
  \centering
  \includegraphics[width=0.9\hsize,clip=true,trim= 0cm 0.8cm 0cm 5.5cm]{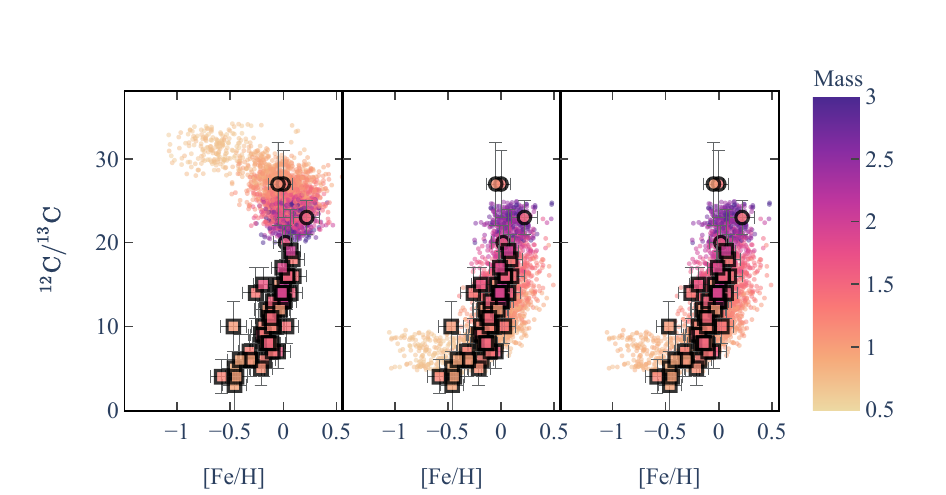}
  \caption{$^{12}$C/$^{13}$C versus [Fe/H] for core He-burning and RGB stars in our sample (squares and circles, respectively) compared with the BGM simulations for core He-burning stars only (dots) of the \textit{Kepler} field and taking into account or not the effects of thermohaline mixing (right and left panels respectively). The stellar mass is colour-coded for both observations (using seismic determinations from PARAM) and simulations (predicted by BGM using STAREVOL stellar evolution models from \citealt{Lagarde17,Lagarde19}). In the right panel, we decrease the mass loss rate on the RGB by a factor of two compare to the left and middle panels (see text for details)}
  \label{c1213vsFeh}
\end{figure*}

\begin{figure*}
  \centering
     \includegraphics[width=0.9\hsize,clip=true,trim= 0cm 0.8cm 0cm 5.5cm]{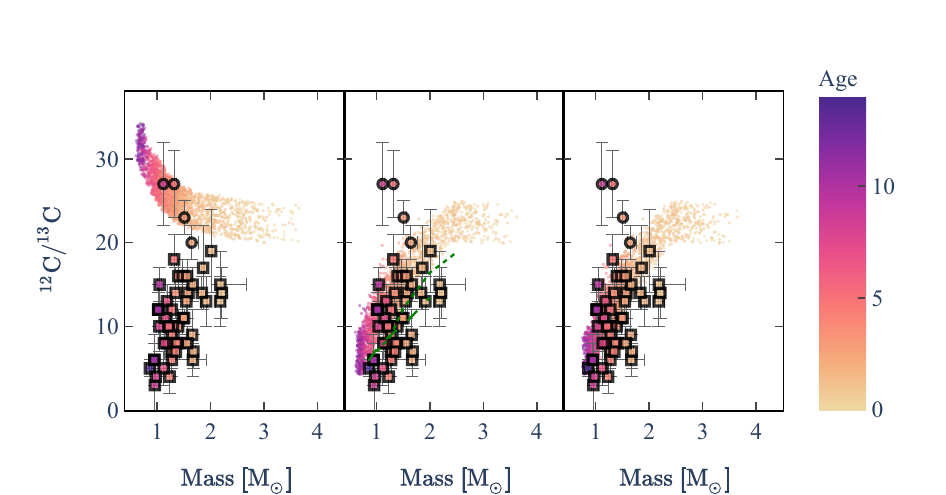}
  \caption{Same as Figure~\ref{c1213vsFeh} for $^{12}$C/$^{13}$C as a function of stellar mass (in the right panel, we decrease the mass loss rate on the RGB by a factor of two compare to the left and right panels; see text for details). The stellar age is colour-coded for both observations and BGM simulations. The green lines in the middle panel show the $^{12}$C/$^{13}$C at the core He-burning phase predicted by stellar evolution models computed by \citet{Lagarde12a} and including both thermohaline and rotation-induced mixing (at V/V$_{crit}\sim$0.3) at [Fe/H]=0 and [Fe/H]=$-0.86$ (dotted and dashed line, respectively).}
  \label{c1213vsMass}
\end{figure*}

\begin{figure*}
  \centering
  \includegraphics[width=0.8\hsize,clip=true,trim= 0cm 0.8cm 0cm 5.5cm]{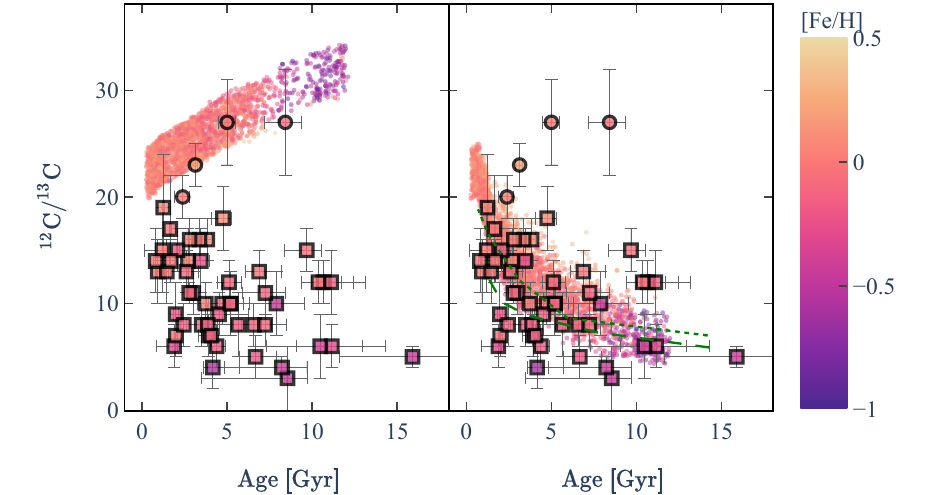}    
  \caption{Same as Figure~\ref{c1213vsFeh} for $^{12}$C/$^{13}$C as a function of stellar age, with the stellar metallicity colour-coded for both observations and simulations. The green lines in the right pannel have the same meaning as in Fig.~\ref{c1213vsMass}.
  }
\label{c1213vsAge}
\end{figure*}

\subsection{$^{12}$C/$^{13}$C predicted and observed in the {\it{Kepler}} field} 
\label{cir BGM vs Kepler}

Figures~\ref{c1213vsFeh}, \ref{c1213vsMass} and \ref{c1213vsAge} show the carbon isotopic ratio as a function of [Fe/H], stellar mass, and age predicted by the BGM based on STAREVOL evolution tracks and compared with the observed abundances of our {\it{Kepler}} golden sample with asteroseismic masses and ages from PARAM. 
The simulations performed with the BGM show only the core He-burning stars, unlike the observations where, in addition to the core He-burning stars (squares), we also show the four stars with a luminosity lower than the RGB-bump and for which we could determine the carbon isotopic ratio are also shown (circles). Since the thermohaline mixing is efficient only above the RGB bump, the $^{12}$C/$^{13}$C observed at the surface of these four first-ascent-RGB stars are well reproduced by simulations that do or do not include thermohaline mixing, within the error bars (those are larger for the less evolved stars that may not have finished the 1DUP, hence are less enriched in $^{13}$C). These stars act as an indicator of the mixing efficiency between two different evolutionary stages (i.e., before the stars reach the RGB bump and the core He-burning phase). As already anticipated, all the core-He burning stars show carbon isotopic ratio lower than predicted by classical models, but they are well reproduced when thermohaline mixing is accounted for. 

Simulations indicate that under the presence of thermohaline mixing, $^{12}$C/$^{13}$C decreases as [Fe/H] decreases, reaching a "plateau" value around 6-7 at the lowest metallicities (see Fig.\ref{c1213vsFeh}). The core He-burning stars of our golden/seismic sample confirm this ``banana-shaped" behaviour already seen in Fig.~\ref{c1213_lit} which combines the various observations published in the literature. The model perfectly reproduces this observational behaviour, confirming the importance of a mixing process such as the thermohaline mixing for understanding the carbon abundance of giant-field stars.
However, since we compare BGM model predictions for ages and masses based on stellar evolution models to data with seismic ages and masses from PARAM, we do not expect a perfect match when it comes to these two quantities. In particular, the predicted masses of the simulated stellar population depend on the mass loss rate prescription used in the stellar evolution models. \citet{Lagarde17,Lagarde19} used \citet{Reimers75} mass loss prescription with the $\eta$ parameter equal to 0.5. We compare in Fig.~\ref{c1213vsMass} the BGM masses for this value and for $\eta=0.25$ (see also Fig.~\ref{c1213vsFeh}). For this, we simply consider that the difference between the initial and the clump masses of each star of the simulated population is decreased by a factor of two. This leads to an ever better agreement with the asteroseismic masses, hence between the observed and predicted carbon isotopic ratio as a function of mass.  

Generally, and as previously underlined by \citet{ChaLag10} and \citet{Lagarde12a}, three different mass domains can be considered to measure the efficiency of thermohaline instability: 

\begin{itemize}
\item For low-mass (M$<$1.25M$_\odot$) and low-metallicity stars, thermohaline mixing is the most efficient transport process, decreasing the surface abundance of $^{12}$C and increasing the surface abundances of $^{13}$C and nitrogen. This is why the simulations presented here reproduce very well the $^{12}$C/$^{13}$C observed in this mass and metallicity domain (see Fig.~\ref{c1213vsMass} and Fig.\ref{c1213vsFeh}). Interestingly, the simulations favour a lower mass loss rate on the RGB than the one that was used in the stellar evolution models. As these stars are also the oldest stars in our Galaxy, the simulations also explain very well the $^{12}$C/$^{13}$C observed at the surface of the oldest stars (see Fig.~\ref{c1213vsAge}).

\item For intermediate-mass stars (1.25$\leq$M$\leq$2.2M$_\odot$) our simulations including the effects of thermohaline mixing have slightly higher $^{12}$C/$^{13}$C than observations, but they come in very good agreement when we consider mass loss rates lower than in the original stellar evolution models (see Fig.~\ref{c1213vsMass}). In addition to thermohaline mixing on the RGB, rotation-induced mixing on the main sequence (not included in the stellar tracks used in the BGM simulations, but see the green lines in the figures) has also been shown to lower the surface carbon isotopic of RGB stars in this particular mass range (see e.g. Figure~17 of \citealt{ChaLag10} to see the impact of different initial velocities). 

\item More massive stars do not experience thermohaline mixing as they do not go through the RGB-bump during their short first ascent of the red giant branch, making thermohaline instability irrelevant for them. The selection of stars for this study was specifically aimed at testing the efficiency of thermohaline mixing, which is why there are no stars in this mass range. 
\end{itemize}

Finally, we show in Fig.~\ref{c1213vsAge} the predicted behaviour of the carbon isotopic ratio as a function of the ages of the stars of the simulated population, and compare with the data for the golden sample. We see that the uncertainty on the ages obtained with PARAM are larger than the uncertainties of the masses \citep[see e.g.][for a discussion]{ChMi13}.

\subsection{CNO abundance ratios}

\begin{figure*}
  \centering
\includegraphics[width=0.8\hsize,clip=true,trim= 0cm 0.5cm 0cm 6.5cm]{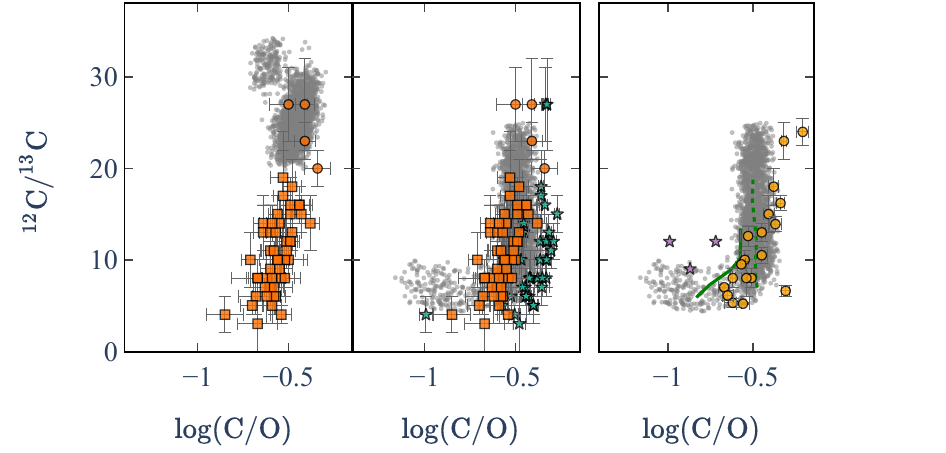}    
  \caption{$^{12}$C/$^{13}$C as a function of log(C/O) for core He-burning 
  and RGB stars 
  in our sample (orange squares and circles, respectively) compared to the BGM simulations for core He-burning stars only. Simulations are taking into account the thermohaline mixing or not (right and middle/left panels, respectively, grey dots). The C/O ratio derived by \citet{APOGEE17} for stars in common is also shown,  using our carbon isotopic ratios (blue stars on middle panel). The abundances derived by \citet{Galan17}, \citet{Galan2023} and \citet{Gratton00} are also compared with the simulation on the right panel. The $^{12}$C/$^{13}$C at the core He-burning phase predicted by stellar evolution models computed by \citet{Lagarde12a} and including both thermohaline and rotation-induced mixing (at V/V$_{crit}\sim$0.3) are shown: at [Fe/H]=0 and [Fe/H]=$-0.86$ (green dotted and solid lines, respectively). }
  \label{CO}
\end{figure*}

\begin{figure*}
  \centering
\includegraphics[width=0.8\hsize,clip=true,trim= 0cm 0.5cm 0cm 6.5cm]{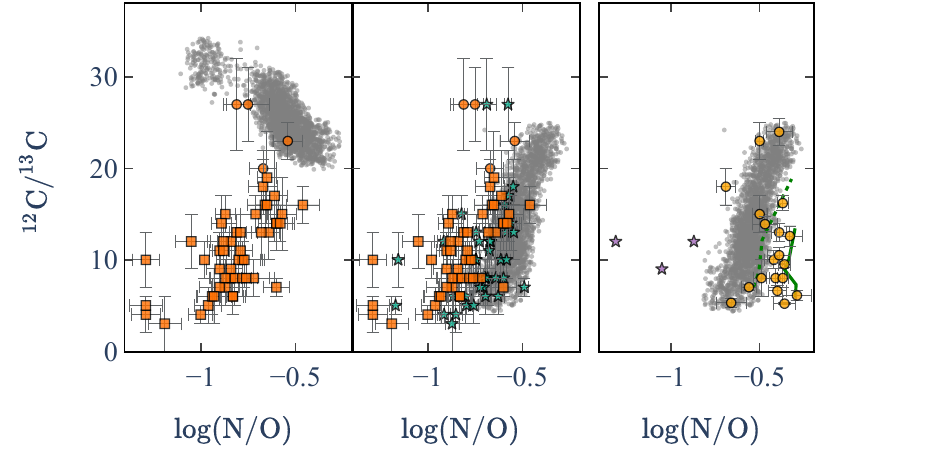}    
  \caption{Same as Figure \ref{CO}. $^{12}$C/$^{13}$C as a function of log(N/O).}
  \label{NO}
\end{figure*}

As shown in Sect.~\ref{thermoh}, the thermohaline mixing has a significant effect on the C and N surface abundances, but leaves the O unchanged. Figures \ref{CO} and \ref{NO} present the $^{12}$C/$^{13}$C as a function of C/O and N/O observed for giants in our sample (left and middle panels). First-ascent RGB stars have a carbon isotopic ratio as well as C/O and N/O in agreement with the post 1DUP value. Observations of core He-burning stars show that a low $^{12}$C/$^{13}$C is associated with low C/O and N/O ratios. 
The predicted behaviour of $^{12}$C/$^{13}$C with C/O and N/O for core He-burning stars is correct, although our simulations predict a higher N/O ratio than our observations. In order to understand this discrepancy in nitrogen abundance derived from our observations and simulations, we also show the APOGEE DR17 data for C/O and N/O for the stars we have in common (blue star symbols on the middle panels in Fig~ \ref{CO} and \ref{NO}, using our carbon isotopic values). As already shown in Sect.\ref{section:comparison} the abundances of [C/Fe] and [N/Fe] are higher in APOGEE DR17 catalogue than in our study, while [O/Fe] is lower. This is why C/O and N/O are higher in the APOGEE DR17 catalogue than in our study. An explanation for these differences might be found in slightly different atmospheric parameters (ours are asteroseismic) and spectral lines used (in the APOGEE survey, the infrared OH, CO, and CN lines were investigated while we investigated optical [O\,I], C$_2$, and CN lines). In any case, the observational trends for $^{12}$C/$^{13}$C versus C/O and N/O are similar in both samples, and it can be explained only when thermohaline mixing is accounted for in the simulations.

In addition to the APOGEE data, we also show in Fig~ \ref{CO} and \ref{NO} the values of $^{12}$C/$^{13}$C as well as CNO abundances \citet{Galan17} determined for a sample of atypical 
stars (yellow circles on the right panels of Figs.\ref{CO} and \ref{NO}).
These are so-called symbiotic stars, which are binaries composed of a star in the later stages of evolution and a stellar remnant. They derived the photospheric composition of the giant stars in the binary system, and found enhanced $^{14}$N, depleted $^{12}$C, and decreased $^{12}$C/$^{13}$C. These observations show a C/O ratio in agreement with our observations and APOGEE DR17, but a much higher N/O ratio. These higher N/O ratio values are well reproduced by BGM simulations and with models including the effects of both thermohaline and  rotation-induced mixing.  

In our sample of stars we also have twelve atypical eruptive stars. Eruptive variables are stars varying in brightness because of violent processes and flares occurring in their chromosphere and coronae. The light changes are usually accompanied by shell events or mass outflow in the form of stellar winds or variable intensity and/or by interactions with the surrounding interstellar medium (\citealt{Good03}). Even though that sometimes flashes according to theoretical predictions may be very energetic and alter the stellar chemical composition (e.g. \citealt{Sackmann74}), in our sample of eruptive stars, the CNO and $^{12}$C/$^{13}$C abundances do not differ from the He-core burning stars.

\begin{figure}
  \centering
\includegraphics[width=\hsize,clip=true,trim= 0cm 0.5cm 1cm 2cm]{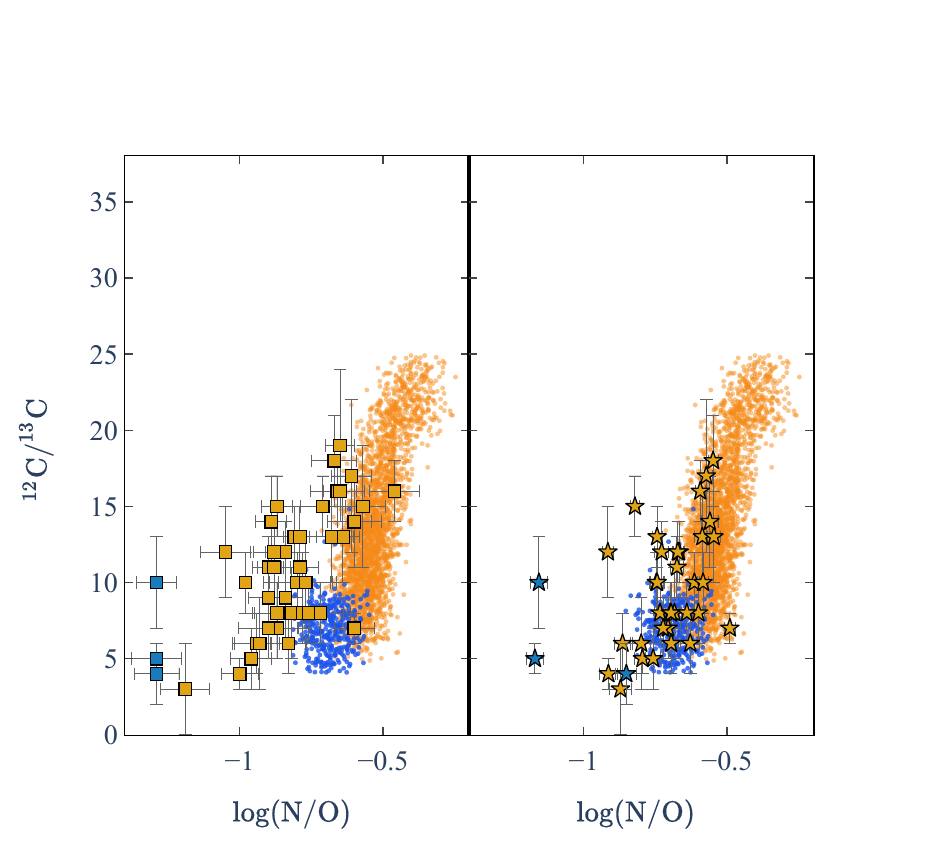}
\caption{$^{12}$C/$^{13}$C versus log(N/O) for core He-burning stars in our sample (left panel) and versus log(N/O) derived by APOGEE DR17 (right panel) compared to the BGM simulations including the effects of thermohaline mixing. Thin and thick discs stars according to their \gaia kinematics are represented by the orange and blue symbols. The same colour code is used in the BGM simulation. }
  \label{NO_pop}
\end{figure}

Finally, using the kinematics from the \gaia satellite, we separate our sample into two populations: stars belonging to the Galactic thin disc and the thick disc (see Sect.\ref{kine}). Figure \ref{NO_pop} shows the carbon isotopic ratio as a function of the N/O ratio for core He-burning stars only for these two populations using both our data and that of APOGEE. Although the $^{12}$C/$^{13}$C vs N/O trends are the same in the simulations and in the observations for both N/O determinations, there is a shift between the observations and the simulation.
Part of it comes from the differences between our values for N/O and those from APOGEE, as discussed above. Furthermore, the thick disc stars have a very low N/O ratio but are also the stars for which the difference in N/O ratio between APOGEE and our determination is the largest. Since our study contains only three  core He-burning thick disc stars with $^{12}$C/$^{13}$C as well as N/O ratio, a more detailed study of the nitrogen and oxygen observed at the surface of a large number of thick disc stars is needed to understand this difference.

\section{Conclusions}
\label{conclu}

Thanks to the rich observational context, the surface properties of stars are derived from spectroscopic observations, such as their photospheric composition, while asteroseismology allows us better determine their masses and ages. One of the key elements in effectively constraining the mixing occurring in giant stars is the carbon isotopic ratio. Only seven studies have derived and published the $^{12}$C/$^{13}$C in field giant stars, with only one using asteroseismic constraints for a couple of objects. The compilation of these studies presented in Sect\ref{compil} show that $^{12}$C/$^{13}$C increases with [Fe/H], the mass, and seems to decrease with age. 
In this study, we have therefore built a catalog of 71 stars in the \textit{Kepler} field with the best properties to constrain the efficiency of the mixing taking place between the end of the first dredge-up and the central He-burning phase. We have carried out spectroscopic follow-up using the FIbre-fed Echelle Spectrograph on the Nordic Optical Telescope for a sample of giant \textit{Kepler} stars. We have determined the atmospheric properties ($T\mathrm{_{eff}}$, log $g$) as well as the chemical abundances in C, N, O, Mg, Fe, O and the $^{12}$C/$^{13}$C isotopic ratio for 71 giant stars (see Sect.\ref{abundance}). We validated some of our determinations by comparing them with those already published in the literature by various surveys (from \gaia \citep{RecioBlanco23}, PASTEL \citep{Soubiran22}, LAMOST DR7 \citep{LAMOSTDR7} and APOGEE DR17 \citep{APOGEE17}). 

As our sample stars were observed by the \textit{Kepler} satellite, we were able to determine their mass and age from seismic observables ($\Delta\nu$, $\nu_{max}$), and we could unambiguously determine whether they are ascending the first ascent RGB stars or sitting in the core-He burning phase 
using $\Delta\Pi_{\ell=1}$ (see Sect.\ref{sismo}). We show that our sample is composed of 9 red giant stars and 62 core He-burning stars. In order to estimate possible uncertainties, we determined the mass and age of stars in our sample using the Stellar Parameters INferred Systematically \citep{SPINS20} and our stellar evolution models \citep{Lagarde17}, showing good consistency between the two determinations.

To harness the complete potential of our extensive catalogue, considering both the Milky Way's evolution and the impact of stellar evolution – which can influence stellar properties to varying degrees – we employed simulations generated through the Besan\c con stellar population synthesis model of our Galaxy (BGM). We built mock catalogues using the BGM in which stellar evolution models taking into account (and not) the effects of thermohaline instability \citep[][]{Lagarde17, Lagarde19}. As of today, thermohaline mixing stands as the sole physical mechanism suggested in the literature to account for the photospheric composition of evolved red giant stars, even with simplistic analytical prescriptionsas those used in our stellar evolution models and that can not yet capture the full picture of this complex double diffusive instability \citep[][]{SenGar18,Harrington19,Fraser22}. In this study, we focused on discussing this non-canonical, extra mixing in red giant stars. This was achieved by investigating the changes in the $^{12}$C/$^{13}$C ratio throughout both the red giant branch and the helium-burning phase. Additionally, we analysed how this ratio varies concerning stellar age, mass, and metallicity. Finally, we investigated CNO abundances as a function of carbon isotopic ratio. 
Our main results can be summarised as follows:

\begin{itemize} 
\item Thanks to asteroseismology, we confirm that the carbon isotopic ratio at the surface of core He-burning stars is lower than that of first ascent RGB stars. Thermohaline instability develops from the luminosity of the RGB-bump, decreasing surface abundances of $^{12}$C and increasing $^{13}$C. Simulations including this mixing therefore show a lower $^{12}$C/$^{13}$C ratio during the core He-burning phase compared with the red giant phase just before the RGB bump (e.g., $^{12}$C/$^{13}$C $\sim$8 and 30, respectively for a 1.0\,$M_\odot$ at [Fe/H]=$-0.54$). 
\item Considering only core He-burning stars, we confirm that the carbon isotopic ratio measured at the surface of these objects increases with [Fe/H] and stellar mass while it decreases with stellar age. Simulations done with the BGM and including the effects of thermohaline mixing explain these trends in an exceptional way (see Sect.\ref{thermoh}). Standard models that do not include the effects of thermohaline mixing show inverse trends that can explain the surface abundances of giant stars just after the 1DUP, but not the abundances of more evolved stars such as core He-burning stars. 
\item  Our spectroscopic analysis shows that low $^{12}$C/$^{13}$C values are correlated with low C/O and N/O ratios. These behaviours are well predicted by simulations that include the effects of thermohaline mixing. As already shown by \citet{ChaLag10} and \citet{Lagarde19}, thermohaline mixing decreases the carbon abundance while nitrogen abundance increases and oxygen stay constant. Although the observed behaviours are well reproduced, with some notable shifts in N/O between our values and values from the literature, it should be noted that there are the stars belonging to the thick disc that show a mismatch between the nitrogen abundances derived from the spectra and those simulated. 
\end{itemize}

The relatively large sample we gathered in our pioneer study 
is still limited due to the limitations of the asteroseismic missions in terms of either their coverage across the Galactic volume or the duration of observations. These limitations subsequently constrain the level of precision attainable for deducing key stellar properties, including the age of stars. However, the Transiting Exoplanet Survey Satellite (TESS, \citealt{Ricker15}) and the forthcoming PLATO \citep{PLATO} space missions are set to offer stellar age estimates as well as evolutionary states across various directions within the Milky Way, encompassing a substantial number of stars. Furthermore, future spectroscopic surveys will provide a complementary chemical vision of our Galaxy, allowing investigation of the CNO -- Mass/Age or metallicity relation and probing all stellar populations in the Milky Way.

\begin{acknowledgements}
     Based on observations made with the Nordic Optical Telescope, owned in collaboration by the University of Turku and Aarhus University, and operated jointly by Aarhus University, the University of Turku and the University of Oslo, representing Denmark, Finland and Norway, the University of Iceland and Stockholm University at the Observatorio del Roque de los Muchachos, La Palma, Spain, of the Instituto de Astrofisica de Canarias. The observational project has received funding from the European Union's Horizon 2020 research and innovation programme under grant agreement No 730890 (OPTICON). N.L., C.C., C.R. acknowledge financial support from "Programme National de Physique Stellaire" (PNPS) and from the "Programme National Cosmology et Galaxies (PNCG)" of CNRS/INSU, France. C.C. thanks the Swiss National Science Foundation (SNF; Project 200020-192039). R.M., A.D., G.T., B.B. acknowledge funding from the Lithuanian Science Council (LMTLT, grant No. P-MIP-23-24).
     The “Stellar Parameters INferred Systematically” (SPInS) code is a spin-off of the “Asteroseismic Inference on a Massive Scale” (AIMS) project, one of the deliverables of the SpaceINN network, funded by the European Community’s Seventh Framework Programme (FP7/2007-2013) under grant agreement no. 312844. SPInS was initially created for the 5th International Young Astronomer School “Scientific Exploitation of Gaia Data” held in Paris, (26 February - 2 March 2018), as a simple tool to estimate stellar ages, as well as other stellar properties.
\end{acknowledgements}

\bibliographystyle{aa}
\bibliography{Reference.bib}

\begin{thebibliography}{131}
\expandafter\ifx\csname natexlab\endcsname\relax\def\natexlab#1{#1}\fi

\bibitem[{{Abdurro'uf} {et~al.}(2022){Abdurro'uf}, {Accetta}, {Aerts}, {Silva
  Aguirre}, {Ahumada}, {Ajgaonkar}, {Filiz Ak}, {Alam}, {Allende Prieto},
  {Almeida}, \& et~al.}]{APOGEEDR17}
{Abdurro'uf}, {Accetta}, K., {Aerts}, C., {et~al.} 2022, \apjs, 259, 35

\bibitem[{{Aguilera-G{\'o}mez} {et~al.}(2023){Aguilera-G{\'o}mez}, {Jones}, \&
  {Chanam{\'e}}}]{AguileraGomez23}
{Aguilera-G{\'o}mez}, C., {Jones}, M.~I., \& {Chanam{\'e}}, J. 2023, \aap, 670,
  A73

\bibitem[{{Alvarez} \& {Plez}(1998)}]{Alvarez1998}
{Alvarez}, R. \& {Plez}, B. 1998, \aap, 330, 1109

\bibitem[{{Amard} {et~al.}(2019){Amard}, {Palacios}, {Charbonnel}, {Gallet},
  {Georgy}, {Lagarde}, \& {Siess}}]{Amard19}
{Amard}, L., {Palacios}, A., {Charbonnel}, C., {et~al.} 2019, \aap, 631, A77

\bibitem[{{Anders} {et~al.}(2017{\natexlab{a}}){Anders}, {Chiappini},
  {Minchev}, {Miglio}, {Montalb{\'a}n}, {Mosser}, {Rodrigues}, {Santiago},
  {Baudin}, {Beers}, {da Costa}, {Garc{\'{\i}}a},
  {Garc{\'{\i}}a-Hern{\'a}ndez}, {Holtzman}, {Maia}, {Majewski}, {Mathur},
  {Noels-Grotsch}, {Pan}, {Schneider}, {Schultheis}, {Steinmetz}, {Valentini},
  \& {Zamora}}]{Anders17b}
{Anders}, F., {Chiappini}, C., {Minchev}, I., {et~al.} 2017{\natexlab{a}},
  \aap, 600, A70

\bibitem[{{Anders} {et~al.}(2017{\natexlab{b}}){Anders}, {Chiappini},
  {Rodrigues}, {Miglio}, {Montalb{\'a}n}, {Mosser}, {Girardi}, {Valentini},
  {Noels}, {Morel}, {Johnson}, {Schultheis}, {Baudin}, {de Assis Peralta},
  {Hekker}, {Theme{\ss}l}, {Kallinger}, {Garc{\'\i}a}, {Mathur}, {Baglin},
  {Santiago}, {Martig}, {Minchev}, {Steinmetz}, {da Costa}, {Maia}, {Allende
  Prieto}, {Cunha}, {Beers}, {Epstein}, {Garc{\'\i}a P{\'e}rez},
  {Garc{\'\i}a-Hern{\'a}ndez}, {Harding}, {Holtzman}, {Majewski},
  {M{\'e}sz{\'a}ros}, {Nidever}, {Pan}, {Pinsonneault}, {Schiavon},
  {Schneider}, {Shetrone}, {Stassun}, {Zamora}, \& {Zasowski}}]{Anders17a}
{Anders}, F., {Chiappini}, C., {Rodrigues}, T.~S., {et~al.} 2017{\natexlab{b}},
  \aap, 597, A30

\bibitem[{{Angelou} {et~al.}(2011){Angelou}, {Church}, {Stancliffe},
  {Lattanzio}, \& {Smith}}]{Angelou11}
{Angelou}, G.~C., {Church}, R.~P., {Stancliffe}, R.~J., {Lattanzio}, J.~C., \&
  {Smith}, G.~H. 2011, \apj, 728, 79

\bibitem[{{Baglin} {et~al.}(2006){Baglin}, {Auvergne}, {Boisnard}, {Lam-Trong},
  {Barge}, {Catala}, {Deleuil}, {Michel}, \& {Weiss}}]{Baglin06}
{Baglin}, A., {Auvergne}, M., {Boisnard}, L., {et~al.} 2006, in 36th COSPAR
  Scientific Assembly, Vol.~36, 3749

\bibitem[{{Bailer-Jones} {et~al.}(2021){Bailer-Jones}, {Rybizki}, {Fouesneau},
  {Demleitner}, \& {Andrae}}]{BailerJones21}
{Bailer-Jones}, C.~A.~L., {Rybizki}, J., {Fouesneau}, M., {Demleitner}, M., \&
  {Andrae}, R. 2021, \aj, 161, 147

\bibitem[{{Bedding} {et~al.}(2011){Bedding}, {Mosser}, {Huber},
  {Montalb{\'a}n}, {Beck}, {Christensen-Dalsgaard}, {Elsworth},
  {Garc{\'{\i}}a}, {Miglio}, {Stello}, {White}, {De Ridder}, {Hekker}, {Aerts},
  {Barban}, {Belkacem}, {Broomhall}, {Brown}, {Buzasi}, {Carrier}, {Chaplin},
  {di Mauro}, {Dupret}, {Frandsen}, {Gilliland}, {Goupil}, {Jenkins},
  {Kallinger}, {Kawaler}, {Kjeldsen}, {Mathur}, {Noels}, {Aguirre}, \&
  {Ventura}}]{Bedding11}
{Bedding}, T.~R., {Mosser}, B., {Huber}, D., {et~al.} 2011, \nat, 471, 608

\bibitem[{{Belkacem} {et~al.}(2011){Belkacem}, {Goupil}, {Dupret}, {Samadi},
  {Baudin}, {Noels}, \& {Mosser}}]{Belkacem11}
{Belkacem}, K., {Goupil}, M.~J., {Dupret}, M.~A., {et~al.} 2011, \aap, 530,
  A142

\bibitem[{{Bell} {et~al.}(1990){Bell}, {Briley}, \& {Smith}}]{Bell90}
{Bell}, R.~A., {Briley}, M.~M., \& {Smith}, G.~H. 1990, \aj, 100, 187

\bibitem[{{Bienaym{\'e}} {et~al.}(2018){Bienaym{\'e}}, {Leca}, \&
  {Robin}}]{Bienayme18}
{Bienaym{\'e}}, O., {Leca}, J., \& {Robin}, A.~C. 2018, \aap, 620, A103

\bibitem[{{Bienaym{\'e}} {et~al.}(2015){Bienaym{\'e}}, {Robin}, \&
  {Famaey}}]{Bienayme15}
{Bienaym{\'e}}, O., {Robin}, A.~C., \& {Famaey}, B. 2015, \aap, 581, A123

\bibitem[{{Blanton} {et~al.}(2017){Blanton}, {Bershady}, {Abolfathi},
  {Albareti}, {Allende Prieto}, {Almeida}, {Alonso-Garc{\'{\i}}a}, {Anders},
  {Anderson}, {Andrews}, \& et~al.}]{APOGEE17}
{Blanton}, M.~R., {Bershady}, M.~A., {Abolfathi}, B., {et~al.} 2017, \aj, 154,
  28

\bibitem[{{Borucki} {et~al.}(2010){Borucki}, {Koch}, {Basri}, {Batalha},
  {Brown}, {Caldwell}, {Caldwell}, {Christensen-Dalsgaard}, {Cochran},
  {DeVore}, {Dunham}, {Dupree}, {Gautier}, {Geary}, {Gilliland}, {Gould},
  {Howell}, {Jenkins}, {Kondo}, {Latham}, {Marcy}, {Meibom}, {Kjeldsen},
  {Lissauer}, {Monet}, {Morrison}, {Sasselov}, {Tarter}, {Boss}, {Brownlee},
  {Owen}, {Buzasi}, {Charbonneau}, {Doyle}, {Fortney}, {Ford}, {Holman},
  {Seager}, {Steffen}, {Welsh}, {Rowe}, {Anderson}, {Buchhave}, {Ciardi},
  {Walkowicz}, {Sherry}, {Horch}, {Isaacson}, {Everett}, {Fischer}, {Torres},
  {Johnson}, {Endl}, {MacQueen}, {Bryson}, {Dotson}, {Haas}, {Kolodziejczak},
  {Van Cleve}, {Chandrasekaran}, {Twicken}, {Quintana}, {Clarke}, {Allen},
  {Li}, {Wu}, {Tenenbaum}, {Verner}, {Bruhweiler}, {Barnes}, \&
  {Prsa}}]{Borucki10}
{Borucki}, W.~J., {Koch}, D., {Basri}, G., {et~al.} 2010, Science, 327, 977

\bibitem[{{Bouret} {et~al.}(2021){Bouret}, {Martins}, {Hillier}, {Marcolino},
  {Rocha-Pinto}, {Georgy}, {Lanz}, \& {Hubeny}}]{Bouret21}
{Bouret}, J.~C., {Martins}, F., {Hillier}, D.~J., {et~al.} 2021, \aap, 647,
  A134

\bibitem[{{Brown} \& {Wallerstein}(1989)}]{BrWa89}
{Brown}, J.~A. \& {Wallerstein}, G. 1989, \aj, 98, 1643

\bibitem[{{Brown} {et~al.}(2013){Brown}, {Garaud}, \& {Stellmach}}]{Brown13}
{Brown}, J.~M., {Garaud}, P., \& {Stellmach}, S. 2013, \apj, 768, 34

\bibitem[{{Brown} {et~al.}(1991){Brown}, {Gilliland}, {Noyes}, \&
  {Ramsey}}]{Brown91}
{Brown}, T.~M., {Gilliland}, R.~L., {Noyes}, R.~W., \& {Ramsey}, L.~W. 1991,
  \apj, 368, 599

\bibitem[{{Buder} {et~al.}(2021){Buder}, {Sharma}, {Kos}, {Amarsi},
  {Nordlander}, {Lind}, {Martell}, {Asplund}, {Bland-Hawthorn}, {Casey}, {de
  Silva}, {D'Orazi}, {Freeman}, {Hayden}, {Lewis}, {Lin}, {Schlesinger},
  {Simpson}, {Stello}, {Zucker}, {Zwitter}, {Beeson}, {Buck}, {Casagrande},
  {Clark}, {{\v{C}}otar}, {da Costa}, {de Grijs}, {Feuillet}, {Horner},
  {Kafle}, {Khanna}, {Kobayashi}, {Liu}, {Montet}, {Nandakumar}, {Nataf},
  {Ness}, {Spina}, {Tepper-Garc{\'\i}a}, {Ting}, {Traven},
  {Vogrin{\v{c}}i{\v{c}}}, {Wittenmyer}, {Wyse}, {{\v{Z}}erjal}, \& {Galah
  Collaboration}}]{GALAHDR3}
{Buder}, S., {Sharma}, S., {Kos}, J., {et~al.} 2021, \mnras, 506, 150

\bibitem[{{Chaplin} \& {Miglio}(2013)}]{ChMi13}
{Chaplin}, W.~J. \& {Miglio}, A. 2013, \araa, 51, 353

\bibitem[{{Charbonnel}(1994)}]{Charbonnel94}
{Charbonnel}, C. 1994, \aap, 282, 811

\bibitem[{{Charbonnel}(1995)}]{Charbonnel95}
{Charbonnel}, C. 1995, \apjl, 453, L41+

\bibitem[{{Charbonnel} \& {Balachandran}(2000)}]{ChBa00}
{Charbonnel}, C. \& {Balachandran}, S.~C. 2000, \aap, 359, 563

\bibitem[{{Charbonnel} {et~al.}(1998){Charbonnel}, {Brown}, \&
  {Wallerstein}}]{Charbonnel98}
{Charbonnel}, C., {Brown}, J.~A., \& {Wallerstein}, G. 1998, \aap, 332, 204

\bibitem[{{Charbonnel} \& {Lagarde}(2010)}]{ChaLag10}
{Charbonnel}, C. \& {Lagarde}, N. 2010, \aap, 522, A10

\bibitem[{{Charbonnel} {et~al.}(2020{\natexlab{a}}){Charbonnel}, {Lagarde},
  {Jasniewicz}, {North}, {Shetrone}, {Krugler Hollek}, {Smith}, {Smiljanic},
  {Palacios}, \& {Ottoni}}]{ChaLag20}
{Charbonnel}, C., {Lagarde}, N., {Jasniewicz}, G., {et~al.} 2020{\natexlab{a}},
  \aap, 633, A34

\bibitem[{{Charbonnel} {et~al.}(2020{\natexlab{b}}){Charbonnel}, {Lagarde},
  {Jasniewicz}, {North}, {Shetrone}, {Krugler Hollek}, {Smith}, {Smiljanic},
  {Palacios}, \& {Ottoni}}]{2020A&A...633A..34C}
{Charbonnel}, C., {Lagarde}, N., {Jasniewicz}, G., {et~al.} 2020{\natexlab{b}},
  \aap, 633, A34

\bibitem[{{Charbonnel} \& {Zahn}(2007{\natexlab{a}})}]{ChaZah07b}
{Charbonnel}, C. \& {Zahn}, J. 2007{\natexlab{a}}, \aap, 476, L29

\bibitem[{{Charbonnel} \& {Zahn}(2007{\natexlab{b}})}]{ChaZah07a}
{Charbonnel}, C. \& {Zahn}, J.-P. 2007{\natexlab{b}}, \aap, 467, L15

\bibitem[{{Creevey} {et~al.}(2013){Creevey}, {Th{\'e}venin}, {Basu}, {Chaplin},
  {Bigot}, {Elsworth}, {Huber}, {Monteiro}, \& {Serenelli}}]{Creevey2013}
{Creevey}, O.~L., {Th{\'e}venin}, F., {Basu}, S., {et~al.} 2013, \mnras, 431,
  2419

\bibitem[{{Czekaj} {et~al.}(2014){Czekaj}, {Robin}, {Figueras}, {Luri}, \&
  {Haywood}}]{Czekaj14}
{Czekaj}, M.~A., {Robin}, A.~C., {Figueras}, F., {Luri}, X., \& {Haywood}, M.
  2014, \aap, 564, A102

\bibitem[{{Deal} {et~al.}(2020){Deal}, {Goupil}, {Marques}, {Reese}, \&
  {Lebreton}}]{Deal20}
{Deal}, M., {Goupil}, M.~J., {Marques}, J.~P., {Reese}, D.~R., \& {Lebreton},
  Y. 2020, \aap, 633, A23

\bibitem[{{Dearborn} {et~al.}(1976){Dearborn}, {Eggleton}, \&
  {Schramm}}]{Dearborn76}
{Dearborn}, D.~S.~P., {Eggleton}, P.~P., \& {Schramm}, D.~N. 1976, \apj, 203,
  455

\bibitem[{{Denissenkov}(2010)}]{Denissenkov10}
{Denissenkov}, P.~A. 2010, \apj, 723, 563

\bibitem[{{Denissenkov} \& {Merryfield}(2011)}]{DenissenkovMerryfield10}
{Denissenkov}, P.~A. \& {Merryfield}, W.~J. 2011, \apjl, 727, L8+

\bibitem[{{Dumont} {et~al.}(2021){Dumont}, {Charbonnel}, {Palacios}, \&
  {Borisov}}]{Dumont21}
{Dumont}, T., {Charbonnel}, C., {Palacios}, A., \& {Borisov}, S. 2021, \aap,
  654, A46

\bibitem[{{Eggleton} {et~al.}(2006){Eggleton}, {Dearborn}, \&
  {Lattanzio}}]{Eggleton06}
{Eggleton}, P.~P., {Dearborn}, D.~S.~P., \& {Lattanzio}, J.~C. 2006, Science,
  314, 1580

\bibitem[{{Fraser} {et~al.}(2022){Fraser}, {Joyce}, {Anders}, {Tayar}, \&
  {Cantiello}}]{Fraser22}
{Fraser}, A.~E., {Joyce}, M., {Anders}, E.~H., {Tayar}, J., \& {Cantiello}, M.
  2022, \apj, 941, 164

\bibitem[{{Fraser} {et~al.}(2023){Fraser}, {Reifenstein}, \&
  {Garaud}}]{Fraser23}
{Fraser}, A.~E., {Reifenstein}, S.~A., \& {Garaud}, P. 2023, arXiv e-prints,
  arXiv:2302.11610

\bibitem[{{Fusi Pecci} {et~al.}(1990){Fusi Pecci}, {Ferraro}, {Crocker},
  {Rood}, \& {Buonanno}}]{FusiPecci90}
{Fusi Pecci}, F., {Ferraro}, F.~R., {Crocker}, D.~A., {Rood}, R.~T., \&
  {Buonanno}, R. 1990, \aap, 238, 95

\bibitem[{{Gaia Collaboration} {et~al.}(2021){Gaia Collaboration}, {Brown},
  {Vallenari}, {Prusti}, {de Bruijne}, {Babusiaux}, {Biermann}, {Creevey},
  {Evans}, {Eyer}, {Hutton}, {Jansen}, {Jordi}, {Klioner}, {Lammers},
  {Lindegren}, {Luri}, {Mignard}, {Panem}, {Pourbaix}, {Randich}, {Sartoretti},
  {Soubiran}, {Walton}, {Arenou}, {Bailer-Jones}, {Bastian}, {Cropper},
  {Drimmel}, {Katz}, {Lattanzi}, {van Leeuwen}, {Bakker}, {Cacciari},
  {Casta{\~n}eda}, {De Angeli}, {Ducourant}, {Fabricius}, {Fouesneau},
  {Fr{\'e}mat}, {Guerra}, {Guerrier}, {Guiraud}, {Jean-Antoine Piccolo},
  {Masana}, {Messineo}, {Mowlavi}, {Nicolas}, {Nienartowicz}, {Pailler},
  {Panuzzo}, {Riclet}, {Roux}, {Seabroke}, {Sordo}, {Tanga}, {Th{\'e}venin},
  {Gracia-Abril}, {Portell}, {Teyssier}, {Altmann}, {Andrae}, {Bellas-Velidis},
  {Benson}, {Berthier}, {Blomme}, {Brugaletta}, {Burgess}, {Busso}, {Carry},
  {Cellino}, {Cheek}, {Clementini}, {Damerdji}, {Davidson}, {Delchambre},
  {Dell'Oro}, {Fern{\'a}ndez-Hern{\'a}ndez}, {Galluccio}, {Garc{\'\i}a-Lario},
  {Garcia-Reinaldos}, {Gonz{\'a}lez-N{\'u}{\~n}ez}, {Gosset}, {Haigron},
  {Halbwachs}, {Hambly}, {Harrison}, {Hatzidimitriou}, {Heiter},
  {Hern{\'a}ndez}, {Hestroffer}, {Hodgkin}, {Holl}, {Jan{\ss}en}, {Jevardat de
  Fombelle}, {Jordan}, {Krone-Martins}, {Lanzafame}, {L{\"o}ffler}, {Lorca},
  {Manteiga}, {Marchal}, {Marrese}, {Moitinho}, {Mora}, {Muinonen}, {Osborne},
  {Pancino}, {Pauwels}, {Petit}, {Recio-Blanco}, {Richards}, {Riello},
  {Rimoldini}, {Robin}, {Roegiers}, {Rybizki}, {Sarro}, {Siopis}, {Smith},
  {Sozzetti}, {Ulla}, {Utrilla}, {van Leeuwen}, {van Reeven}, {Abbas}, {Abreu
  Aramburu}, {Accart}, {Aerts}, {Aguado}, {Ajaj}, {Altavilla}, {{\'A}lvarez},
  {{\'A}lvarez Cid-Fuentes}, {Alves}, {Anderson}, {Anglada Varela}, {Antoja},
  {Audard}, {Baines}, {Baker}, {Balaguer-N{\'u}{\~n}ez}, {Balbinot}, {Balog},
  {Barache}, {Barbato}, {Barros}, {Barstow}, {Bartolom{\'e}}, {Bassilana},
  {Bauchet}, {Baudesson-Stella}, {Becciani}, {Bellazzini}, {Bernet}, {Bertone},
  {Bianchi}, {Blanco-Cuaresma}, {Boch}, {Bombrun}, {Bossini}, {Bouquillon},
  {Bragaglia}, {Bramante}, {Breedt}, {Bressan}, {Brouillet}, {Bucciarelli},
  {Burlacu}, {Busonero}, {Butkevich}, {Buzzi}, {Caffau}, {Cancelliere},
  {C{\'a}novas}, {Cantat-Gaudin}, {Carballo}, {Carlucci}, {Carnerero},
  {Carrasco}, {Casamiquela}, {Castellani}, {Castro-Ginard}, {Castro Sampol},
  {Chaoul}, {Charlot}, {Chemin}, {Chiavassa}, {Cioni}, {Comoretto}, {Cooper},
  {Cornez}, {Cowell}, {Crifo}, {Crosta}, {Crowley}, {Dafonte}, {Dapergolas},
  {David}, {David}, {de Laverny}, {De Luise}, {De March}, {De Ridder}, {de
  Souza}, {de Teodoro}, {de Torres}, {del Peloso}, {del Pozo}, {Delbo},
  {Delgado}, {Delgado}, {Delisle}, {Di Matteo}, {Diakite}, {Diener},
  {Distefano}, {Dolding}, {Eappachen}, {Edvardsson}, {Enke}, {Esquej}, {Fabre},
  {Fabrizio}, {Faigler}, {Fedorets}, {Fernique}, {Fienga}, {Figueras},
  {Fouron}, {Fragkoudi}, {Fraile}, {Franke}, {Gai}, {Garabato},
  {Garcia-Gutierrez}, {Garc{\'\i}a-Torres}, {Garofalo}, {Gavras}, {Gerlach},
  {Geyer}, {Giacobbe}, {Gilmore}, {Girona}, {Giuffrida}, {Gomel}, {Gomez},
  {Gonzalez-Santamaria}, {Gonz{\'a}lez-Vidal}, {Granvik},
  {Guti{\'e}rrez-S{\'a}nchez}, {Guy}, {Hauser}, {Haywood}, {Helmi}, {Hidalgo},
  {Hilger}, {H{\l}adczuk}, {Hobbs}, {Holland}, {Huckle}, {Jasniewicz},
  {Jonker}, {Juaristi Campillo}, {Julbe}, {Karbevska}, {Kervella}, {Khanna},
  {Kochoska}, {Kontizas}, {Kordopatis}, {Korn}, {Kostrzewa-Rutkowska},
  {Kruszy{\'n}ska}, {Lambert}, {Lanza}, {Lasne}, {Le Campion}, {Le Fustec},
  {Lebreton}, {Lebzelter}, {Leccia}, {Leclerc}, {Lecoeur-Taibi}, {Liao},
  {Licata}, {Lindstr{\o}m}, {Lister}, {Livanou}, {Lobel}, {Madrero Pardo},
  {Managau}, {Mann}, {Marchant}, {Marconi}, {Marcos Santos}, {Marinoni},
  {Marocco}, {Marshall}, {Martin Polo}, {Mart{\'\i}n-Fleitas}, {Masip},
  {Massari}, {Mastrobuono-Battisti}, {Mazeh}, {McMillan}, {Messina},
  {Michalik}, {Millar}, {Mints}, {Molina}, {Molinaro}, {Moln{\'a}r},
  {Montegriffo}, {Mor}, {Morbidelli}, {Morel}, {Morris}, {Mulone}, {Munoz},
  {Muraveva}, {Murphy}, {Musella}, {Noval}, {Ord{\'e}novic}, {Orr{\`u}},
  {Osinde}, {Pagani}, {Pagano}, {Palaversa}, {Palicio}, {Panahi}, {Pawlak},
  {Pe{\~n}alosa Esteller}, {Penttil{\"a}}, {Piersimoni}, {Pineau}, {Plachy},
  {Plum}, {Poggio}, {Poretti}, {Poujoulet}, {Pr{\v{s}}a}, {Pulone}, {Racero},
  {Ragaini}, {Rainer}, {Raiteri}, {Rambaux}, {Ramos}, {Ramos-Lerate}, {Re
  Fiorentin}, {Regibo}, {Reyl{\'e}}, {Ripepi}, {Riva}, {Rixon}, {Robichon},
  {Robin}, {Roelens}, {Rohrbasser}, {Romero-G{\'o}mez}, {Rowell}, {Royer},
  {Rybicki}, {Sadowski}, {Sagrist{\`a} Sell{\'e}s}, {Sahlmann}, {Salgado},
  {Salguero}, {Samaras}, {Sanchez Gimenez}, {Sanna}, {Santove{\~n}a},
  {Sarasso}, {Schultheis}, {Sciacca}, {Segol}, {Segovia}, {S{\'e}gransan},
  {Semeux}, {Shahaf}, {Siddiqui}, {Siebert}, {Siltala}, {Slezak}, {Smart},
  {Solano}, {Solitro}, {Souami}, {Souchay}, {Spagna}, {Spoto}, {Steele},
  {Steidelm{\"u}ller}, {Stephenson}, {S{\"u}veges}, {Szabados}, {Szegedi-Elek},
  {Taris}, {Tauran}, {Taylor}, {Teixeira}, {Thuillot}, {Tonello}, {Torra},
  {Torra}, {Turon}, {Unger}, {Vaillant}, {van Dillen}, {Vanel}, {Vecchiato},
  {Viala}, {Vicente}, {Voutsinas}, {Weiler}, {Wevers}, {Wyrzykowski}, {Yoldas},
  {Yvard}, {Zhao}, {Zorec}, {Zucker}, {Zurbach}, \& {Zwitter}}]{GaiaEDR3}
{Gaia Collaboration}, {Brown}, A.~G.~A., {Vallenari}, A., {et~al.} 2021, \aap,
  650, C3

\bibitem[{{Gaia Collaboration} {et~al.}(2016){Gaia Collaboration}, {Brown},
  {Vallenari}, {Prusti}, {de Bruijne}, {Mignard}, {Drimmel}, {Babusiaux},
  {Bailer-Jones}, {Bastian}, {Biermann}, {Evans}, {Eyer}, {Jansen}, {Jordi},
  {Katz}, {Klioner}, {Lammers}, {Lindegren}, {Luri}, {O'Mullane}, {Panem},
  {Pourbaix}, {Randich}, {Sartoretti}, {Siddiqui}, {Soubiran}, {Valette}, {van
  Leeuwen}, {Walton}, {Aerts}, {Arenou}, {Cropper}, {H{\o}g}, {Lattanzi},
  {Grebel}, {Holland}, {Huc}, {Passot}, {Perryman}, {Bramante}, {Cacciari},
  {Casta{\~n}eda}, {Chaoul}, {Cheek}, {De Angeli}, {Fabricius}, {Guerra},
  {Hern{\'a}ndez}, {Jean-Antoine-Piccolo}, {Masana}, {Messineo}, {Mowlavi},
  {Nienartowicz}, {Ord{\'o}{\~n}ez-Blanco}, {Panuzzo}, {Portell}, {Richards},
  {Riello}, {Seabroke}, {Tanga}, {Th{\'e}venin}, {Torra}, {Els},
  {Gracia-Abril}, {Comoretto}, {Garcia-Reinaldos}, {Lock}, {Mercier},
  {Altmann}, {Andrae}, {Astraatmadja}, {Bellas-Velidis}, {Benson}, {Berthier},
  {Blomme}, {Busso}, {Carry}, {Cellino}, {Clementini}, {Cowell}, {Creevey},
  {Cuypers}, {Davidson}, {De Ridder}, {de Torres}, {Delchambre}, {Dell'Oro},
  {Ducourant}, {Fr{\'e}mat}, {Garc{\'\i}a-Torres}, {Gosset}, {Halbwachs},
  {Hambly}, {Harrison}, {Hauser}, {Hestroffer}, {Hodgkin}, {Huckle}, {Hutton},
  {Jasniewicz}, {Jordan}, {Kontizas}, {Korn}, {Lanzafame}, {Manteiga},
  {Moitinho}, {Muinonen}, {Osinde}, {Pancino}, {Pauwels}, {Petit},
  {Recio-Blanco}, {Robin}, {Sarro}, {Siopis}, {Smith}, {Smith}, {Sozzetti},
  {Thuillot}, {van Reeven}, {Viala}, {Abbas}, {Abreu Aramburu}, {Accart},
  {Aguado}, {Allan}, {Allasia}, {Altavilla}, {{\'A}lvarez}, {Alves},
  {Anderson}, {Andrei}, {Anglada Varela}, {Antiche}, {Antoja}, {Ant{\'o}n},
  {Arcay}, {Bach}, {Baker}, {Balaguer-N{\'u}{\~n}ez}, {Barache}, {Barata},
  {Barbier}, {Barblan}, {Barrado y Navascu{\'e}s}, {Barros}, {Barstow},
  {Becciani}, {Bellazzini}, {Bello Garc{\'\i}a}, {Belokurov}, {Bendjoya},
  {Berihuete}, {Bianchi}, {Bienaym{\'e}}, {Billebaud}, {Blagorodnova},
  {Blanco-Cuaresma}, {Boch}, {Bombrun}, {Borrachero}, {Bouquillon}, {Bourda},
  {Bouy}, {Bragaglia}, {Breddels}, {Brouillet}, {Br{\"u}semeister},
  {Bucciarelli}, {Burgess}, {Burgon}, {Burlacu}, {Busonero}, {Buzzi}, {Caffau},
  {Cambras}, {Campbell}, {Cancelliere}, {Cantat-Gaudin}, {Carlucci},
  {Carrasco}, {Castellani}, {Charlot}, {Charnas}, {Chiavassa}, {Clotet},
  {Cocozza}, {Collins}, {Costigan}, {Crifo}, {Cross}, {Crosta}, {Crowley},
  {Dafonte}, {Damerdji}, {Dapergolas}, {David}, {David}, {De Cat}, {de Felice},
  {de Laverny}, {De Luise}, {De March}, {de Martino}, {de Souza}, {Debosscher},
  {del Pozo}, {Delbo}, {Delgado}, {Delgado}, {Di Matteo}, {Diakite},
  {Distefano}, {Dolding}, {Dos Anjos}, {Drazinos}, {Duran}, {Dzigan},
  {Edvardsson}, {Enke}, {Evans}, {Eynard Bontemps}, {Fabre}, {Fabrizio},
  {Faigler}, {Falc{\~a}o}, {Farr{\`a}s Casas}, {Federici}, {Fedorets},
  {Fern{\'a}ndez-Hern{\'a}ndez}, {Fernique}, {Fienga}, {Figueras}, {Filippi},
  {Findeisen}, {Fonti}, {Fouesneau}, {Fraile}, {Fraser}, {Fuchs}, {Gai},
  {Galleti}, {Galluccio}, {Garabato}, {Garc{\'\i}a-Sedano}, {Garofalo},
  {Garralda}, {Gavras}, {Gerssen}, {Geyer}, {Gilmore}, {Girona}, {Giuffrida},
  {Gomes}, {Gonz{\'a}lez-Marcos}, {Gonz{\'a}lez-N{\'u}{\~n}ez},
  {Gonz{\'a}lez-Vidal}, {Granvik}, {Guerrier}, {Guillout}, {Guiraud},
  {G{\'u}rpide}, {Guti{\'e}rrez-S{\'a}nchez}, {Guy}, {Haigron},
  {Hatzidimitriou}, {Haywood}, {Heiter}, {Helmi}, {Hobbs}, {Hofmann}, {Holl},
  {Holland }, {Hunt}, {Hypki}, {Icardi}, {Irwin}, {Jevardat de Fombelle},
  {Jofr{\'e}}, {Jonker}, {Jorissen}, {Julbe}, {Karampelas}, {Kochoska},
  {Kohley}, {Kolenberg}, {Kontizas}, {Koposov}, {Kordopatis}, {Koubsky},
  {Krone-Martins}, {Kudryashova}, {Kull}, {Bachchan}, {Lacoste-Seris}, {Lanza},
  {Lavigne}, {Le Poncin-Lafitte}, {Lebreton}, {Lebzelter}, {Leccia}, {Leclerc},
  {Lecoeur-Taibi}, {Lemaitre}, {Lenhardt}, {Leroux}, {Liao}, {Licata},
  {Lindstr{\o}m}, {Lister}, {Livanou}, {Lobel}, {L{\"o}ffler}, {L{\'o}pez},
  {Lorenz}, {MacDonald}, {Magalh{\~a}es Fernandes}, {Managau}, {Mann},
  {Mantelet}, {Marchal}, {Marchant}, {Marconi}, {Marinoni}, {Marrese},
  {Marschalk{\'o}}, {Marshall}, {Mart{\'\i}n-Fleitas}, {Martino}, {Mary},
  {Matijevi{\v{c}}}, {Mazeh}, {McMillan}, {Messina}, {Michalik}, {Millar},
  {Mirand a}, {Molina}, {Molinaro}, {Molinaro}, {Moln{\'a}r}, {Moniez},
  {Montegriffo}, {Mor}, {Mora}, {Morbidelli}, {Morel}, {Morgenthaler},
  {Morris}, {Mulone}, {Muraveva}, {Musella}, {Narbonne}, {Nelemans},
  {Nicastro}, {Noval}, {Ord{\'e}novic}, {Ordieres-Mer{\'e}}, {Osborne},
  {Pagani}, {Pagano}, {Pailler}, {Palacin}, {Palaversa}, {Parsons}, {Pecoraro},
  {Pedrosa}, {Pentik{\"a}inen}, {Pichon}, {Piersimoni}, {Pineau}, {Plachy},
  {Plum}, {Poujoulet}, {Pr{\v{s}}a}, {Pulone}, {Ragaini}, {Rago}, {Rambaux},
  {Ramos-Lerate}, {Ranalli}, {Rauw}, {Read}, {Regibo}, {Reyl{\'e}}, {Ribeiro},
  {Rimoldini}, {Ripepi}, {Riva}, {Rixon}, {Roelens}, {Romero-G{\'o}mez},
  {Rowell}, {Royer}, {Ruiz-Dern}, {Sadowski}, {Sagrist{\`a} Sell{\'e}s},
  {Sahlmann}, {Salgado}, {Salguero}, {Sarasso}, {Savietto}, {Schultheis},
  {Sciacca}, {Segol}, {Segovia}, {Segransan}, {Shih}, {Smareglia}, {Smart},
  {Solano}, {Solitro}, {Sordo}, {Soria Nieto}, {Souchay}, {Spagna}, {Spoto},
  {Stampa}, {Steele}, {Steidelm{\"u}ller}, {Stephenson}, {Stoev}, {Suess},
  {S{\"u}veges}, {Surdej}, {Szabados}, {Szegedi-Elek}, {Tapiador}, {Taris},
  {Tauran}, {Taylor}, {Teixeira}, {Terrett}, {Tingley}, {Trager}, {Turon},
  {Ulla}, {Utrilla}, {Valentini}, {van Elteren}, {Van Hemelryck}, {van
  Leeuwen}, {Varadi}, {Vecchiato}, {Veljanoski}, {Via}, {Vicente}, {Vogt},
  {Voss}, {Votruba}, {Voutsinas}, {Walmsley}, {Weiler}, {Weingrill}, {Wevers},
  {Wyrzykowski}, {Yoldas}, {{\v{Z}}erjal}, {Zucker}, {Zurbach}, {Zwitter},
  {Alecu}, {Allen}, {Allende Prieto}, {Amorim}, {Anglada-Escud{\'e}},
  {Arsenijevic}, {Azaz}, {Balm}, {Beck}, {Bernstein}, {Bigot}, {Bijaoui},
  {Blasco}, {Bonfigli}, {Bono}, {Boudreault}, {Bressan}, {Brown}, {Brunet},
  {Bunclark}, {Buonanno}, {Butkevich}, {Carret}, {Carrion}, {Chemin},
  {Ch{\'e}reau}, {Corcione}, {Darmigny}, {de Boer}, {de Teodoro}, {de Zeeuw},
  {Delle Luche}, {Domingues}, {Dubath}, {Fodor}, {Fr{\'e}zouls}, {Fries},
  {Fustes}, {Fyfe}, {Gallardo}, {Gallegos}, {Gardiol}, {Gebran}, {Gomboc},
  {G{\'o}mez}, {Grux}, {Gueguen}, {Heyrovsky}, {Hoar}, {Iannicola}, {Isasi
  Parache}, {Janotto}, {Joliet}, {Jonckheere}, {Keil}, {Kim}, {Klagyivik},
  {Klar}, {Knude}, {Kochukhov}, {Kolka}, {Kos}, {Kutka}, {Lainey}, {LeBouquin},
  {Liu}, {Loreggia}, {Makarov}, {Marseille}, {Martayan}, {Martinez-Rubi},
  {Massart}, {Meynadier}, {Mignot}, {Munari}, {Nguyen}, {Nordlander}, {Ocvirk},
  {O'Flaherty}, {Olias Sanz}, {Ortiz}, {Osorio}, {Oszkiewicz}, {Ouzounis},
  {Palmer}, {Park}, {Pasquato}, {Peltzer}, {Peralta}, {P{\'e}turaud},
  {Pieniluoma}, {Pigozzi}, {Poels}, {Prat}, {Prod'homme}, {Raison}, {Rebordao},
  {Risquez}, {Rocca-Volmerange}, {Rosen}, {Ruiz-Fuertes}, {Russo}, {Sembay},
  {Serraller Vizcaino}, {Short}, {Siebert}, {Silva}, {Sinachopoulos}, {Slezak},
  {Soffel}, {Sosnowska}, {Strai{\v{z}}ys}, {ter Linden}, {Terrell}, {Theil},
  {Tiede}, {Troisi}, {Tsalmantza}, {Tur}, {Vaccari}, {Vachier}, {Valles}, {Van
  Hamme}, {Veltz}, {Virtanen}, {Wallut}, {Wichmann}, {Wilkinson}, {Ziaeepour},
  \& {Zschocke}}]{Gaia16}
{Gaia Collaboration}, {Brown}, A.~G.~A., {Vallenari}, A., {et~al.} 2016, \aap,
  595, A2

\bibitem[{{Gaia Collaboration} {et~al.}(2023{\natexlab{a}}){Gaia
  Collaboration}, {Recio-Blanco}, {Kordopatis}, {de Laverny}, {Palicio},
  {Spagna}, {Spina}, {Katz}, {Re Fiorentin}, {Poggio}, {McMillan}, {Vallenari},
  {Lattanzi}, {Seabroke}, {Casamiquela}, {Bragaglia}, {Antoja}, {Bailer-Jones},
  {Schultheis}, {Andrae}, {Fouesneau}, {Cropper}, {Cantat-Gaudin}, {Bijaoui},
  {Heiter}, {Brown}, {Prusti}, {de Bruijne}, {Arenou}, {Babusiaux}, {Biermann},
  {Creevey}, {Ducourant}, {Evans}, {Eyer}, {Guerra}, {Hutton}, {Jordi},
  {Klioner}, {Lammers}, {Lindegren}, {Luri}, {Mignard}, {Panem}, {Pourbaix},
  {Randich}, {Sartoretti}, {Soubiran}, {Tanga}, {Walton}, {Bastian}, {Drimmel},
  {Jansen}, {van Leeuwen}, {Bakker}, {Cacciari}, {Casta{\~n}eda}, {De Angeli},
  {Fabricius}, {Fr{\'e}mat}, {Galluccio}, {Guerrier}, {Masana}, {Messineo},
  {Mowlavi}, {Nicolas}, {Nienartowicz}, {Pailler}, {Panuzzo}, {Riclet}, {Roux},
  {Sordo}, {Th{\'e}venin}, {Gracia-Abril}, {Portell}, {Teyssier}, {Altmann},
  {Audard}, {Bellas-Velidis}, {Benson}, {Berthier}, {Blomme}, {Burgess},
  {Busonero}, {Busso}, {C{\'a}novas}, {Carry}, {Cellino}, {Cheek},
  {Clementini}, {Damerdji}, {Davidson}, {de Teodoro}, {Nu{\~n}ez Campos},
  {Delchambre}, {Dell'Oro}, {Esquej}, {Fern{\'a}ndez-Hern{\'a}ndez}, {Fraile},
  {Garabato}, {Garc{\'\i}a-Lario}, {Gosset}, {Haigron}, {Halbwachs}, {Hambly},
  {Harrison}, {Hern{\'a}ndez}, {Hestroffer}, {Hodgkin}, {Holl}, {Jan{\ss}en},
  {Jevardat de Fombelle}, {Jordan}, {Krone-Martins}, {Lanzafame},
  {L{\"o}ffler}, {Marchal}, {Marrese}, {Moitinho}, {Muinonen}, {Osborne},
  {Pancino}, {Pauwels}, {Reyl{\'e}}, {Riello}, {Rimoldini}, {Roegiers},
  {Rybizki}, {Sarro}, {Siopis}, {Smith}, {Sozzetti}, {Utrilla}, {van Leeuwen},
  {Abbas}, {{\'A}brah{\'a}m}, {Abreu Aramburu}, {Aerts}, {Aguado}, {Ajaj},
  {Aldea-Montero}, {Altavilla}, {{\'A}lvarez}, {Alves}, {Anders}, {Anderson},
  {Anglada Varela}, {Baines}, {Baker}, {Balaguer-N{\'u}{\~n}ez}, {Balbinot},
  {Balog}, {Barache}, {Barbato}, {Barros}, {Barstow}, {Bartolom{\'e}},
  {Bassilana}, {Bauchet}, {Becciani}, {Bellazzini}, {Berihuete}, {Bernet},
  {Bertone}, {Bianchi}, {Binnenfeld}, {Blanco-Cuaresma}, {Boch}, {Bombrun},
  {Bossini}, {Bouquillon}, {Bramante}, {Breedt}, {Bressan}, {Brouillet},
  {Brugaletta}, {Bucciarelli}, {Burlacu}, {Butkevich}, {Buzzi}, {Caffau},
  {Cancelliere}, {Carballo}, {Carlucci}, {Carnerero}, {Carrasco}, {Castellani},
  {Castro-Ginard}, {Chaoul}, {Charlot}, {Chemin}, {Chiaramida}, {Chiavassa},
  {Chornay}, {Comoretto}, {Contursi}, {Cooper}, {Cornez}, {Cowell}, {Crifo},
  {Crosta}, {Crowley}, {Dafonte}, {Dapergolas}, {David}, {De Luise}, {De
  March}, {De Ridder}, {de Souza}, {de Torres}, {del Peloso}, {del Pozo},
  {Delbo}, {Delgado}, {Delisle}, {Demouchy}, {Dharmawardena}, {Di Matteo},
  {Diakite}, {Diener}, {Distefano}, {Dolding}, {Edvardsson}, {Enke}, {Fabre},
  {Fabrizio}, {Faigler}, {Fedorets}, {Fernique}, {Figueras}, {Fournier},
  {Fouron}, {Fragkoudi}, {Gai}, {Garcia-Gutierrez}, {Garcia-Reinaldos},
  {Garc{\'\i}a-Torres}, {Garofalo}, {Gavel}, {Gavras}, {Gerlach}, {Geyer},
  {Giacobbe}, {Gilmore}, {Girona}, {Giuffrida}, {Gomel}, {Gomez},
  {Gonz{\'a}lez-N{\'u}{\~n}ez}, {Gonz{\'a}lez-Santamar{\'\i}a},
  {Gonz{\'a}lez-Vidal}, {Granvik}, {Guillout}, {Guiraud},
  {Guti{\'e}rrez-S{\'a}nchez}, {Guy}, {Hatzidimitriou}, {Hauser}, {Haywood},
  {Helmer}, {Helmi}, {Sarmiento}, {Hidalgo}, {H{\l}adczuk}, {Hobbs}, {Holland},
  {Huckle}, {Jardine}, {Jasniewicz}, {Jean-Antoine Piccolo},
  {Jim{\'e}nez-Arranz}, {Juaristi Campillo}, {Julbe}, {Karbevska}, {Kervella},
  {Khanna}, {Korn}, {K{\'o}sp{\'a}l}, {Kostrzewa-Rutkowska}, {Kruszy{\'n}ska},
  {Kun}, {Laizeau}, {Lambert}, {Lanza}, {Lasne}, {Le Campion}, {Lebreton},
  {Lebzelter}, {Leccia}, {Leclerc}, {Lecoeur-Taibi}, {Liao}, {Licata},
  {Lindstr{\o}m}, {Lister}, {Livanou}, {Lobel}, {Lorca}, {Loup}, {Madrero
  Pardo}, {Magdaleno Romeo}, {Managau}, {Mann}, {Manteiga}, {Marchant},
  {Marconi}, {Marcos}, {Marcos Santos}, {Mar{\'\i}n Pina}, {Marinoni},
  {Marocco}, {Marshall}, {Martin Polo}, {Mart{\'\i}n-Fleitas}, {Marton},
  {Mary}, {Masip}, {Massari}, {Mastrobuono-Battisti}, {Mazeh}, {Messina},
  {Michalik}, {Millar}, {Mints}, {Molina}, {Molinaro}, {Moln{\'a}r}, {Monari},
  {Mongui{\'o}}, {Montegriffo}, {Montero}, {Mor}, {Mora}, {Morbidelli},
  {Morel}, {Morris}, {Muraveva}, {Murphy}, {Musella}, {Nagy}, {Noval},
  {Oca{\~n}a}, {Ogden}, {Ordenovic}, {Osinde}, {Pagani}, {Pagano}, {Palaversa},
  {Pallas-Quintela}, {Panahi}, {Payne-Wardenaar}, {Pe{\~n}alosa Esteller},
  {Penttil{\"a}}, {Pichon}, {Piersimoni}, {Pineau}, {Plachy}, {Plum},
  {Pr{\v{s}}a}, {Pulone}, {Racero}, {Ragaini}, {Rainer}, {Raiteri}, {Ramos},
  {Ramos-Lerate}, {Regibo}, {Richards}, {Rios Diaz}, {Ripepi}, {Riva}, {Rix},
  {Rixon}, {Robichon}, {Robin}, {Robin}, {Roelens}, {Rogues}, {Rohrbasser},
  {Romero-G{\'o}mez}, {Rowell}, {Royer}, {Ruz Mieres}, {Rybicki}, {Sadowski},
  {S{\'a}ez N{\'u}{\~n}ez}, {Sagrist{\`a} Sell{\'e}s}, {Sahlmann}, {Salguero},
  {Samaras}, {Sanchez Gimenez}, {Sanna}, {Santove{\~n}a}, {Sarasso}, {Sciacca},
  {Segol}, {Segovia}, {S{\'e}gransan}, {Semeux}, {Shahaf}, {Siddiqui},
  {Siebert}, {Siltala}, {Silvelo}, {Slezak}, {Slezak}, {Smart}, {Snaith},
  {Solano}, {Solitro}, {Souami}, {Souchay}, {Spoto}, {Steele},
  {Steidelm{\"u}ller}, {Stephenson}, {S{\"u}veges}, {Surdej}, {Szabados},
  {Szegedi-Elek}, {Taris}, {Taylor}, {Teixeira}, {Tolomei}, {Tonello}, {Torra},
  {Torra}, {Torralba Elipe}, {Trabucchi}, {Tsounis}, {Turon}, {Ulla}, {Unger},
  {Vaillant}, {van Dillen}, {van Reeven}, {Vanel}, {Vecchiato}, {Viala},
  {Vicente}, {Voutsinas}, {Weiler}, {Wevers}, {Wyrzykowski}, {Yoldas}, {Yvard},
  {Zhao}, {Zorec}, {Zucker}, \& {Zwitter}}]{RecioBlanco22}
{Gaia Collaboration}, {Recio-Blanco}, A., {Kordopatis}, G., {et~al.}
  2023{\natexlab{a}}, \aap, 674, A38

\bibitem[{{Gaia Collaboration} {et~al.}(2023{\natexlab{b}}){Gaia
  Collaboration}, {Vallenari}, {Brown}, {Prusti}, {de Bruijne}, {Arenou},
  {Babusiaux}, {Biermann}, {Creevey}, {Ducourant}, {Evans}, {Eyer}, {Guerra},
  {Hutton}, {Jordi}, {Klioner}, {Lammers}, {Lindegren}, {Luri}, {Mignard},
  {Panem}, {Pourbaix}, {Randich}, {Sartoretti}, {Soubiran}, {Tanga}, {Walton},
  {Bailer-Jones}, {Bastian}, {Drimmel}, {Jansen}, {Katz}, {Lattanzi}, {van
  Leeuwen}, {Bakker}, {Cacciari}, {Casta{\~n}eda}, {De Angeli}, {Fabricius},
  {Fouesneau}, {Fr{\'e}mat}, {Galluccio}, {Guerrier}, {Heiter}, {Masana},
  {Messineo}, {Mowlavi}, {Nicolas}, {Nienartowicz}, {Pailler}, {Panuzzo},
  {Riclet}, {Roux}, {Seabroke}, {Sordo}, {Th{\'e}venin}, {Gracia-Abril},
  {Portell}, {Teyssier}, {Altmann}, {Andrae}, {Audard}, {Bellas-Velidis},
  {Benson}, {Berthier}, {Blomme}, {Burgess}, {Busonero}, {Busso},
  {C{\'a}novas}, {Carry}, {Cellino}, {Cheek}, {Clementini}, {Damerdji},
  {Davidson}, {de Teodoro}, {Nu{\~n}ez Campos}, {Delchambre}, {Dell'Oro},
  {Esquej}, {Fern{\'a}ndez-Hern{\'a}ndez}, {Fraile}, {Garabato},
  {Garc{\'\i}a-Lario}, {Gosset}, {Haigron}, {Halbwachs}, {Hambly}, {Harrison},
  {Hern{\'a}ndez}, {Hestroffer}, {Hodgkin}, {Holl}, {Jan{\ss}en}, {Jevardat de
  Fombelle}, {Jordan}, {Krone-Martins}, {Lanzafame}, {L{\"o}ffler}, {Marchal},
  {Marrese}, {Moitinho}, {Muinonen}, {Osborne}, {Pancino}, {Pauwels},
  {Recio-Blanco}, {Reyl{\'e}}, {Riello}, {Rimoldini}, {Roegiers}, {Rybizki},
  {Sarro}, {Siopis}, {Smith}, {Sozzetti}, {Utrilla}, {van Leeuwen}, {Abbas},
  {{\'A}brah{\'a}m}, {Abreu Aramburu}, {Aerts}, {Aguado}, {Ajaj},
  {Aldea-Montero}, {Altavilla}, {{\'A}lvarez}, {Alves}, {Anders}, {Anderson},
  {Anglada Varela}, {Antoja}, {Baines}, {Baker}, {Balaguer-N{\'u}{\~n}ez},
  {Balbinot}, {Balog}, {Barache}, {Barbato}, {Barros}, {Barstow},
  {Bartolom{\'e}}, {Bassilana}, {Bauchet}, {Becciani}, {Bellazzini},
  {Berihuete}, {Bernet}, {Bertone}, {Bianchi}, {Binnenfeld}, {Blanco-Cuaresma},
  {Blazere}, {Boch}, {Bombrun}, {Bossini}, {Bouquillon}, {Bragaglia},
  {Bramante}, {Breedt}, {Bressan}, {Brouillet}, {Brugaletta}, {Bucciarelli},
  {Burlacu}, {Butkevich}, {Buzzi}, {Caffau}, {Cancelliere}, {Cantat-Gaudin},
  {Carballo}, {Carlucci}, {Carnerero}, {Carrasco}, {Casamiquela}, {Castellani},
  {Castro-Ginard}, {Chaoul}, {Charlot}, {Chemin}, {Chiaramida}, {Chiavassa},
  {Chornay}, {Comoretto}, {Contursi}, {Cooper}, {Cornez}, {Cowell}, {Crifo},
  {Cropper}, {Crosta}, {Crowley}, {Dafonte}, {Dapergolas}, {David}, {David},
  {de Laverny}, {De Luise}, {De March}, {De Ridder}, {de Souza}, {de Torres},
  {del Peloso}, {del Pozo}, {Delbo}, {Delgado}, {Delisle}, {Demouchy},
  {Dharmawardena}, {Di Matteo}, {Diakite}, {Diener}, {Distefano}, {Dolding},
  {Edvardsson}, {Enke}, {Fabre}, {Fabrizio}, {Faigler}, {Fedorets}, {Fernique},
  {Fienga}, {Figueras}, {Fournier}, {Fouron}, {Fragkoudi}, {Gai},
  {Garcia-Gutierrez}, {Garcia-Reinaldos}, {Garc{\'\i}a-Torres}, {Garofalo},
  {Gavel}, {Gavras}, {Gerlach}, {Geyer}, {Giacobbe}, {Gilmore}, {Girona},
  {Giuffrida}, {Gomel}, {Gomez}, {Gonz{\'a}lez-N{\'u}{\~n}ez},
  {Gonz{\'a}lez-Santamar{\'\i}a}, {Gonz{\'a}lez-Vidal}, {Granvik}, {Guillout},
  {Guiraud}, {Guti{\'e}rrez-S{\'a}nchez}, {Guy}, {Hatzidimitriou}, {Hauser},
  {Haywood}, {Helmer}, {Helmi}, {Sarmiento}, {Hidalgo}, {Hilger},
  {H{\l}adczuk}, {Hobbs}, {Holland}, {Huckle}, {Jardine}, {Jasniewicz},
  {Jean-Antoine Piccolo}, {Jim{\'e}nez-Arranz}, {Jorissen}, {Juaristi
  Campillo}, {Julbe}, {Karbevska}, {Kervella}, {Khanna}, {Kontizas},
  {Kordopatis}, {Korn}, {K{\'o}sp{\'a}l}, {Kostrzewa-Rutkowska},
  {Kruszy{\'n}ska}, {Kun}, {Laizeau}, {Lambert}, {Lanza}, {Lasne}, {Le
  Campion}, {Lebreton}, {Lebzelter}, {Leccia}, {Leclerc}, {Lecoeur-Taibi},
  {Liao}, {Licata}, {Lindstr{\o}m}, {Lister}, {Livanou}, {Lobel}, {Lorca},
  {Loup}, {Madrero Pardo}, {Magdaleno Romeo}, {Managau}, {Mann}, {Manteiga},
  {Marchant}, {Marconi}, {Marcos}, {Marcos Santos}, {Mar{\'\i}n Pina},
  {Marinoni}, {Marocco}, {Marshall}, {Martin Polo}, {Mart{\'\i}n-Fleitas},
  {Marton}, {Mary}, {Masip}, {Massari}, {Mastrobuono-Battisti}, {Mazeh},
  {McMillan}, {Messina}, {Michalik}, {Millar}, {Mints}, {Molina}, {Molinaro},
  {Moln{\'a}r}, {Monari}, {Mongui{\'o}}, {Montegriffo}, {Montero}, {Mor},
  {Mora}, {Morbidelli}, {Morel}, {Morris}, {Muraveva}, {Murphy}, {Musella},
  {Nagy}, {Noval}, {Oca{\~n}a}, {Ogden}, {Ordenovic}, {Osinde}, {Pagani},
  {Pagano}, {Palaversa}, {Palicio}, {Pallas-Quintela}, {Panahi},
  {Payne-Wardenaar}, {Pe{\~n}alosa Esteller}, {Penttil{\"a}}, {Pichon},
  {Piersimoni}, {Pineau}, {Plachy}, {Plum}, {Poggio}, {Pr{\v{s}}a}, {Pulone},
  {Racero}, {Ragaini}, {Rainer}, {Raiteri}, {Rambaux}, {Ramos}, {Ramos-Lerate},
  {Re Fiorentin}, {Regibo}, {Richards}, {Rios Diaz}, {Ripepi}, {Riva}, {Rix},
  {Rixon}, {Robichon}, {Robin}, {Robin}, {Roelens}, {Rogues}, {Rohrbasser},
  {Romero-G{\'o}mez}, {Rowell}, {Royer}, {Ruz Mieres}, {Rybicki}, {Sadowski},
  {S{\'a}ez N{\'u}{\~n}ez}, {Sagrist{\`a} Sell{\'e}s}, {Sahlmann}, {Salguero},
  {Samaras}, {Sanchez Gimenez}, {Sanna}, {Santove{\~n}a}, {Sarasso},
  {Schultheis}, {Sciacca}, {Segol}, {Segovia}, {S{\'e}gransan}, {Semeux},
  {Shahaf}, {Siddiqui}, {Siebert}, {Siltala}, {Silvelo}, {Slezak}, {Slezak},
  {Smart}, {Snaith}, {Solano}, {Solitro}, {Souami}, {Souchay}, {Spagna},
  {Spina}, {Spoto}, {Steele}, {Steidelm{\"u}ller}, {Stephenson}, {S{\"u}veges},
  {Surdej}, {Szabados}, {Szegedi-Elek}, {Taris}, {Taylor}, {Teixeira},
  {Tolomei}, {Tonello}, {Torra}, {Torra}, {Torralba Elipe}, {Trabucchi},
  {Tsounis}, {Turon}, {Ulla}, {Unger}, {Vaillant}, {van Dillen}, {van Reeven},
  {Vanel}, {Vecchiato}, {Viala}, {Vicente}, {Voutsinas}, {Weiler}, {Wevers},
  {Wyrzykowski}, {Yoldas}, {Yvard}, {Zhao}, {Zorec}, {Zucker}, \&
  {Zwitter}}]{Vallenari23}
{Gaia Collaboration}, {Vallenari}, A., {Brown}, A.~G.~A., {et~al.}
  2023{\natexlab{b}}, \aap, 674, A1

\bibitem[{{Ga{\l}an} {et~al.}(2017){Ga{\l}an}, {Miko{\l}ajewska}, {Hinkle}, \&
  {Joyce}}]{Galan17}
{Ga{\l}an}, C., {Miko{\l}ajewska}, J., {Hinkle}, K.~H., \& {Joyce}, R.~R. 2017,
  \mnras, 466, 2194

\bibitem[{{Ga{\l}an} {et~al.}(2023){Ga{\l}an}, {Miko{\l}ajewska}, {Hinkle}, \&
  {Joyce}}]{Galan2023}
{Ga{\l}an}, C., {Miko{\l}ajewska}, J., {Hinkle}, K.~H., \& {Joyce}, R.~R. 2023,
  \mnras, 526, 918

\bibitem[{{Gilmore} {et~al.}(2022){Gilmore}, {Randich}, {Worley}, {Hourihane},
  {Gonneau}, {Sacco}, {Lewis}, {Magrini}, {Fran{\c{c}}ois}, {Jeffries},
  {Koposov}, {Bragaglia}, {Alfaro}, {Allende Prieto}, {Blomme}, {Korn},
  {Lanzafame}, {Pancino}, {Recio-Blanco}, {Smiljanic}, {Van Eck}, {Zwitter},
  {Bensby}, {Flaccomio}, {Irwin}, {Franciosini}, {Morbidelli}, {Damiani},
  {Bonito}, {Friel}, {Vink}, {Prisinzano}, {Abbas}, {Hatzidimitriou}, {Held},
  {Jordi}, {Paunzen}, {Spagna}, {Jackson}, {Ma{\'\i}z Apell{\'a}niz},
  {Asplund}, {Bonifacio}, {Feltzing}, {Binney}, {Drew}, {Ferguson}, {Micela},
  {Negueruela}, {Prusti}, {Rix}, {Vallenari}, {Bergemann}, {Casey}, {de
  Laverny}, {Frasca}, {Hill}, {Lind}, {Sbordone}, {Sousa}, {Adibekyan},
  {Caffau}, {Daflon}, {Feuillet}, {Gebran}, {Gonzalez Hernandez}, {Guiglion},
  {Herrero}, {Lobel}, {Merle}, {Mikolaitis}, {Montes}, {Morel}, {Ruchti},
  {Soubiran}, {Tabernero}, {Tautvai{\v{s}}ien{\.{e}}}, {Traven}, {Valentini},
  {Van der Swaelmen}, {Villanova}, {Viscasillas V{\'a}zquez}, {Bayo}, {Biazzo},
  {Carraro}, {Edvardsson}, {Heiter}, {Jofr{\'e}}, {Marconi}, {Martayan},
  {Masseron}, {Monaco}, {Walton}, {Zaggia}, {Aguirre B{\o}rsen-Koch}, {Alves},
  {Balaguer-Nunez}, {Barklem}, {Barrado}, {Bellazzini}, {Berlanas}, {Binks},
  {Bressan}, {Capuzzo-Dolcetta}, {Casagrande}, {Casamiquela}, {Collins},
  {D'Orazi}, {Dantas}, {Debattista}, {Delgado-Mena}, {Di Marcantonio},
  {Drazdauskas}, {Evans}, {Famaey}, {Franchini}, {Fr{\'e}mat}, {Fu}, {Geisler},
  {Gerhard}, {Gonz{\'a}lez Solares}, {Grebel}, {Guti{\'e}rrez Albarr{\'a}n},
  {Jim{\'e}nez-Esteban}, {J{\"o}nsson}, {Khachaturyants}, {Kordopatis}, {Kos},
  {Lagarde}, {Ludwig}, {Mahy}, {Mapelli}, {Marfil}, {Martell}, {Messina},
  {Miglio}, {Minchev}, {Moitinho}, {Montalban}, {Monteiro}, {Morossi},
  {Mowlavi}, {Mucciarelli}, {Murphy}, {Nardetto}, {Ortolani}, {Paletou},
  {Palou{\v{s}}}, {Pickering}, {Quirrenbach}, {Re Fiorentin}, {Read}, {Romano},
  {Ryde}, {Sanna}, {Santos}, {Seabroke}, {Spina}, {Steinmetz}, {Stonkut{\'e}},
  {Sutorius}, {Th{\'e}venin}, {Tosi}, {Tsantaki}, {Wright}, {Wyse}, {Zoccali},
  {Zorec}, \& {Zucker}}]{GilmoreGES22}
{Gilmore}, G., {Randich}, S., {Worley}, C.~C., {et~al.} 2022, \aap, 666, A120

\bibitem[{{Gilmore} \& {Reid}(1983)}]{GiRe83}
{Gilmore}, G. \& {Reid}, N. 1983, \mnras, 202, 1025

\bibitem[{{Gilroy} \& {Brown}(1991)}]{GiBr91}
{Gilroy}, K.~K. \& {Brown}, J.~A. 1991, \apj, 371, 578

\bibitem[{{Good}(2003)}]{Good03}
{Good}, G.~A. 2003, {Observing Variable Stars}

\bibitem[{{Gratton} {et~al.}(2000){Gratton}, {Sneden}, {Carretta}, \&
  {Bragaglia}}]{Gratton00}
{Gratton}, R.~G., {Sneden}, C., {Carretta}, E., \& {Bragaglia}, A. 2000, \aap,
  354, 169

\bibitem[{{Grevesse} {et~al.}(2007){Grevesse}, {Asplund}, \&
  {Sauval}}]{Grevesse07}
{Grevesse}, N., {Asplund}, M., \& {Sauval}, A.~J. 2007, \ssr, 130, 105

\bibitem[{{Guer{\c{c}}o} {et~al.}(2022){Guer{\c{c}}o}, {Smith}, {Cunha},
  {Ekstr{\"o}m}, {Abia}, {Plez}, {Meynet}, {Ramirez}, {Prantzos}, {Sellgren},
  {Hayes}, \& {Majewski}}]{Guerco22}
{Guer{\c{c}}o}, R., {Smith}, V.~V., {Cunha}, K., {et~al.} 2022, \mnras, 516,
  2801

\bibitem[{{Gustafsson} {et~al.}(2008){Gustafsson}, {Edvardsson}, {Eriksson},
  {J{\o}rgensen}, {Nordlund}, \& {Plez}}]{Gustafsson2008}
{Gustafsson}, B., {Edvardsson}, B., {Eriksson}, K., {et~al.} 2008, \aap, 486,
  951

\bibitem[{{Harrington} \& {Garaud}(2019)}]{Harrington19}
{Harrington}, P.~Z. \& {Garaud}, P. 2019, \apjl, 870, L5

\bibitem[{{Haywood}(2006)}]{Haywood06}
{Haywood}, M. 2006, \mnras, 371, 1760

\bibitem[{{Heiter} {et~al.}(2015){Heiter}, {Lind}, {Asplund}, {Barklem},
  {Bergemann}, {Magrini}, {Masseron}, {Mikolaitis}, {Pickering}, \&
  {Ruffoni}}]{Heiter2015}
{Heiter}, U., {Lind}, K., {Asplund}, M., {et~al.} 2015, \physscr, 90, 054010

\bibitem[{{Henkel} {et~al.}(2017){Henkel}, {Karakas}, \&
  {Lattanzio}}]{Henkel17}
{Henkel}, K., {Karakas}, A.~I., \& {Lattanzio}, J.~C. 2017, \mnras, 469, 4600

\bibitem[{{Iben}(1967)}]{Iben67}
{Iben}, Jr., I. 1967, \apj, 147, 624

\bibitem[{{Johansson} {et~al.}(2003){Johansson}, {Litz{\'e}n}, {Lundberg}, \&
  {Zhang}}]{Johansson03}
{Johansson}, S., {Litz{\'e}n}, U., {Lundberg}, H., \& {Zhang}, Z. 2003, \apjl,
  584, L107

\bibitem[{{Khan} {et~al.}(2019){Khan}, {Miglio}, {Mosser}, {Arenou},
  {Belkacem}, {Brown}, {Katz}, {Casagrand e}, {Chaplin}, {Davies}, {Rendle},
  {Rodrigues}, {Bossini}, {Cantat-Gaudin}, {Elsworth}, {Girardi}, {North}, \&
  {Vallenari}}]{Khan19}
{Khan}, S., {Miglio}, A., {Mosser}, B., {et~al.} 2019, \aap, 628, A35

\bibitem[{{Kippenhahn} {et~al.}(1980){Kippenhahn}, {Ruschenplatt}, \&
  {Thomas}}]{Kippen80}
{Kippenhahn}, R., {Ruschenplatt}, G., \& {Thomas}, H.-C. 1980, \aap, 91, 175

\bibitem[{{Kjeldsen} \& {Bedding}(1995)}]{KjeBed95}
{Kjeldsen}, H. \& {Bedding}, T.~R. 1995, \aap, 293, 87

\bibitem[{{Kordopatis} {et~al.}(2013){Kordopatis}, {Gilmore}, {Steinmetz},
  {Boeche}, {Seabroke}, {Siebert}, {Zwitter}, {Binney}, {de Laverny},
  {Recio-Blanco}, {Williams}, {Piffl}, {Enke}, {Roeser}, {Bijaoui}, {Wyse},
  {Freeman}, {Munari}, {Carrillo}, {Anguiano}, {Burton}, {Campbell}, {Cass},
  {Fiegert}, {Hartley}, {Parker}, {Reid}, {Ritter}, {Russell}, {Stupar},
  {Watson}, {Bienaym{\'e}}, {Bland -Hawthorn}, {Gerhard}, {Gibson}, {Grebel},
  {Helmi}, {Navarro}, {Conrad}, {Famaey}, {Faure}, {Just}, {Kos},
  {Matijevi{\v{c}}}, {McMillan}, {Minchev}, {Scholz}, {Sharma}, {Siviero}, {de
  Boer}, \& {{\v{Z}}erjal}}]{Kordopatis13a}
{Kordopatis}, G., {Gilmore}, G., {Steinmetz}, M., {et~al.} 2013, \aj, 146, 134

\bibitem[{{Kroupa}(2001)}]{Kroupa01}
{Kroupa}, P. 2001, \mnras, 322, 231

\bibitem[{{Kroupa} {et~al.}(2013){Kroupa}, {Weidner}, {Pflamm-Altenburg},
  {Thies}, {Dabringhausen}, {Marks}, \& {Maschberger}}]{Kroupa13}
{Kroupa}, P., {Weidner}, C., {Pflamm-Altenburg}, J., {et~al.} 2013, in Planets,
  Stars and Stellar Systems. Volume 5: Galactic Structure and Stellar
  Populations, ed. T.~D. {Oswalt} \& G.~{Gilmore}, Vol.~5, 115

\bibitem[{{Kurucz}(2005)}]{Kurucz05}
{Kurucz}, R.~L. 2005, Memorie della Societa Astronomica Italiana Supplementi,
  8, 189

\bibitem[{{Lagarde} {et~al.}(2011){Lagarde}, {Charbonnel}, {Decressin}, \&
  {Hagelberg}}]{Lagarde11}
{Lagarde}, N., {Charbonnel}, C., {Decressin}, T., \& {Hagelberg}, J. 2011,
  \aap, 536, A28

\bibitem[{{Lagarde} {et~al.}(2012{\natexlab{a}}){Lagarde}, {Decressin},
  {Charbonnel}, {Eggenberger}, {Ekstr{\"o}m}, \& {Palacios}}]{Lagarde12a}
{Lagarde}, N., {Decressin}, T., {Charbonnel}, C., {et~al.} 2012{\natexlab{a}},
  \aap, 543, A108

\bibitem[{{Lagarde} {et~al.}(2015){Lagarde}, {Miglio}, {Eggenberger}, {Morel},
  {Montalb{\'a}n}, {Mosser}, {Rodrigues}, {Girardi}, {Rainer}, {Poretti},
  {Barban}, {Hekker}, {Kallinger}, {Valentini}, {Carrier}, {Hareter},
  {Mantegazza}, {Elsworth}, {Michel}, \& {Baglin}}]{Lagarde15}
{Lagarde}, N., {Miglio}, A., {Eggenberger}, P., {et~al.} 2015, \aap, 580, A141

\bibitem[{{Lagarde} {et~al.}(2021){Lagarde}, {Reyl{\'e}}, {Chiappini}, {Mor},
  {Anders}, {Figueras}, {Miglio}, {Romero-G{\'o}mez}, {Antoja}, {Cabral},
  {Salomon}, {Robin}, {Bienaym{\'e}}, {Soubiran}, {Cornu}, \&
  {Montillaud}}]{Lagarde21}
{Lagarde}, N., {Reyl{\'e}}, C., {Chiappini}, C., {et~al.} 2021, \aap, 654, A13

\bibitem[{{Lagarde} {et~al.}(2019){Lagarde}, {Reyl{\'e}}, {Robin},
  {Tautvai{\v{s}}ien{\.{e}}}, {Drazdauskas}, {Mikolaitis},
  {Minkevi{\v{c}}i{\={u}}t{\.{e}}}, {Stonkut{\.{e}}}, {Chorniy}, {Bagdonas},
  {Miglio}, {Nasello}, {Gilmore}, {Randich}, {Bensby}, {Bragaglia},
  {Flaccomio}, {Francois}, {Korn}, {Pancino}, {Smiljanic}, {Bayo}, {Carraro},
  {Costado}, {Jim{\'e}nez-Esteban}, {Jofr{\'e}}, {Martell}, {Masseron},
  {Monaco}, {Morbidelli}, {Sbordone}, {Sousa}, \& {Zaggia}}]{Lagarde19}
{Lagarde}, N., {Reyl{\'e}}, C., {Robin}, A.~C., {et~al.} 2019, \aap, 621, A24

\bibitem[{{Lagarde} {et~al.}(2017){Lagarde}, {Robin}, {Reyl{\'e}}, \&
  {Nasello}}]{Lagarde17}
{Lagarde}, N., {Robin}, A.~C., {Reyl{\'e}}, C., \& {Nasello}, G. 2017, \aap,
  601, A27

\bibitem[{{Lagarde} {et~al.}(2012{\natexlab{b}}){Lagarde}, {Romano},
  {Charbonnel}, {Tosi}, {Chiappini}, \& {Matteucci}}]{Lagarde12b}
{Lagarde}, N., {Romano}, D., {Charbonnel}, C., {et~al.} 2012{\natexlab{b}},
  \aap, 542, A62

\bibitem[{{Lallement} {et~al.}(2018){Lallement}, {Capitanio}, {Ruiz-Dern},
  {Danielski}, {Babusiaux}, {Vergely}, {Elyajouri}, {Arenou}, \&
  {Leclerc}}]{Lallement2018}
{Lallement}, R., {Capitanio}, L., {Ruiz-Dern}, L., {et~al.} 2018, \aap, 616,
  A132

\bibitem[{{Lebreton} \& {Reese}(2020)}]{SPINS20}
{Lebreton}, Y. \& {Reese}, D.~R. 2020, \aap, 642, A88

\bibitem[{{Limongi} \& {Chieffi}(2018)}]{LiCh18}
{Limongi}, M. \& {Chieffi}, A. 2018, \apjs, 237, 13

\bibitem[{{Luo} {et~al.}(2022){Luo}, {Zhao}, {Zhao}, \& {et al.}}]{LAMOSTDR7}
{Luo}, A.~L., {Zhao}, Y.~H., {Zhao}, G., \& {et al.} 2022, VizieR Online Data
  Catalog, V/156

\bibitem[{{Maeder} {et~al.}(2014){Maeder}, {Przybilla}, {Nieva}, {Georgy},
  {Meynet}, {Ekstr{\"o}m}, \& {Eggenberger}}]{Maeder14}
{Maeder}, A., {Przybilla}, N., {Nieva}, M.-F., {et~al.} 2014, \aap, 565, A39

\bibitem[{{Magrini} {et~al.}(2021){Magrini}, {Lagarde}, {Charbonnel},
  {Franciosini}, {Randich}, {Smiljanic}, {Casali}, {Viscasillas V{\'a}zquez},
  {Spina}, {Biazzo}, {Pasquini}, {Bragaglia}, {Van der Swaelmen},
  {Tautvai{\v{s}}ien{\.{e}}}, {Inno}, {Sanna}, {Prisinzano}, {Degl'Innocenti},
  {Prada Moroni}, {Roccatagliata}, {Tognelli}, {Monaco}, {de Laverny},
  {Delgado-Mena}, {Baratella}, {D'Orazi}, {Vallenari}, {Gonneau}, {Worley},
  {Jim{\'e}nez-Esteban}, {Jofre}, {Bensby}, {Fran{\c{c}}ois}, {Guiglion},
  {Bayo}, {Jeffries}, {Binks}, {Gilmore}, {Damiani}, {Korn}, {Pancino},
  {Sacco}, {Hourihane}, {Morbidelli}, \& {Zaggia}}]{Magrini21}
{Magrini}, L., {Lagarde}, N., {Charbonnel}, C., {et~al.} 2021, \aap, 651, A84

\bibitem[{{Majewski} {et~al.}(2017){Majewski}, {Schiavon}, {Frinchaboy},
  {Allende Prieto}, {Barkhouser}, {Bizyaev}, {Blank}, {Brunner}, {Burton},
  {Carrera}, {Chojnowski}, {Cunha}, {Epstein}, {Fitzgerald}, {Garc{\'\i}a
  P{\'e}rez}, {Hearty}, {Henderson}, {Holtzman}, {Johnson}, {Lam}, {Lawler},
  {Maseman}, {M{\'e}sz{\'a}ros}, {Nelson}, {Nguyen}, {Nidever}, {Pinsonneault},
  {Shetrone}, {Smee}, {Smith}, {Stolberg}, {Skrutskie}, {Walker}, {Wilson},
  {Zasowski}, {Anders}, {Basu}, {Beland}, {Blanton}, {Bovy}, {Brownstein},
  {Carlberg}, {Chaplin}, {Chiappini}, {Eisenstein}, {Elsworth}, {Feuillet},
  {Fleming}, {Galbraith-Frew}, {Garc{\'\i}a}, {Garc{\'\i}a-Hern{\'a}ndez},
  {Gillespie}, {Girardi}, {Gunn}, {Hasselquist}, {Hayden}, {Hekker}, {Ivans},
  {Kinemuchi}, {Klaene}, {Mahadevan}, {Mathur}, {Mosser}, {Muna}, {Munn},
  {Nichol}, {O'Connell}, {Parejko}, {Robin}, {Rocha-Pinto}, {Schultheis},
  {Serenelli}, {Shane}, {Silva Aguirre}, {Sobeck}, {Thompson}, {Troup},
  {Weinberg}, \& {Zamora}}]{APOGEE}
{Majewski}, S.~R., {Schiavon}, R.~P., {Frinchaboy}, P.~M., {et~al.} 2017, \aj,
  154, 94

\bibitem[{{Marshall} {et~al.}(2006){Marshall}, {Robin}, {Reyl{\'e}},
  {Schultheis}, \& {Picaud}}]{Marshall2006}
{Marshall}, D.~J., {Robin}, A.~C., {Reyl{\'e}}, C., {Schultheis}, M., \&
  {Picaud}, S. 2006, \aap, 453, 635

\bibitem[{{McCormick} {et~al.}(2023){McCormick}, {Majewski}, {Smith}, {Hayes},
  {Cunha}, {Masseron}, {Weiss}, {Shetrone}, {Almeida}, {Frinchaboy},
  {Garc{\'\i}a-Hern{\'a}ndez}, \& {Nitschelm}}]{McCormick23}
{McCormick}, C., {Majewski}, S.~R., {Smith}, V.~V., {et~al.} 2023, \mnras, 524,
  4418

\bibitem[{{Miglio} {et~al.}(2021){Miglio}, {Chiappini}, {Mackereth}, {Davies},
  {Brogaard}, {Casagrande}, {Chaplin}, {Girardi}, {Kawata}, {Khan}, {Izzard},
  {Montalb{\'a}n}, {Mosser}, {Vincenzo}, {Bossini}, {Noels}, {Rodrigues},
  {Valentini}, \& {Mandel}}]{Miglio21}
{Miglio}, A., {Chiappini}, C., {Mackereth}, J.~T., {et~al.} 2021, \aap, 645,
  A85

\bibitem[{{Mikolaitis} {et~al.}(2010){Mikolaitis}, {Tautvai{\v s}ien{\.e}},
  {Gratton}, {Bragaglia}, \& {Carretta}}]{Mikolaitis10}
{Mikolaitis}, {\v S}., {Tautvai{\v s}ien{\.e}}, G., {Gratton}, R., {Bragaglia},
  A., \& {Carretta}, E. 2010, \mnras, 407, 1866

\bibitem[{{Montalb{\'a}n} {et~al.}(2021){Montalb{\'a}n}, {Mackereth}, {Miglio},
  {Vincenzo}, {Chiappini}, {Buldgen}, {Mosser}, {Noels}, {Scuflaire}, {Vrard},
  {Willett}, {Davies}, {Hall}, {Bo Nielsen}, {Khan}, {Rendle}, {van Rossem},
  {Ferguson}, \& {Chaplin}}]{Montalban21}
{Montalb{\'a}n}, J., {Mackereth}, J.~T., {Miglio}, A., {et~al.} 2021, Nature
  Astronomy [\eprint[arXiv]{2006.01783}]

\bibitem[{{Mor} {et~al.}(2018){Mor}, {Robin}, {Figueras}, \& {Antoja}}]{Mor18}
{Mor}, R., {Robin}, A.~C., {Figueras}, F., \& {Antoja}, T. 2018, \aap, 620, A79

\bibitem[{{Morel} \& {Miglio}(2012)}]{Morel2012}
{Morel}, T. \& {Miglio}, A. 2012, \mnras, 419, L34

\bibitem[{{Morel} {et~al.}(2014){Morel}, {Miglio}, {Lagarde}, {Montalb{\'a}n},
  {Rainer}, {Poretti}, {Eggenberger}, {Hekker}, {Kallinger}, {Mosser},
  {Valentini}, {Carrier}, {Hareter}, \& {Mantegazza}}]{Morel14}
{Morel}, T., {Miglio}, A., {Lagarde}, N., {et~al.} 2014, \aap, 564, A119

\bibitem[{{Mosser} {et~al.}(2011){Mosser}, {Barban}, {Montalb{\'a}n}, {Beck},
  {Miglio}, {Belkacem}, {Goupil}, {Hekker}, {De Ridder}, {Dupret}, {Elsworth},
  {Noels}, {Baudin}, {Michel}, {Samadi}, {Auvergne}, {Baglin}, \&
  {Catala}}]{Mosser11}
{Mosser}, B., {Barban}, C., {Montalb{\'a}n}, J., {et~al.} 2011, \aap, 532, A86

\bibitem[{{Mosser} {et~al.}(2014){Mosser}, {Benomar}, {Belkacem}, {Goupil},
  {Lagarde}, {Michel}, {Lebreton}, {Stello}, {Vrard}, {Barban}, {Bedding},
  {Deheuvels}, {Chaplin}, {De Ridder}, {Elsworth}, {Montalban}, {Noels},
  {Ouazzani}, {Samadi}, {White}, \& {Kjeldsen}}]{Mosser14}
{Mosser}, B., {Benomar}, O., {Belkacem}, K., {et~al.} 2014, \aap, 572, L5

\bibitem[{{Mosser} {et~al.}(2012){Mosser}, {Goupil}, {Belkacem}, {Michel},
  {Stello}, {Marques}, {Elsworth}, {Barban}, {Beck}, {Bedding}, {De Ridder},
  {Garc{\'{\i}}a}, {Hekker}, {Kallinger}, {Samadi}, {Stumpe}, {Barclay}, \&
  {Burke}}]{Mosser12a}
{Mosser}, B., {Goupil}, M.~J., {Belkacem}, K., {et~al.} 2012, \aap, 540, A143

\bibitem[{{Palacios} {et~al.}(2003){Palacios}, {Talon}, {Charbonnel}, \&
  {Forestini}}]{Palacios03}
{Palacios}, A., {Talon}, S., {Charbonnel}, C., \& {Forestini}, M. 2003, \aap,
  399, 603

\bibitem[{{Pavlenko} {et~al.}(2012){Pavlenko}, {Jenkins}, {Jones}, {Ivanyuk},
  \& {Pinfield}}]{Pavlenko2012}
{Pavlenko}, Y.~V., {Jenkins}, J.~S., {Jones}, H.~R.~A., {Ivanyuk}, O., \&
  {Pinfield}, D.~J. 2012, \mnras, 422, 542

\bibitem[{{Pignatari} {et~al.}(2016){Pignatari}, {Herwig}, {Hirschi},
  {Bennett}, {Rockefeller}, {Fryer}, {Timmes}, {Ritter}, {Heger}, {Jones},
  {Battino}, {Dotter}, {Trappitsch}, {Diehl}, {Frischknecht}, {Hungerford},
  {Magkotsios}, {Travaglio}, \& {Young}}]{Pignatari16}
{Pignatari}, M., {Herwig}, F., {Hirschi}, R., {et~al.} 2016, \apjs, 225, 24

\bibitem[{{Pinsonneault} {et~al.}(2018){Pinsonneault}, {Elsworth}, {Tayar},
  {Serenelli}, {Stello}, {Zinn}, {Mathur}, {Garc{\'\i}a}, {Johnson}, {Hekker},
  {Huber}, {Kallinger}, {M{\'e}sz{\'a}ros}, {Mosser}, {Stassun}, {Girardi},
  {Rodrigues}, {Silva Aguirre}, {An}, {Basu}, {Chaplin}, {Corsaro}, {Cunha},
  {Garc{\'\i}a-Hern{\'a}ndez}, {Holtzman}, {J{\"o}nsson}, {Shetrone}, {Smith},
  {Sobeck}, {Stringfellow}, {Zamora}, {Beers}, {Fern{\'a}ndez-Trincado},
  {Frinchaboy}, {Hearty}, \& {Nitschelm}}]{Pinsonneault18}
{Pinsonneault}, M.~H., {Elsworth}, Y.~P., {Tayar}, J., {et~al.} 2018, \apjs,
  239, 32

\bibitem[{{Randich} {et~al.}(2022){Randich}, {Gilmore}, {Magrini}, {Sacco},
  {Jackson}, {Jeffries}, {Worley}, {Hourihane}, {Gonneau}, {Viscasillas
  Vazquez}, {Franciosini}, {Lewis}, {Alfaro}, {Allende Prieto}, {Bensby},
  {Blomme}, {Bragaglia}, {Flaccomio}, {Fran{\c{c}}ois}, {Irwin}, {Koposov},
  {Korn}, {Lanzafame}, {Pancino}, {Recio-Blanco}, {Smiljanic}, {Van Eck},
  {Zwitter}, {Asplund}, {Bonifacio}, {Feltzing}, {Binney}, {Drew}, {Ferguson},
  {Micela}, {Negueruela}, {Prusti}, {Rix}, {Vallenari}, {Bayo}, {Bergemann},
  {Biazzo}, {Carraro}, {Casey}, {Damiani}, {Frasca}, {Heiter}, {Hill},
  {Jofr{\'e}}, {de Laverny}, {Lind}, {Marconi}, {Martayan}, {Masseron},
  {Monaco}, {Morbidelli}, {Prisinzano}, {Sbordone}, {Sousa}, {Zaggia},
  {Adibekyan}, {Bonito}, {Caffau}, {Daflon}, {Feuillet}, {Gebran}, {Gonzalez
  Hernandez}, {Guiglion}, {Herrero}, {Lobel}, {Maiz Apellaniz}, {Merle},
  {Mikolaitis}, {Montes}, {Morel}, {Soubiran}, {Spina}, {Tabernero},
  {Tautvai{\v{s}}iene}, {Traven}, {Valentini}, {Van der Swaelmen}, {Villanova},
  {Wright}, {Abbas}, {Aguirre B{\o}rsen-Koch}, {Alves}, {Balaguer-Nunez},
  {Barklem}, {Barrado}, {Berlanas}, {Binks}, {Bressan}, {Capuzzo-Dolcetta},
  {Casagrande}, {Casamiquela}, {Collins}, {D'Orazi}, {Dantas}, {Debattista},
  {Delgado-Mena}, {Di Marcantonio}, {Drazdauskas}, {Evans}, {Famaey},
  {Franchini}, {Fr{\'e}mat}, {Friel}, {Fu}, {Geisler}, {Gerhard}, {Gonzalez
  Solares}, {Grebel}, {Gutierrez Albarran}, {Hatzidimitriou}, {Held},
  {Jim{\'e}nez-Esteban}, {J{\"o}nsson}, {Jordi}, {Khachaturyants},
  {Kordopatis}, {Kos}, {Lagarde}, {Mahy}, {Mapelli}, {Marfil}, {Martell},
  {Messina}, {Miglio}, {Minchev}, {Moitinho}, {Montalban}, {Monteiro},
  {Morossi}, {Mowlavi}, {Mucciarelli}, {Murphy}, {Nardetto}, {Ortolani},
  {Paletou}, {Palou{\v{s}}}, {Paunzen}, {Pickering}, {Quirrenbach}, {Re
  Fiorentin}, {Read}, {Romano}, {Ryde}, {Sanna}, {Santos}, {Seabroke},
  {Spagna}, {Steinmetz}, {Stonkut{\'e}}, {Sutorius}, {Th{\'e}venin}, {Tosi},
  {Tsantaki}, {Vink}, {Wright}, {Wyse}, {Zoccali}, {Zorec}, {Zucker}, \&
  {Walton}}]{RandichGES22}
{Randich}, S., {Gilmore}, G., {Magrini}, L., {et~al.} 2022, \aap, 666, A121

\bibitem[{{Rauer} {et~al.}(2014){Rauer}, {Catala}, {Aerts}, {Appourchaux},
  {Benz}, {Brandeker}, {Christensen-Dalsgaard}, {Deleuil}, {Gizon}, {Goupil},
  {G{\"u}del}, {Janot-Pacheco}, {Mas-Hesse}, {Pagano}, {Piotto}, {Pollacco},
  {Santos}, {Smith}, {Su{\'a}rez}, {Szab{\'o}}, {Udry}, {Adibekyan}, {Alibert},
  {Almenara}, {Amaro-Seoane}, {Eiff}, {Asplund}, {Antonello}, {Barnes},
  {Baudin}, {Belkacem}, {Bergemann}, {Bihain}, {Birch}, {Bonfils}, {Boisse},
  {Bonomo}, {Borsa}, {Brand{\~a}o}, {Brocato}, {Brun}, {Burleigh}, {Burston},
  {Cabrera}, {Cassisi}, {Chaplin}, {Charpinet}, {Chiappini}, {Church},
  {Csizmadia}, {Cunha}, {Damasso}, {Davies}, {Deeg}, {D{\'{\i}}az}, {Dreizler},
  {Dreyer}, {Eggenberger}, {Ehrenreich}, {Eigm{\"u}ller}, {Erikson}, {Farmer},
  {Feltzing}, {de Oliveira Fialho}, {Figueira}, {Forveille}, {Fridlund},
  {Garc{\'{\i}}a}, {Giommi}, {Giuffrida}, {Godolt}, {Gomes da Silva},
  {Granzer}, {Grenfell}, {Grotsch-Noels}, {G{\"u}nther}, {Haswell}, {Hatzes},
  {H{\'e}brard}, {Hekker}, {Helled}, {Heng}, {Jenkins}, {Johansen},
  {Khodachenko}, {Kislyakova}, {Kley}, {Kolb}, {Krivova}, {Kupka}, {Lammer},
  {Lanza}, {Lebreton}, {Magrin}, {Marcos-Arenal}, {Marrese}, {Marques},
  {Martins}, {Mathis}, {Mathur}, {Messina}, {Miglio}, {Montalban}, {Montalto},
  {Monteiro}, {Moradi}, {Moravveji}, {Mordasini}, {Morel}, {Mortier},
  {Nascimbeni}, {Nelson}, {Nielsen}, {Noack}, {Norton}, {Ofir}, {Oshagh},
  {Ouazzani}, {P{\'a}pics}, {Parro}, {Petit}, {Plez}, {Poretti}, {Quirrenbach},
  {Ragazzoni}, {Raimondo}, {Rainer}, {Reese}, {Redmer}, {Reffert},
  {Rojas-Ayala}, {Roxburgh}, {Salmon}, {Santerne}, {Schneider}, {Schou},
  {Schuh}, {Schunker}, {Silva-Valio}, {Silvotti}, {Skillen}, {Snellen}, {Sohl},
  {Sousa}, {Sozzetti}, {Stello}, {Strassmeier}, {{\v S}vanda}, {Szab{\'o}},
  {Tkachenko}, {Valencia}, {Van Grootel}, {Vauclair}, {Ventura}, {Wagner},
  {Walton}, {Weingrill}, {Werner}, {Wheatley}, \& {Zwintz}}]{PLATO}
{Rauer}, H., {Catala}, C., {Aerts}, C., {et~al.} 2014, Experimental Astronomy,
  38, 249

\bibitem[{{Recio-Blanco} {et~al.}(2014){Recio-Blanco}, {de Laverny},
  {Kordopatis}, {Helmi}, {Hill}, {Gilmore}, {Wyse}, {Adibekyan}, {Randich},
  {Asplund}, {Feltzing}, {Jeffries}, {Micela}, {Vallenari}, {Alfaro}, {Allende
  Prieto}, {Bensby}, {Bragaglia}, {Flaccomio}, {Koposov}, {Korn}, {Lanzafame},
  {Pancino}, {Smiljanic}, {Jackson}, {Lewis}, {Magrini}, {Morbidelli},
  {Prisinzano}, {Sacco}, {Worley}, {Hourihane}, {Bergemann}, {Costado},
  {Heiter}, {Joffre}, {Lardo}, {Lind}, \& {Maiorca}}]{RecioBlanco14}
{Recio-Blanco}, A., {de Laverny}, P., {Kordopatis}, G., {et~al.} 2014, \aap,
  567, A5

\bibitem[{{Recio-Blanco} {et~al.}(2023){Recio-Blanco}, {de Laverny}, {Palicio},
  {Kordopatis}, {{\'A}lvarez}, {Schultheis}, {Contursi}, {Zhao}, {Torralba
  Elipe}, {Ordenovic}, {Manteiga}, {Dafonte}, {Oreshina-Slezak}, {Bijaoui},
  {Fr{\'e}mat}, {Seabroke}, {Pailler}, {Spitoni}, {Poggio}, {Creevey}, {Abreu
  Aramburu}, {Accart}, {Andrae}, {Bailer-Jones}, {Bellas-Velidis}, {Brouillet},
  {Brugaletta}, {Burlacu}, {Carballo}, {Casamiquela}, {Chiavassa}, {Cooper},
  {Dapergolas}, {Delchambre}, {Dharmawardena}, {Drimmel}, {Edvardsson},
  {Fouesneau}, {Garabato}, {Garc{\'\i}a-Lario}, {Garc{\'\i}a-Torres}, {Gavel},
  {Gomez}, {Gonz{\'a}lez-Santamar{\'\i}a}, {Hatzidimitriou}, {Heiter},
  {Jean-Antoine Piccolo}, {Kontizas}, {Korn}, {Lanzafame}, {Lebreton}, {Le
  Fustec}, {Licata}, {Lindstr{\o}m}, {Livanou}, {Lobel}, {Lorca}, {Magdaleno
  Romeo}, {Marocco}, {Marshall}, {Mary}, {Nicolas}, {Pallas-Quintela}, {Panem},
  {Pichon}, {Riclet}, {Robin}, {Rybizki}, {Santove{\~n}a}, {Silvelo}, {Smart},
  {Sarro}, {Sordo}, {Soubiran}, {S{\"u}veges}, {Ulla}, {Vallenari}, {Zorec},
  {Utrilla}, \& {Bakker}}]{RecioBlanco23}
{Recio-Blanco}, A., {de Laverny}, P., {Palicio}, P.~A., {et~al.} 2023, \aap,
  674, A29

\bibitem[{{Reimers}(1975)}]{Reimers75}
{Reimers}, D. 1975, Memoires of the Societe Royale des Sciences de Liege, 8,
  369

\bibitem[{{Ricker} {et~al.}(2015){Ricker}, {Winn}, {Vanderspek}, {Latham},
  {Bakos}, {Bean}, {Berta-Thompson}, {Brown}, {Buchhave}, {Butler}, {Butler},
  {Chaplin}, {Charbonneau}, {Christensen-Dalsgaard}, {Clampin}, {Deming},
  {Doty}, {De Lee}, {Dressing}, {Dunham}, {Endl}, {Fressin}, {Ge}, {Henning},
  {Holman}, {Howard}, {Ida}, {Jenkins}, {Jernigan}, {Johnson}, {Kaltenegger},
  {Kawai}, {Kjeldsen}, {Laughlin}, {Levine}, {Lin}, {Lissauer}, {MacQueen},
  {Marcy}, {McCullough}, {Morton}, {Narita}, {Paegert}, {Palle}, {Pepe},
  {Pepper}, {Quirrenbach}, {Rinehart}, {Sasselov}, {Sato}, {Seager},
  {Sozzetti}, {Stassun}, {Sullivan}, {Szentgyorgyi}, {Torres}, {Udry}, \&
  {Villasenor}}]{Ricker15}
{Ricker}, G.~R., {Winn}, J.~N., {Vanderspek}, R., {et~al.} 2015, Journal of
  Astronomical Telescopes, Instruments, and Systems, 1, 014003

\bibitem[{{Robin} {et~al.}(2017){Robin}, {Bienaym{\'e}},
  {Fern{\'a}ndez-Trincado}, \& {Reyl{\'e}}}]{Robin17}
{Robin}, A.~C., {Bienaym{\'e}}, O., {Fern{\'a}ndez-Trincado}, J.~G., \&
  {Reyl{\'e}}, C. 2017, \aap, 605, A1

\bibitem[{{Robin} {et~al.}(2014){Robin}, {Reyl{\'e}}, {Fliri}, {Czekaj},
  {Robert}, \& {Martins}}]{Robin14}
{Robin}, A.~C., {Reyl{\'e}}, C., {Fliri}, J., {et~al.} 2014, \aap, 569, A13

\bibitem[{{Rodrigues} {et~al.}(2017){Rodrigues}, {Bossini}, {Miglio},
  {Girardi}, {Montalb{\'a}n}, {Noels}, {Trabucchi}, {Coelho}, \&
  {Marigo}}]{Rodrigues17}
{Rodrigues}, T.~S., {Bossini}, D., {Miglio}, A., {et~al.} 2017, \mnras, 467,
  1433

\bibitem[{{Sackmann} {et~al.}(1974){Sackmann}, {Smith}, \&
  {Despain}}]{Sackmann74}
{Sackmann}, I.~J., {Smith}, R.~L., \& {Despain}, K.~H. 1974, \apj, 187, 555

\bibitem[{{Schonhut-Stasik} {et~al.}(2023){Schonhut-Stasik}, {Zinn}, {Stassun},
  {Pinsonneault}, {Johnson}, {Warfield}, {Stello}, {Elsworth}, {Garcia},
  {Marhur}, {Mosser}, {Tayar}, {Stringfellow}, {Beaton}, {Jonsson}, \&
  {Minniti}}]{APOK2}
{Schonhut-Stasik}, J., {Zinn}, J.~C., {Stassun}, K.~G., {et~al.} 2023, arXiv
  e-prints, arXiv:2304.10654

\bibitem[{{Sengupta} \& {Garaud}(2018)}]{SenGar18}
{Sengupta}, S. \& {Garaud}, P. 2018, ArXiv e-prints
  [\eprint[arXiv]{1804.04258}]

\bibitem[{{Smiljanic} {et~al.}(2009){Smiljanic}, {Gauderon}, {North}, {Barbuy},
  {Charbonnel}, \& {Mowlavi}}]{Smiljanic09}
{Smiljanic}, R., {Gauderon}, R., {North}, P., {et~al.} 2009, \aap, 502, 267

\bibitem[{{Smiljanic} {et~al.}(2010){Smiljanic}, {Pasquini}, {Charbonnel}, \&
  {Lagarde}}]{Smiljanic10}
{Smiljanic}, R., {Pasquini}, L., {Charbonnel}, C., \& {Lagarde}, N. 2010, \aap,
  510, A50+

\bibitem[{{Smith} \& {Suntzeff}(1989)}]{SmSu89}
{Smith}, V.~V. \& {Suntzeff}, N.~B. 1989, \aj, 97, 1699

\bibitem[{{Sneden}(1973)}]{Sneden1973}
{Sneden}, C.~A. 1973, PhD thesis, THE UNIVERSITY OF TEXAS AT AUSTIN.

\bibitem[{{Sofue}(2015)}]{Sofue15}
{Sofue}, Y. 2015, \pasj, 67, 75

\bibitem[{{Soubiran} {et~al.}(2022){Soubiran}, {Brouillet}, \&
  {Casamiquela}}]{Soubiran22}
{Soubiran}, C., {Brouillet}, N., \& {Casamiquela}, L. 2022, \aap, 663, A4

\bibitem[{{Stetson} \& {Pancino}(2008)}]{Stetson2008}
{Stetson}, P.~B. \& {Pancino}, E. 2008, \pasp, 120, 1332

\bibitem[{{Takeda} {et~al.}(2019){Takeda}, {Omiya}, {Harakawa}, \&
  {Sato}}]{Takeda19}
{Takeda}, Y., {Omiya}, M., {Harakawa}, H., \& {Sato}, B. 2019, \pasj, 71, 119

\bibitem[{{Talon} {et~al.}(2006){Talon}, {Richard}, \& {Michaud}}]{Talon06}
{Talon}, S., {Richard}, O., \& {Michaud}, G. 2006, \apj, 645, 634

\bibitem[{{Tautvai{\v s}ien{\.e}} {et~al.}(2010){Tautvai{\v s}ien{\.e}},
  {Barisevi{\v c}ius}, {Berdyugina}, {Chorniy}, \& {Ilyin}}]{Tautvaisiene10}
{Tautvai{\v s}ien{\.e}}, G., {Barisevi{\v c}ius}, G., {Berdyugina}, S.,
  {Chorniy}, Y., \& {Ilyin}, I. 2010, Baltic Astronomy, 19, 95

\bibitem[{{Tautvai{\v s}ien{\.e}} {et~al.}(2013){Tautvai{\v s}ien{\.e}},
  {Barisevi{\v c}ius}, {Chorniy}, {Ilyin}, \& {Puzeras}}]{Tautvaisiene13}
{Tautvai{\v s}ien{\.e}}, G., {Barisevi{\v c}ius}, G., {Chorniy}, Y., {Ilyin},
  I., \& {Puzeras}, E. 2013, \mnras, 430, 621

\bibitem[{{Tayar} \& {Joyce}(2022)}]{Tayar22}
{Tayar}, J. \& {Joyce}, M. 2022, \apjl, 935, L30

\bibitem[{{Telting} {et~al.}(2014){Telting}, {Avila}, {Buchhave}, {Frandsen},
  {Gandolfi}, {Lindberg}, {Stempels}, {Prins}, \& {NOT staff}}]{Telting2014}
{Telting}, J.~H., {Avila}, G., {Buchhave}, L., {et~al.} 2014, Astronomische
  Nachrichten, 335, 41

\bibitem[{{Traxler} {et~al.}(2011){Traxler}, {Garaud}, \&
  {Stellmach}}]{Traxleretal11}
{Traxler}, A., {Garaud}, P., \& {Stellmach}, S. 2011, \apjl, 728, L29+

\bibitem[{{Ulrich}(1972)}]{Ulrich72}
{Ulrich}, R.~K. 1972, \apj, 172, 165

\bibitem[{{Ulrich}(1986)}]{Ulrich86}
{Ulrich}, R.~K. 1986, \apjl, 306, L37

\bibitem[{{Valentini} {et~al.}(2019){Valentini}, {Chiappini}, {Bossini},
  {Miglio}, {Davies}, {Mosser}, {Elsworth}, {Mathur}, {Garc{\'\i}a}, {Girardi},
  {Rodrigues}, {Steinmetz}, \& {Vallenari}}]{Valentini19}
{Valentini}, M., {Chiappini}, C., {Bossini}, D., {et~al.} 2019, \aap, 627, A173

\bibitem[{{Valentini} {et~al.}(2016){Valentini}, {Chiappini}, {Miglio},
  {Montalb{\'a}n}, {Rodrigues}, {Mosser}, {Anders}, {CoRoT RG Group}, \& {GES
  Consortium}}]{Valentini16}
{Valentini}, M., {Chiappini}, C., {Miglio}, A., {et~al.} 2016, Astronomische
  Nachrichten, 337, 970

\bibitem[{{Vauclair} {et~al.}(1978){Vauclair}, {Vauclair}, {Schatzman}, \&
  {Michaud}}]{Vauclair78}
{Vauclair}, S., {Vauclair}, G., {Schatzman}, E., \& {Michaud}, G. 1978, \apj,
  223, 567

\bibitem[{{Vrard} {et~al.}(2016){Vrard}, {Mosser}, \& {Samadi}}]{Vrard16}
{Vrard}, M., {Mosser}, B., \& {Samadi}, R. 2016, \aap, 588, A87

\bibitem[{{Yoshii}(1982)}]{Yoshii82}
{Yoshii}, Y. 1982, \pasj, 34, 365

\end{thebibliography}

\begin{appendix}
\section{Queries of Gaia data samples}
\label{ADQL}
The following ADQL queries to Gaia DR3 catalog through the Gaia archive can be used to retrived the data sample used in this study. We follow the recommendations given in \citet{RecioBlanco23} and using the definition given in \citet{RecioBlanco22}

\lstset{language=SQL}

\begin{lstlisting}[caption={\texttt{ADQL} query with simple cuts in the limiting parameters for the \textbf{General sample}},captionpos=b]

FROM gaiadr3.astrophysical_parameters
WHERE (teff_gspspec>3500 AND teff_gspspec<7000) AND
(logg_gspspec>0 AND logg_gspspec<5) AND
(flags_gspspec LIKE '0%') AND (flags_gspspec
LIKE '_0%') AND (flags_gspspec LIKE '__0%') AND
(flags_gspspec LIKE '___0%') AND (flags_gspspec
LIKE '____0%') AND (flags_gspspec LIKE
'_____0%') AND (flags_gspspec LIKE '______0%')
AND ((flags_gspspec LIKE '_______0%') OR
(flags_gspspec LIKE '_______1%') OR
(flags_gspspec LIKE '_______2%') ) AND
((flags_gspspec LIKE '____________0%') OR
(flags_gspspec LIKE '____________1%') ) AND
(flags_gspspec LIKE '________0%') AND
(flags_gspspec LIKE '_________0%') AND
 (flags_gspspec LIKE '__________0%') AND
(flags_gspspec LIKE '___________0%')  
\end{lstlisting}

\begin{lstlisting}[caption={\texttt{ADQL} query for the Nitrogen abundance sample},captionpos=b]

GENERAL SAMPLE AND
((nfe_gspspec_upper-nfe_gspspec_lower)<0.30) AND
((flags_gspspec LIKE '_____________0%') OR
(flags_gspspec LIKE '_____________1%')) AND
(flags_gspspec LIKE '______________0%') AND
(nfe_gspspec_nlines>=2) AND
(nfe_gspspec_linescatter<0.1) 
\end{lstlisting}

\begin{lstlisting}[caption={\texttt{ADQL} query for the Magnesium abundance sample},captionpos=b]
GENERAL SAMPLE AND
((Mgfe_gspspec_upper-Mgfe_gspspec_lower)<0.40) AND
((flags_gspspec LIKE '_______________0%') OR
(flags_gspspec LIKE '_______________1%')) AND
(flags_gspspec LIKE '________________0%') 
\end{lstlisting}

\begin{lstlisting}[caption={\texttt{ADQL} query for the Iron abundance sample},captionpos=b]
GENERAL SAMPLE AND
((mh_gspspec_upper-mh_gspspec_lower)<.4) AND 
((flags_gspspec LIKE '___________________________0%') OR
(flags_gspspec LIKE '___________________________1%')) AND
(flags_gspspec LIKE '____________________________0%')
\end{lstlisting}

We also calibrated the Gaia abundances of iron, nitrogen and magnesium as well as log\,$g$ using the following formulae: 
\noindent 
\begin{multline}
\text{[Fe/H]}_{\text{{calibrated}}}=\text{[Fe/H]}_{\text{published}} + 0.3699 -0.068 \times \text{{log\,$g$}}_{\text{published}} \\
   \hspace{1cm}   +0.0028 \times \text{{log\,$g$}}_{\text{published}}^2 -0.0004\times\text{{log\,$g$}}_{\text{published}}^3 \\  
\text{where [Fe/H]$_{\text{published}}$=[FeI/M]$_{\text{published}}$+[M/H]$_{\text{published}}$}
\end{multline}
\begin{multline}
\text{[N/Fe]}_{\text{{calibrated}}}=\text{[N/Fe]}_{\text{published}} + 0.0975 -0.0293 \times \text{{log\,$g$}}_{\text{published}} \\
   \hspace{1cm}   +0.0238 \times \text{{log\,$g$}}_{\text{published}}^2 -0.0071\times\text{{log\,$g$}}_{\text{published}}^3 
\end{multline}

\begin{multline}
\text{[Mg/Fe]}_{\text{{calibrated}}}=\text{[Mg/Fe]}_{\text{published}} - 0.7244 +0.3779 \times \text{{log\,$g$}}_{\text{published}} \\
   \hspace{1cm}   -0.0421 \times \text{{log\,$g$}}_{\text{published}}^2 -0.0038\times\text{{log\,$g$}}_{\text{published}}^3 
\end{multline}

\begin{multline}
\text{{log\,$g$}}_{\text{{calibrated}}}=\text{{log\,$g$}}_{\text{published}} + 0.4496 - 0.0036 \times \text{{log\,$g$}}_{\text{published}} \\
   \hspace{1cm}   -0.0224 \times \text{{log\,$g$}}_{\text{published}}^2  
\end{multline}

\begin{landscape}
\begin{table}[h]
\caption{Informations about the giants stars in our sample.Mass$_{SPiNS,1}$ and Mass$_{SPiNS,2}$ indicate the mass determinations with SPiNS  considering spectroscopic observables and  both spectroscopic and asteroseismic observables, respectively. The column `` Evol" indicates the stellar evolutionnary state (1 for first ascent RGB stars and 2 for clump stars).  }
\label{observations_table}
\begin{center}
       \scalebox{0.6}{
	\begin{tabular}{|l|l|l|r|r|r|r|r|r|r|r|r|r|r|r|r|r|r|r|r|r|r|r|r|r|}
 	\hline
            KIC & RA & DEC & T$_{\rm eff}$ & eT$_{\rm eff}$ & log\,$g$ &elog\,$g$ & [Fe/H] &e[Fe/H] & A(C) &eA(C) & A(N) & eA(N) & A(O) & eA(O) & $^{12}$C/$^{13}$C & e$^{12}$C/$^{13}$C & Mass$_{PARAM}$ &  Mass$_{PARAM, 68L}$ & Mass$_{PARAM, 68U}$ & Mass$_{SPiNS,1}$ & eMass$_{SPiNS,1}$ &  Mass$_{SPiNS,2}$ & eMass$_{SPiNS,2}$ &  Evol \\
	\hline
	     KIC2696955 & 19:08:33.73 & +37:54:36.8 & 4727 & 61.0 & 2.93 & 0.19 & 0.02 & 0.11 & 8.42 & 0.05 & 8.09 & 0.05 & 8.76 & 0.05 & 20.0 & 2.0 & 1.64 & 1.52 & 1.69 & 1.43 & 0.41 & 1.72 & 0.08 & 1\\
              KIC3744043 & 19:22:27.46 & +38:50:52.9 & 4917 & 52.0 & 2.96 & 0.24 & -0.31 & 0.11 & 8.01 & 0.02 & 7.59 & 0.04 & 8.52 & 0.05 &  &  & 1.17 & 1.14 & 1.21 & 1.37 & 0.38 & 1.26 & 0.04 & 1\\
              KIC4262505 & 19:26:30.55 & +39:19:24.7 & 4846 & 50.0 & 2.85 & 0.22 & -0.18 & 0.11 & 8.09 & 0.05 & 7.7 & 0.06 & 8.72 & 0.05 &  &  & 1.34 & 1.30 & 1.38 & 1.45 & 0.42 & 1.42 & 0.04 & 1\\
              KIC5006817 & 19:21:49.43 & +40:08:44.6 & 4767 & 49.0 & 3.06 & 0.17 & 0.0 & 0.09 & 8.36 & 0.03 & 8.11 & 0.05 & 8.86 & 0.1 & 27.0 & 4.0 & 1.32 & 1.28 & 1.34 & 1.41 & 0.32 & 1.32 & 0.03 & 1\\
              KIC5530598 & 19:23:16.88 & +40:42:21.5 & 4552 & 70.0 & 2.92 & 0.2 & 0.22 & 0.12 & 8.5 & 0.03  & 8.37 & 0.06 & 8.91 & 0.05 & 23.0 & 2.0 & 1.51 & 1.46 & 1.52 & 1.29 & 0.28 & 1.45 & 0.05 & 1\\
              KIC6144777 & 19:49:36.53 & +41:24:14.2 & 4758 & 49.0 & 3.01 & 0.17 & -0.05 & 0.09 & 8.31 & 0.01 & 7.91 & 0.05 & 8.72 & 0.05 & 27.0 & 5.0 & 1.12 & 1.08 & 1.14 & 1.37 & 0.30 & 1.11 & 0.03 & 1\\
              KIC8718745 & 20:03:39.49 & +44:49:04.5 & 4795 & 52.5 & 3.03 & 0.22 & -0.35 & 0.11 & 8.12 & 0.03 & 7.54 & 0.03 & 8.64 & 0.05 &  &  & 0.98 & 0.95 & 1.00 & 1.11 & 0.20 & 1.09 & 0.02 & 1\\
              KIC11550492 & 19:06:59.90 & +49:32:24.4 & 4723 & 65.0 & 2.88 & 0.24 & -0.13 & 0.13 & 8.34 & 0.05 & 7.87 & 0.05 & 8.83 & 0.1 &  &  & 1.08 & 1.06 & 1.12 & 1.25 & 0.30 & 1.13 & 0.03 & 1\\
              KIC11618103 & 19:41:35.35 & +49:41:43.3 & 4873 & 53.5 & 2.94 & 0.23 & -0.22 & 0.12 & 8.11 & 0.01 & 7.85 & 0.04 & 8.66 & 0.05 &  &  & 1.27& 1.23& 1.31 & 1.46 & 0.46 & 1.30 & 0.03 & 1\\
              KIC1161618 & 19:24:26.15 & +36:48:47.9 & 4648 & 52.0 & 2.44 & 0.18 & -0.14 & 0.1 & 8.25 & 0.04 & 7.95 & 0.04 & 8.79 & 0.1 & 8.0 & 2.0 & 1.18 & 1.11 & 1.21 & 1.25 & 0.34 & 1.10 & 0.045 & 2\\
              KIC1726211 & 19:30:01.07 & +37:17:33.9 & 4960 & 42.0 & 2.39 & 0.22 & -0.58 & 0.11 & 7.77 & 0.09 & 7.33 & 0.06 & 8.62 & 0.05 & 4.0 & 2.0 & 1.23 & 1.18 & 1.43 & 1.13 & 0.33 & 1.21 & 0.13 & 2\\
              KIC2303367 & 19:26:08.81 & +37:40:24.4 & 4621 & 65.0 & 2.43 & 0.22 & -0.03 & 0.12 & 8.38 & 0.01 & 8.05 & 0.06 & 8.86 & 0.05 & 13.0 & 2.0 & 1.18 & 1.12 & 1.25 & 1.40 & 0.47 & 1.13 & 0.06 & 2\\
              KIC2714397 & 19:26:54.91 & +37:56:29.9 & 4831 & 48.0 & 2.43 & 0.24 & -0.47 & 0.12 & 7.92 & 0.01 & 7.34 & 0.05 & 8.63 & 0.05 & 10.0 & 3.0 & 1.04 & 0.96 & 1.09 & 1.23 & 0.36 & 0.97 & 0.05 & 2\\
              KIC3098716 & 19:04:45.13 & +38:17:31.0 & 4931 & 55.0 & 2.39 & 0.25 & -0.36 & 0.12 & 7.92 & 0.01 & 7.45 & 0.08 & 8.59 & 0.05 &  &  & 0.95 & 0.87 & 1.00 & 1.03 & 0.41 & 0.70 & 0.02 & 2\\
              KIC3730953 & 19:04:43.38 & +38:53:53.6 & 4908 & 50.0 & 2.61 & 0.23 & -0.21 & 0.12 & 8.06 & 0.02 & 7.82 & 0.03 & 8.66 & 0.05 & 9.0 & 3.0 & 1.65 & 1.63 & 1.72 & 1.14 & 0.50& 1.84 & 0.19 & 2\\
              KIC3854605 & 19:30:22.02 & +38:55:54.7 & 4506 & 78.0 & 2.47 & 0.21 & 0.1 & 0.12 & 8.35 & 0.02 & 8.33 & 0.07 & 8.79 & 0.05 & 16.0 & 2.0 & 1.39 & 1.35 & 1.44 & 1.36 & 0.40 & 1.63 & 0.11 & 2\\
              KIC4042951 & 19:08:54.47 & +39:08:03.7 & 5004 & 44.0 & 2.41 & 0.24 & -0.46 & 0.11 & 7.89 & 0.1 & 7.37 & 0.07 & 8.56 & 0.05 & 3.0 & 3.0 & 0.96 & 0.86 & 0.99 & 1.07 & 0.46 & 0.71 & 0.01 & 2\\
              KIC4446181 & 19:01:13.09 & +39:30:13.1 & 4830 & 47.0 & 2.39 & 0.24 & -0.46 & 0.12 & 7.93 & 0.04 & 7.34 & 0.07 & 8.63 & 0.05 & 5.0 & 1.0 & 0.86 & 0.83 & 0.93 & 1.21 & 0.33& 1.11& 0.08 & 2\\
              KIC4457200 & 19:18:02.30 & +39:31:16.7 & 4613 & 65.0 & 2.54 & 0.2 & 0.07 & 0.11 & 8.4 & 0.01 & 8.18 & 0.08 & 8.78 & 0.05 & 14.0 & 3.0 & 1.57 & 1.54 & 1.60 & 1.59 & 0.48 & 1.81 & 0.14 & 2\\
              KIC4571402 & 19:37:13.48 & +39:41:30.7 & 4902 & 51.0 & 2.46 & 0.26 & -0.26 & 0.12 & 7.89 & 0.01 & 7.64 & 0.02 & 8.53 & 0.05 & 14.0 & 3.0 & 1.34 & 1.28 & 1.43 & 1.09 & 0.50 & 1.80 & 0.083 & 2\\
              KIC4575645 & 19:41:12.76 & +39:37:21.6 & 4540 & 65.0 & 2.48 & 0.2 & 0.09 & 0.11 & 8.33 & 0.01 & 8.14 & 0.06 & 8.81 & 0.05 & 18.0 & 3.0 & 1.32 & 1.26 & 1.37 & 1.38 & 0.41 & 1.53 & 0.12 & 2\\
               KIC4755467 & 19:36:51.74 & +39:49:54.5 & 4834 & 48.0 & 2.49 & 0.22 & -0.18 & 0.11 & 8.07 & 0.01 & 7.81 & 0.03 & 8.64 & 0.05 & 6.0 & 2.0 & 1.28 & 1.25& 1.32 & 1.07 & 0.48 & 1.49 & 0.15 & 2\\
               KIC4770846 & 19:50:04.58 & +39:52:21.6 & 4826 & 56.0 & 2.65 & 0.23 & 0.03 & 0.12 & 8.24 & 0.01 & 8.03 & 0.06 & 8.83 & 0.05 &  &  & 1.84 & 1.82 & 1.87 & 1.18& 0.63 & 1.59 & 0.07 & 2\\
               KIC5000307 & 19:13:39.11 & +40:11:04.6 & 4992 & 48.0 & 2.54 & 0.25 & -0.23 & 0.12 &  &  &  &  &  &  &  &  & 1.52 & 1.46 & 1.55 & 1.032 & 0.51 & 1.27 & 0.07 & 2\\
               KIC5095946 & 19:23:07.47 & +40:17:48.6 & 4658 & 62.0 & 2.46 & 0.23 & -0.1 & 0.13 & 8.24 & 0.01 & 7.85 & 0.05 & 8.73 & 0.05 & 12.0 & 2.0 & 1.27 & 1.19 & 1.32 & 1.31 & 0.42 & 1.08 & 0.07 & 2\\
               KIC5121605 & 19:49:06.26 & +40:12:31.0 & 4629 & 62.0 & 2.42 & 0.23 & -0.13 & 0.13 & 8.3 & 0.01 & 7.93 & 0.03 & 8.81 & 0.05 & 11.0 & 2.0 & 1.16& 1.07 & 1.19 & 1.25 & 0.36 & 1.10 & 0.07 & 2\\
               KIC5128171 & 19:53:08.54 & +40:07:11.4 & 4809 & 60.0 & 2.65 & 0.24 & -0.01 & 0.13 & 8.13 & 0.05 & 8.14 & 0.04 & 8.72 & 0.05 & 14.0 & 2.0 & 1.83 & 1.8 & 1.97 & 1.14 & 0.51 & 1.80 & 0.08 & 2\\
               KIC5392657 & 19:56:06.75 & +40:35:010.0 & 4413 & 71.0 & 2.55 & 0.17 & 0.14 & 0.1 & 8.44 & 0.05 & 8.32 & 0.05 & 8.92 & 0.05 &  &  & -99.9 & -99.9 & -99.9 & 1.26 & 0.28 & 1.85& 0.14 & 2\\
               KIC5700368 & 19:20:38.71 & +40:55:45.6 & 4792 & 54.0 & 2.48 & 0.24 & -0.18 & 0.13 & 8.09 & 0.05 & 7.74 & 0.05 & 8.64 & 0.05 & 9.0 & 2.0 & 1.28 & 1.23 & 1.50 & 1.23 & 0.47 & 1.54 & 0.04 & 2\\
               KIC5709564 & 19:32:18.56 & +40:58:21.9 & 4661 & 50.0 & 2.34 & 0.21 & -0.34 & 0.11 & 8.12 & 0.05 & 7.55 & 0.08 & 8.71 &  &  &  & 0.93 & 0.86 & 1.00 & 1.24 & 0.36 & 0.94 & 0.06 & 2\\
               KIC5770923 & 18:56:52.79 & +41:01:15.3 & 5024 & 50.0 & 2.82 & 0.24 & 0.01 & 0.11 & 8.07 & 0.02 & 8.07 & 0.03 & 8.71 & 0.05 & 13.0 & 3.0 & 2.18 & 2.07 & 2.27 & 2.61 & 0.38 & 2.33 & 0.06 & 2\\
               KIC5782127 & 19:16:04.98 & +41:02:35.2 & 4732 & 56.0 & 2.36 & 0.24 & -0.17 & 0.12 & 8.04 & 0.02 & 7.81 & 0.04 & 8.63 & 0.05 & 8.0 & 3.0 & 1.21 & 1.10 & 1.26 & 1.18 & 0.40 & 1.06 & 0.07 & 2\\
               KIC5795626 & 19:33:08.67 & +41:03:08.6 & 5025 & 47.0 & 2.5 & 0.21 & -0.67 & 0.1 & 7.7 & 0.1 & 7.16 & 0.04 & 8.4 & 0.1 &  &  & 1.33 & 1.29 & 1.38 & 1.10 & 0.37 & 1.58 & 0.11 & 2\\
               KIC5872509 & 19:22:08.48 & +41:08:07.2 & 4755 & 50.0 & 2.48 & 0.22 & -0.14 & 0.11 & 8.06 & 0.04 & 7.86 & 0.07 & 8.73 & 0.05 & 8.0 & 2.0 & 1.36 & 1.31 & 1.40 & 1.27 & 0.48 & 1.60 & 0.11 & 2\\
               KIC5983299 & 19:50:53.65 & +41:14:35.5 & 4871 & 53.0 & 2.72 & 0.23 & 0.0 & 0.12 & 8.14 & 0.03 & 8.06 & 0.06 & 8.74 & 0.05 & 13.0 & 3.0 & 1.92 & 1.89 & 2.03 & 1.10 & 0.59 & 1.95 & 0.15 & 2\\
               KIC6780490 & 19:31:01.22 & +42:16:59.6 & 4710 & 53.0 & 2.48 & 0.2 & -0.1 & 0.1 & 8.04 & 0.04 & 8.07 & 0.05 & 8.67 & 0.05 & 7.0 & 1.0 & 1.35 & 1.31 & 1.38& 1.16 & 0.391 & 1.46 & 0.11& 2\\
               KIC6838707 & 18:49:46.48 & +42:22:24.7 & 4672 & 61.0 & 2.41 & 0.23 & -0.11 & 0.13 & 8.19 & 0.01 & 7.93 & 0.06 & 8.71 & 0.05 & 8.0 & 2.0 & 1.13 & 1.03 & 1.18 & 1.25 & 0.40 & 1.34 & 0.010 & 2\\
               KIC6849167 & 19:08:44.60 & +42:19:48.2 & 4799 & 48.0 & 2.45 & 0.23 & -0.21 & 0.11 & 8.1 & 0.01 & 7.73 & 0.05 & 8.69 & 0.05 & 5.0 & 2.0 & 1.12 & 1.07 & 1.19 & 1.10 & 0.43 & 1.03 & 0.08 & 2\\
               KIC6854095 & 19:16:11.94 & +42:20:36.2 & 4834 & 45.0 & 2.38 & 0.23 & -0.31 & 0.11 & 7.95 & 0.04 & 7.59 & 0.06 & 8.53 & 0.05 & 6.0 & 2.0 & 0.95 & 0.91 & 1.03 & 1.02 & 0.35 & 1.19 & 0.05 & 2\\
               KIC7366121 & 19:31:18.14 & +42:59:13.2 & 4806 & 51.0 & 2.56 & 0.23 & -0.1 & 0.12 & 8.15 & 0.01 & 7.93 & 0.03 & 8.72 & 0.05 & 13.0 & 2.0 & 1.55 & 1.48 & 1.75 & 1.09 & 0.48 & 1.92 & 0.07 & 2\\
               KIC7374855 & 19:41:08.37 & +42:54:59.4 & 4674 & 56.0 & 2.42 & 0.19 & 0.01 & 0.1 & 8.2 & 0.01 & 8.12 & 0.06 & 8.69 & 0.05 & 15.0 & 4.0 & 2.18 & 1.69 & 2.35 & 1.06 & 0.40 & 1.93 & 0.19 & 2\\
               KIC7734065 & 18:53:12.44 & +43:29:22.3 & 4894 & 40.0 & 2.33 & 0.22 & -0.44 & 0.1 & 7.89 & 0.03 & 7.43 & 0.04 & 8.43 & 0.05 & 4.0 & 1.0 & 0.98 & 0.90 & 1.04 & 1.12 & 0.41 & 0.89 & 0.08 & 2\\
               KIC7820638 & 19:24:48.69 & +43:31:18.5 & 4800 & 53.0 & 2.59 & 0.18 & -0.19 & 0.09 & 8.18 & 0.01 & 8.03 & 0.06 & 8.74 & 0.05 & 15.0 & 2.0 & 1.67 & 1.63 & 1.70& 1.63 & 0.50 & 1.56 & 0.13 & 2\\
               KIC7867854 & 18:42:12.12 & +43:37:42.1 & 4906 & 47.0 & 2.42 & 0.25 & -0.41 & 0.12 & 7.95 & 0.06 & 7.64 & 0.08 & 8.57 & 0.05 & 6.0 & 3.0 & 0.95 & 0.89 & 1.00 & 1.21 & 0.42 & 0.71 & 0.01 & 2\\
               KIC8039416 & 19:41:31.65 & +43:49:21.1 & 4650 & 49.0 & 2.4 & 0.17 & -0.03 & 0.09 & 8.37 & 0.02 & 7.96 & 0.03 &  & 0.1 & 12.0 & 2.0 & 1.02 & 0.99 & 1.08 & 1.46 & 0.50 & 1.23 & 0.03 & 2\\
               KIC8078554 & 18:52:38.48 & +43:55:51.9 & 4778 & 56.0 & 2.46 & 0.24 & -0.11 & 0.12 & 8.03 & 0.04 & 7.82 & 0.08 & 8.59 & 0.05 & 10.0 & 3.0 & 1.22 & 1.13 & 1.29 & 1.11 & 0.48 & 1.51 & 0.08 & 2\\
               KIC8088244 & 19:13:06.65 & +43:55:50.6 & 4882 & 46.0 & 2.38 & 0.25 & -0.48 & 0.12 & 7.91 & 0.07 & 7.26 & 0.03 & 8.56 & 0.05 &  &  & 0.98 & 0.86 & 1.04 & 1.18 & 0.38 & 1.30 & 0.12 & 2\\
               KIC8210100 & 18:42:15.04 & +44:10:14.1 & 4525 & 51.0 & 2.51 & 0.14 & 0.04 & 0.08 & 8.32 & 0.02 & 8.15 & 0.08 & 8.81 & 0.05 & 16.0 & 2.0 & 1.48& 1.46 & 1.50 & 1.42 & 0.34 & 1.67 & 0.14 & 2\\
               KIC8246635 & 19:45:05.21 & +44:06:38.8 & 4595 & 54.0 & 2.52 & 0.17 & 0.04 & 0.1 & 8.3 & 0.05 & 8.09 & 0.03 & 8.74 & 0.05 & 16.0 & 2.0 & 1.56 & 1.52 & 1.57 & 1.50 & 0.44 & 1.70 & 0.14 & 2\\
               KIC8258311 & 19:57:02.19 & +44:08:32.7 & 4976 & 45.0 & 2.48 & 0.24 & -0.39 & 0.11 & 7.85 & 0.01 & 7.61 & 0.12 & 8.53 & 0.05 &  &  & 1.14 & 1.10 & 1.21 & 1.07 & 0.49 & 1.38 & 0.05 & 2\\
               KIC8686058 & 19:22:25.76 & +44:53:49.5 & 4844 & 48.0 & 2.56 & 0.21 & -0.14 & 0.1 & 7.99 & 0.1 & 7.9 & 0.07 & 8.66 & 0.1 & 8.0 & 2.0 & 1.56& 1.52 & 1.59 & 1.06 & 0.50 & 1.85 & 0.13 & 2\\
               KIC8753234 & 19:26:46.80 & +44:54:36.8 & 4791 & 50.0 & 2.65 & 0.19 & -0.01 & 0.1 & 8.19 & 0.01 & 8.11 & 0.05 & 8.72 & 0.05 & 17.0 & 5.0 & 1.86 & 1.83 & 1.88 & 2.07 & 0.584 & 1.75 & 0.08 & 2\\
               KIC8813946 & 19:19:55.06 & +45:01:55.5 & 4858 & 48.0 & 2.77 & 0.18 & 0.07 & 0.09 & 8.2 & 0.05 & 8.08 & 0.01 & 8.73 & 0.05 & 19.0 & 5.0 & 2.01 & 1.98 & 2.07 & 2.43 & 0.51 & 1.87 & 0.12 & 2\\
               KIC8936084 & 18:57:19.73 & +45:16:12.2 & 5039 & 44.0 & 2.41 & 0.25 & -0.49 & 0.11 & 7.74 & 0.15 & 7.48 & 0.1 & 8.4 & 0.05 &  &  & 0.95 & 0.84& 0.98 & 1.05 & 0.45 & 0.72 & 0.02 & 2\\
               KIC9173371 & 19:52:48.35 & +45:31:44.2 & 5059 & 49.0 & 2.91 & 0.24 & -0.02 & 0.11 & 8.06 & 0.05 & 8.01 & 0.03 & 8.61 & 0.05 & 14.0 & 3.0 & 2.21 & 2.11 & 2.30& 2.56 & 0.39 & 2.38 & 0.05 & 2\\
               KIC9652494 & 19:32:44.23 & +46:18:38.2 & 4779 & 56.0 & 2.54 & 0.24 & -0.11 & 0.12 & 8.09 & 0.05 & 7.88 & 0.04 & 8.67 & 0.05 & 11.0 & 1.0 & 1.50 & 1.47 & 1.55 & 1.16 & 0.50 & 1.72 & 0.15& 2\\
               KIC9705687 & 19:19:37.58 & +46:29:39.6 & 5143 & 48.0 & 2.78 & 0.25 & -0.22 & 0.11 & 7.91 & 0.05 & 7.83 & 0.01 & 8.45 & 0.05 &  &  & 1.95& 1.92 & 2.02 & 2.56 & 0.31 & 1.83& 0.10 & 2\\
               KIC10186053 & 18:43:19.55 & +47:13:010.0 & 4560 & 58.0 & 2.4 & 0.19 & -0.08 & 0.11 & 8.35 & 0.01 & 8.0 & 0.05 & 8.84 & 0.05 & 12.0 & 2.0 & 1.02 & 0.99 & 1.05 & 1.20 & 0.36 & 1.19 & 0.07 & 2\\
               KIC10404994 & 19:27:45.48 & +47:32:13.0 & 4757 & 41.0 & 2.52 & 0.17 & -0.13 & 0.08 & 8.09 & 0.02 & 7.79 & 0.02 & 8.69 & 0.05 & 11.0 & 2.0 & 1.48 & 1.43 & 1.53 & 1.60 & 0.47 & 1.77 & 0.07 & 2\\
               KIC10426854 & 19:56:41.70 & +47:33:29.7 & 4976 & 46.0 & 2.52 & 0.25 & -0.31 & 0.12 & 7.87 & 0.08 & 7.72 & 0.04 & 8.55 & 0.05 & 6.0 & 2.0 & 1.66 & 1.41 & 1.68 & 1.03 & 0.48 & 1.72 & 0.07 & 2\\
               KIC10604460 & 19:34:39.50 & +47:49:56.7 & 4548 & 56.0 & 2.39 & 0.17 & -0.04 & 0.1 & 8.43 & 0.01 & 7.99 & 0.02 & 8.86 & 0.05 & 15.0 & 2.0 & 1.04 & 1.01 & 1.09 & 1.25 & 0.39 & 1.15 & 0.05 & 2\\
               KIC10716853 & 19:01:06.55 & +48:02:08.0 & 4854 & 52.0 & 2.6 & 0.23 & -0.05 & 0.12 & 8.08 & 0.05 & 7.81 & 0.05 & 8.68 & 0.05 & 7.0 & 2.0 & 1.68 & 1.65 & 1.71 & 1.05 & 0.56 & 1.71 & 0.18 & 2\\
               KIC11188067 & 19:22:08.09 & +48:50:04.4 & 4552 & 59.0 & 2.48 & 0.19 & 0.03 & 0.11 & 8.34 & 0.02 & 8.06 & 0.06 & 8.84 & 0.1 & 10.0 & 2.0 & 1.39 & 1.37 & 1.44 & 1.36 & 0.39 & 1.25 & 0.08 & 2\\
               KIC11247049 & 19:33:12.36 & +48:54:45.8 & 4727 & 54.0 & 2.42 & 0.21 & -0.15 & 0.11 & 8.11 & 0.02 & 8.01 & 0.04 & 8.73 & 0.1 & 8.0 & 2.0 & 1.41 & 1.26 & 1.50 & 1.21 & 0.42 & 1.35 & 0.04 & 2\\
               KIC11352756 & 19:32:19.00 & +49:10:47.7 & 4552 & 60.0 & 2.33 & 0.21 & -0.15 & 0.12 & 8.3 & 0.03 & 7.76 & 0.07 & 8.81 & 0.05 & 12.0 & 3.0 & 1.02 & 0.99 & 1.07 & 1.18 & 0.33 & 1.10 & 0.06& 2\\
               KIC11410549 & 19:42:03.28 & +49:17:05.0 & 4678 & 49.0 & 2.44 & 0.17 & -0.09 & 0.09 & 8.19 & 0.02 & 7.91 & 0.06 & 8.71 & 0.1 & 10.0 & 2.0 & 1.29 & 1.23 & 1.36 & 1.47 & 0.49 & 1.56 & 0.05 & 2\\
               KIC11469401 & 19:52:24.05 & +49:23:09.4 & 4326 & 101.0 & 2.35 & 0.19 & 0.32 & 0.12 & 8.65 & 0.01 & 8.48 & 0.1 & 8.99 & 0.05 &  &  & 1.08 & 1.03 & 1.11& 1.31 & 0.39 & 1.30 & 0.05 & 2\\
               KIC11757739 & 19:15:08.39 & +49:58:39.5 & 4863 & 43.0 & 2.48 & 0.21 & -0.32 & 0.1 & 7.98 & 0.02 & 7.66 & 0.03 & 8.56 & 0.1 & 7.0 & 2.0 & 1.27 & 1.23 & 1.32 & 1.08 & 0.41 & 1.49 & 0.12 & 2\\
               KIC11819760 & 19:38:54.60 & +50:05:57.1 & 4748 & 51.0 & 2.37 & 0.22 & -0.19 & 0.11 & 8.07 & 0.02 & 7.65 & 0.04 & 8.63 &  & 10.0 & 2.0 & 1.21 & 1.11 & 1.31& 1.21 & 0.44 & 1.10 & 0.07 & 2\\
               KIC12884274 & 19:18:40.89 & +52:14:50.6 & 4670 & 54.0 & 2.52 & 0.17 & 0.01 & 0.1 & 8.31 & 0.01 & 8.17 & 0.08 & 8.82 & 0.05 & 13.0 & 2.0 & -99.99 & -99.99 & -99.99 & 1.684 & 0.51 & -99.99 & -99.99 & 2\\
               KIC5612549 & 19:18:08.25 & +40:51:16.8 & 4866 & 44.0 & 2.35 & 0.23 & -0.37 & 0.11 & 7.89 & 0.01 & 7.64 & 0.06 & 8.55 & 0.05 & 5.0 & 1.0 & -99.99 & -99.99 & -99.99 & 1.07 & 0.37 & -99.99 & -99.99 & 2\\
		 \hline
	\end{tabular}
	}
	\label{observations_table}
	\end{center}
\end{table}
\end{landscape}

\end{appendix}
\end{document}